\documentclass[aps,pra,reprint,twocolumn,superscriptaddress,showpacs,showkeys,a4paper,longbibliography]{revtex4-1}

\usepackage{amsmath,amssymb,amstext}
\usepackage[usenames,dvipsnames]{color}
\usepackage{graphicx,subfigure,relsize}
\usepackage{bm,bbold,braket}
\usepackage{natbib}

\usepackage{umoline}

\definecolor{mygreen}{rgb}{0,0.5,0}
\definecolor{myblue}{rgb}{0,0,0.75}
\definecolor{mymagenta}{cmyk}{0,1,0,0.12}

\newcommand{\eq}[1]{\begin{equation} #1 \end{equation}}
\newcommand{\eqa}[1]{\begin{eqnarray} #1 \end{eqnarray}}

\newcommand{\vect}[1]{\ensuremath{\bm{#1}}}

\newcommand{\ue}{\mathrm{e}}

\newcommand{\sgn}{\mathrm{sgn}}

\newcommand{\Nm}{N_{\mathrm{m}}}
\newcommand{\tH}{\tilde{H}}
\newcommand{\tG}{\tilde{G}}
\newcommand{\tS}{\tilde{S}}
\newcommand{\tsigma}{\tilde{\sigma}}

\newcommand*\widebar[1]{%
  \hbox{%
    \vbox{%
      \hrule height 0.5pt % The actual bar
      \kern0.3ex%0.5ex%         % Distance between bar and symbol
      \hbox{%
        \kern-0.1em%      % Shortening on the left side
        \ensuremath{#1}%
        \kern-0.1em%      % Shortening on the right side
      }%
    }%
  }%
}

\begin{document}

%Title of paper
\title{Quantum simulation of a lattice Schwinger model in a chain of trapped ions}

\author{P.~Hauke}
    \email{philipp.hauke@uibk.ac.at}    
    \affiliation{Institute for Quantum Optics and Quantum Information of the Austrian Academy of Sciences, A-6020 Innsbruck, Austria}
\author{D.~Marcos}
    \affiliation{Institute for Quantum Optics and Quantum Information of the Austrian Academy of Sciences, A-6020 Innsbruck, Austria}
\author{M.~Dalmonte}
    \affiliation{Institute for Quantum Optics and Quantum Information of the Austrian Academy of Sciences, A-6020 Innsbruck, Austria}
        \affiliation{Institute for Theoretical Physics, University of Innsbruck, A-6020 Innsbruck, Austria}
\author{P.~Zoller}
    \affiliation{Institute for Quantum Optics and Quantum Information of the Austrian Academy of Sciences, A-6020 Innsbruck, Austria}
    \affiliation{Institute for Theoretical Physics, University of Innsbruck, A-6020 Innsbruck, Austria}
\thanks{}
%\altaffiliation{}

\date{\today}

\begin{abstract}
We discuss how a lattice Schwinger model can be realized in a linear ion trap, allowing a detailed study of the physics of Abelian lattice gauge theories related to one-dimensional quantum electrodynamics. 
Relying on the rich quantum-simulation toolbox available in state-of-the-art trapped-ion experiments, we show how one can engineer an effectively gauge-invariant dynamics by imposing energetic constraints, provided by strong Ising-like interactions. 
Applying exact diagonalization to ground-state and time-dependent properties, we study the underlying microscopic model, and discuss undesired interaction terms and other imperfections. 
As our analysis shows, the proposed scheme allows for the observation in realistic setups of spontaneous parity- and charge-symmetry breaking, as well as false-vacuum decay.
Besides an implementation aimed at larger ion chains, we also discuss a minimal setting, consisting of only four ions in a simpler experimental setup, which enables to probe basic physical phenomena related to the full many-body problem. The proposal opens a new route for analog quantum simulation of high-energy and condensed-matter models where gauge symmetries play a prominent role.
\end{abstract}



\maketitle

\section{Introduction}

At present, trapped ions are one of the physical systems in quantum-information science with the largest number of achievements \cite{Haeffner2008,Monroe2013}. 
By coupling pseudo-spins represented by internal atomic states to collective lattice vibrations, high-fidelity entangling gates are now routinely performed in the lab \cite{Benhelm2008,Monz2009,Lanyon2011,Khromova2012}.
A natural step forward undertaken in recent years was to exploit these technological possibilities for the quantum simulation of spin models \cite{Johanning2009,Schneider2012,Blatt2012}, in both digital \cite{Lanyon2011,Barreiro2011} and analog \cite{Friedenauer2008,Kim2009,Kim2010,Kim2011,Islam2011,Islam2012,Britton2012,Richerme2013a} protocols. 
The underlying idea of such quantum simulations is to take advantage of accurate control of a physical quantum system, and to thus engineer an effective dynamics that mimics a quantum many-body model of interest  \cite{Cirac2012,Hauke2011d,Lewenstein2012,Bloch2012,AspuruGuzik2012,Houck2012}.
First extensions into the domain of high-energy physics have also been undertaken, e.g., by simulating the Dirac equation \cite{Gerritsma2010}, or with proposals for simulation of coupled quantum fields \cite{Casanova2011} or the Majorana equation \cite{Casanova2011a}.  
But it remains an outstanding challenge in this field to extend these ideas to lattice gauge theories, whose many facets, from static to dynamical properties, present several technical difficulties for classical computations \cite{Montvay1994,Creutz1997,DeGrand2006,Gattringer2010}. 
In the context of ultracold neutral atoms, several proposals for the quantum simulation of lattice gauge theories have been made recently  \cite{Kapit2011,Zohar2011,Banerjee2012,Tagliacozzo2012,Zohar2012,Kasamatsu2012,Banerjee2013,Tagliacozzo2013,Zohar2013,Zohar2013a}, but the realization of gauge symmetries in alternative atomic or optical systems is currently unexplored \footnote{Very recently, a proposal for superconducting qubits has been put forward \cite{Marcos2013}}. 
The aim of this article is to show that the excellent control of the microscopic dynamics reached in ion traps may offer interesting perspectives in this direction. 
Concretely, we propose -- relying on existing trapped-ion technology -- a quantum simulation of the lattice Schwinger model \cite{Schwinger1962}, which is a one-dimensional (1D) version of quantum electrodynamics. 

Currently, quantum simulation of gauge theories is receiving an increasing degree of interest, as these theories represent one of the most solid and elegant theoretical frameworks able to capture a variety of physical phenomena. 
For example, in condensed matter physics, gauge theories play a prominent role in frustrated spin systems, where the identification of emergent degrees of freedom in terms of gauge fields has provided a deeper insight into the physics of quantum spin liquids and exotic insulators \cite{Kogut1979,Lee2006,Lacroix2010}. 
At a microscopic level, in the standard model of particle physics, gauge theories provide a natural description of interactions between fundamental constituents of matter~\cite{Montvay1994,Creutz1997,DeGrand2006,Gattringer2010}. 
Despite their apparent simplicity, solving gauge theories is generally challenging, a long-standing example being the theory of strong interactions, quantum chromodynamics \cite{Montvay1994,Creutz1997,Gattringer2010,DeGrand2006}. 
One of the main reasons for the difficulties in solving gauge theories is that they may present non-perturbative effects that are hard to capture within diagrammatic expansions, preventing the application of unbiased analytical approaches to the many-body problem.

Some of these limitations can be circumvented in the framework of lattice gauge theories (LGTs) \cite{Wilson1974,Montvay1994,Creutz1997,DeGrand2006,Gattringer2010}. 
Here, by means of Monte Carlo simulation of the corresponding lattice action, a broad regime of interaction parameters  
becomes accessible, thus allowing the investigation of non-perturbative effects with controlled numerical techniques \cite{Lacroix2010,Gattringer2010}. 
Nevertheless, classical simulations are severely limited by the sign problem, which prevents an accurate description of finite-density regimes (as relevant, e.g., for the core of dense neutron stars) and out-of-equilibrium dynamics (which is realized in heavy-ion collider experiments). 
Given these difficulties, it becomes particularly attractive to develop a quantum simulator of LGTs.

\begin{figure}
\centering
\includegraphics[width=1\columnwidth]{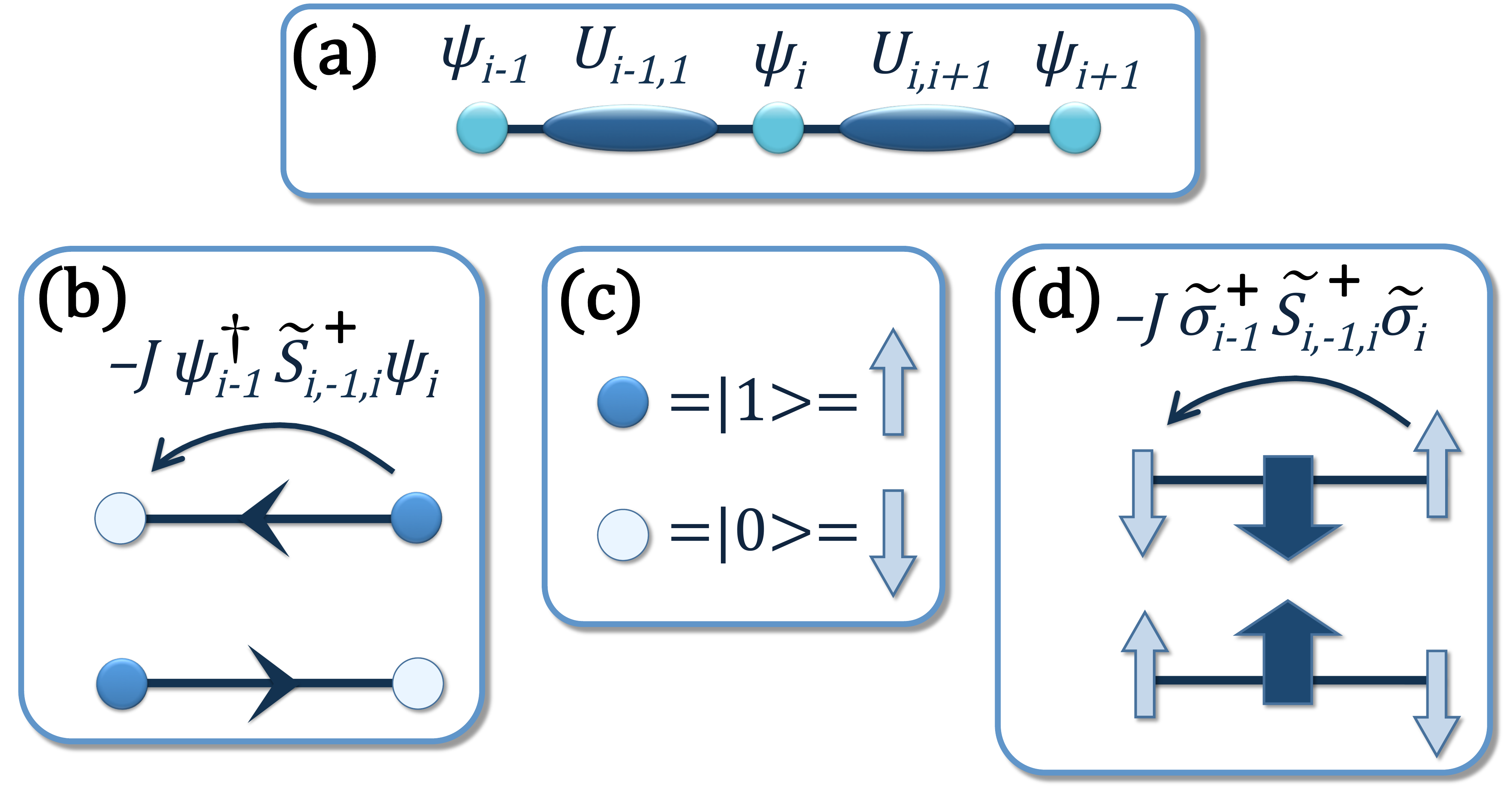}
\caption{ 
{\bf (a) Lattice gauge theory.} Fermions with annihilation operators $\psi_i$, living on lattice sites $i$, couple to gauge fields, represented by the operators $U_{i,i+1}$, living on links $i,i+1$. 
{\bf (b)} A fermion tunnels from site $i$ to $i-1$ and flips the gauge field $\tS_{i-1,i}$ on the link in between. 
{\bf (c)} A filled (empty) bullet denotes an occupied (empty) site. 
These can be mapped to spin up / down. 
{\bf (d)} The tunneling process of panel b expressed in spin language. 
}
\label{fig:tunneling_process}
\end{figure}

The basic elements of a LGT are (typically fermionic) matter fields $\psi_i$ occupying lattice sites $i$, coupled to bosonic degrees of freedom, the gauge fields $U_{ij}$, which live on the links between neighboring sites (see Fig.~\ref{fig:tunneling_process}a). 
In the context of a quantum simulator consisting of ultracold neutral atoms in an optical lattice, the fermionic fields can be naturally implemented with fermionic atoms.
In contrast, the basic degrees of freedom in a trapped-ion quantum simulator are spins and bosons, constituted by internal states and collective ion vibrations, respectively. 
To make fermionic matter fields accessible to an ion setup, we therefore consider one-dimensional ion chains, allowing to use the Jordan--Wigner transformation to map fermionic degrees of freedom to pseudo-spins. 
Linear chains have the additional advantage that they are the natural geometry implemented in linear Paul traps. 

Regarding the gauge fields, recent ideas to simulate {\it classical} gauge fields have emerged in the context of trapped ions \cite{Bermudez2011a,Bermudez2012,Shi2013}, paralleling an exciting development in optical-lattice experiments \cite{Dalibard2011,Aidelsburger2011,Jimenez2012,Struck2012,Struck2013}. 
Such classical fields have no dynamics of its own and may be described by a simple phase picked up during a tunneling process, $U_{ij}=\ue^{i\alpha_{ij}}$, with $\alpha\in\mathbb{R}$. 
Here, in contrast, we are interested in {\it dynamical} gauge fields, where $U_{ij}$ becomes a quantum field. 
In the Wilson formulation of LGTs \cite{Wilson1974} the $U_{ij}$ are characterized by continuous degrees of freedom. 
In a trapped-ion setup with discrete degrees of freedom, a natural formalism to implement the gauge fields is provided by the so called {\it quantum link models} (QLMs) \cite{Horn1981,Orland1990,Chandrasekharan1997,Brower1999,Wiese2013}, which represent a generic class of models that capture gauge invariance and at the same time facilitate quantum simulations. 
In this framework, Abelian gauge fields can be represented by spin-$S$ quantum operators, where the limit $S\to \infty$ recovers the corresponding continuous-variable gauge theory such as quantum electrodynamics or quantum chromodynamics. 
For lower spin representations, QLMs retain exactly the featured gauge symmetry and still share many features with the corresponding continuum gauge theories, such as the physics of confinement, string breaking, or false vacuum decay \cite{Wiese2013}. This last one is the example that we will analyze in the present article. 
The goal of this article is \textsl{(i)} to propose a realistic architecture, suitable for large ion chains, to realize such a QLM, and \textsl{(ii)} to illustrate how minimal instances of LGTs can be accessed even in basic setups. These points thus track a possible experimental roadmap where steps of increasing difficulty demonstrate significant signatures of many-body physics. 

The present article is organized as follows. 
First, we introduce a QLM version of the Schwinger model  (Sec.~\ref{cha:idealQLM}), focusing especially on the spontaneous charge-conjugation and parity symmetry breaking encountered in this model.
Then, in Sec.~\ref{cha:MicroscopicModel}, we explain how engineered spin--spin interactions can be used to enforce gauge invariance and to construct a microscopic Hamiltonian that -- to second-order perturbation theory -- generates the dynamics of the ideal QLM. 
In Sec.~\ref{cha:numericsShortChains}, we show the validity of the effective model by studying static and dynamic properties of the microscopic Hamiltonian, using exact diagonalizations of finite chains. 
Since experiments will likely start with very few ions, we present in Sec.~\ref{cha:minimalModel} a simpler (although, in contrast to Sec.~\ref{cha:MicroscopicModel}, not scalable to larger chains) implementation with four ions. We show that this simple system already exhibits features related to the full many-body problem. 
In Sec.~\ref{cha:errors}, we outline a possible experimental sequence and discuss the most prominent error sources. 
Finally, in Sec.~\ref{sec:outlook}, we present our conclusions and an outlook on future directions.

\section{One-dimensional U(1) lattice gauge theories as quantum link models\label{cha:idealQLM}}
In this article, we focus on a conceptually simple instance of a gauge theory,
a one-dimensional U(1) QLM with staggered fermions. 
This model is chosen as a good compromise between feasibility and fundamental interest, allowing to demonstrate basic phenomena such as string breaking \cite{Hebenstreit2013} shared with other, more complex LGTs~\cite{Banerjee2012}. 
It can be understood as a QLM version of the Schwinger model, which represents quantum electrodynamics in one dimension. 
The Hamiltonian describing the dynamics of this QLM is
\eqa{
\label{eq:QLM_fermionic}
\tH &=& -\frac{J}{2} \sum_{i=1}^{\Nm} \left( \psi_i^{\dagger} U_{i,i+1} \psi_{i+1} + \mathrm{h.c.}\right) \nonumber\\
	&+& m \sum_{i=1}^{\Nm} \left(-1\right)^i \psi_i^{\dagger}\psi_i + \frac {g^2} 2 \sum_{i=1}^{\Nm} \left(  E_{i,i+1} \right)^2\\
&=& \tH_J+\tH_m+\tH_g
\,.
}
The field operators $\psi_i^{\dagger}$ and $\psi_i$ at the lattice sites $i$ are creation and annihilation operators, respectively, of `staggered fermion' particles. Staggered fermions are characterized by a mass term $\tH_m$ with alternating sign, and provide an elegant way of simultaneously incorporating matter and antimatter fields by using a single degree of freedom on bipartite lattices. 
In an associated quark picture, a fermion on even sites corresponds to a quark ($q$) and the absence of a fermion on odd sites to an anti-quark ($\bar{q}$); the opposite configurations denote the absence of quarks / anti-quarks. In order to keep the symmetry between matter and antimatter, the number of lattice sites $\Nm$ has to be even.

In Hamiltonian $\tH$, the term $\tH_g$ describes the energy of the gauge field, with $E_{i,i+1}$ being an electric field operator associated with the link $i,i+1$. 
While in the Wilson formulation of LGTs, the link variables $U_{i,i+1}$ are parallel transporters spanning an infinite-dimensional Hilbert space, in the QLM formulation they may be defined as spin operators of arbitrary representation $S$, by setting $U_{i,i+1}=\tS^{+}_{i,i+1}$ and $E_{i,i+1}\propto \tS^{z}_{i,i+1}$. 
This choice of `quantum links' retains the commutation relations $\left[E_{i,i+1},U_{i,i+1}\right] = U_{i,i+1}$, required to assure gauge covariance of the gauge fields. 
The term $\tH_J$, finally, couples the matter and gauge fields via an assisted tunneling (see Fig.~\ref{fig:tunneling_process}b-c),
and represents a key ingredient of any gauge theory in presence of matter fields \cite{Kogut1979,Montvay1994,Creutz1997,DeGrand2006,Gattringer2010}.

\begin{figure*}
\centering
\includegraphics[width=1.95\columnwidth]{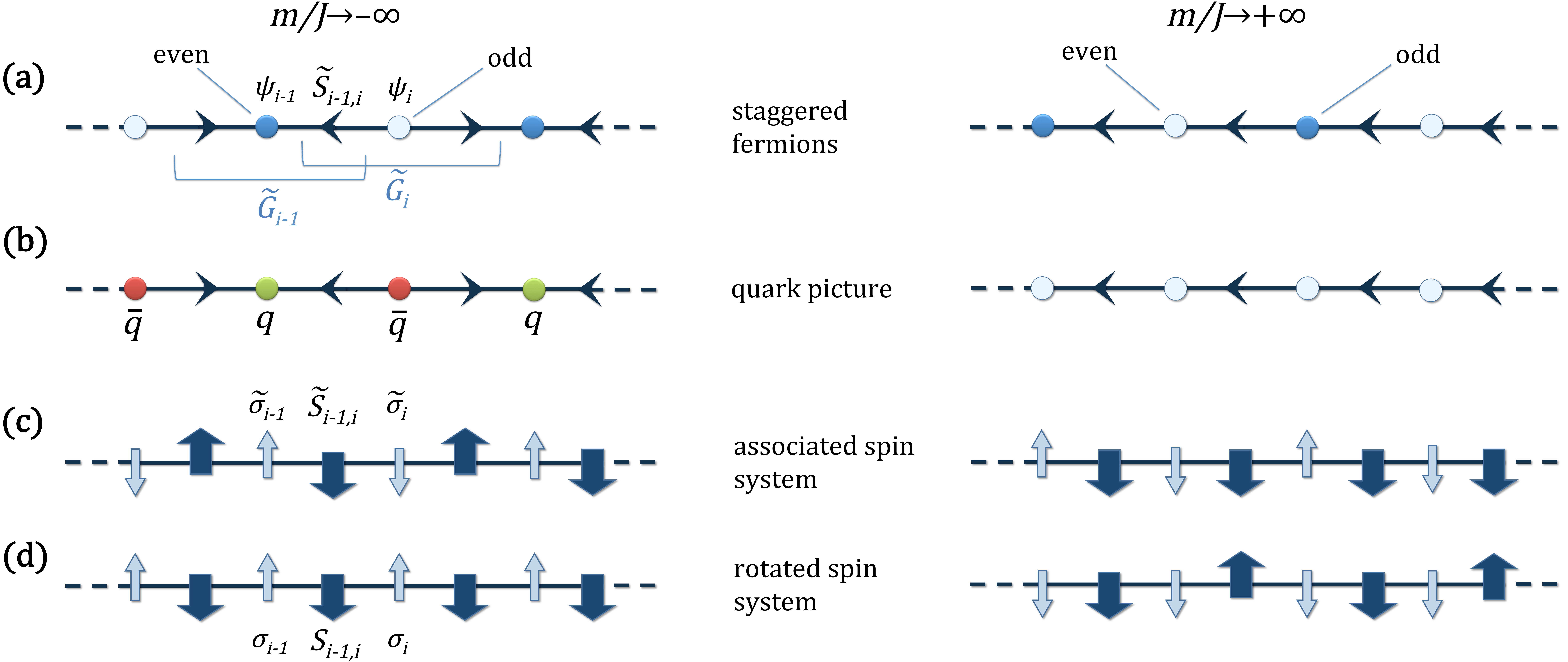}
\caption{ 
{\bf Ground states of the quantum link model} (QLM) in several equivalent representations, for $m/J\to -\infty$ (left) and $m/J\to +\infty$ (right, only one of two degenerate solutions is drawn). 
Sketches are for vanishing quantum fluctuations ($J=0^+$), periodic boundary conditions, and gauge fields represented by spins 1/2. 
{\bf (a) Staggered-fermion QLM.} 
Physical states are invariant under gauge transformations with generators $\tG_i$, which span one matter and two gauge fields. 
{\bf (b) Associated quark picture.} 
Fermions on even sites correspond to quarks ($q$, light green) and vacancies on odd sites to anti-quarks ($\bar{q}$, dark red); the opposite configurations denote the absence of (anti-)quarks. 
For $m/J\to-\infty$, alternating sources and drains of flux completely cover the chain, and the ground state is charge (C) and parity (P) invariant. 
For $m/J\to+\infty$, the system is empty of matter and has an un-broken flux string, breaking C and P symmetry. 
In this case, the Gauss law \eqref{eq:generator_fermionic} requires the same orientation for all gauge fields, allowing a double energy degeneracy, corresponding to the two directions of electric flux (only one of which is visualized here). 
{\bf (c) Equivalent spin model} $\tH_{\mathrm{QLM}}$. 
Fermions are mapped to spins via the Jordan--Wigner transformation. 
For $S=1/2$ quantum links, right (left) flowing flux translates to an up (down) gauge-field spin. 
{\bf (d) Rotated spin model $H_{\mathrm{QLM}}$ for experminental implementation.} 
At $m/J\to-\infty$, the order parameters are characterized by $\sum_i\langle S_{i,i+1}^z \rangle = -1$ and $\sum_i\langle \sigma_{i}^z \rangle =1$. In the C and P breaking state at $m/J\to-\infty$, they take the values $\sum_i\langle S_{i,i+1}^z \rangle = 0$ and $\sum_i\langle \sigma_{i}^z \rangle =-1$. 
}
\label{fig:systems}
\end{figure*}

The most distinctive feature of gauge theories such as the QLM of Eq.~\eqref{eq:QLM_fermionic} is the presence of local 
({\it gauge}) symmetries. 

These imply that, 
given a set of local symmetry generators $\{ \tilde{G}_i \}$, the system Hamiltonian is invariant under gauge transformations of the form 
\begin{equation}
\label{eq:gauge_symmetry}
\tilde{H}^\prime=\prod_{i} \exp\{ i\alpha_i \tilde G_i \} \; \tilde H \; \prod_i \exp\{ -i\alpha_i \tilde G_i \}\,,
\end{equation}
where $\{ \alpha_i \}$ is an arbitrary set of parameters. 
For the gauge theory we are interested in, the generators read 
\begin{equation}
\label{eq:generator_fermionic}
\tilde G_i  = E_{i-1,i} - E_{i,i+1} + \psi^{\dagger}_i \psi_i +\frac{(-1)^i - 1}{2}\,, 
\end{equation} 
where the constant term is due to the use of staggered fermions. 

The gauge invariance expressed in Eq.~\eqref{eq:gauge_symmetry} is equivalent to the condition $[\tilde H, \tilde G_i]=0\; \forall i$, implying that the Hamiltonian does not mix eigenstates of $\tilde G_i$ with different eigenvalues $\lambda_{\tilde G_i}$. 
Thus, gauge invariance can be imposed by restricting the Hilbert space to a sector with fixed $\lambda_{\tilde G_i}$. 
In the case of the present U(1) gauge theory, the sector we are interested in is the one of balanced matter, i.e., equal number of quarks and anti-quarks, and we will restrict our considerations to this subspace. 
This subspace is defined by $\lambda_{\tilde G_i}=0\; \forall\, i$, i.e., it consists of all states satisfying the so called {\it Gauss law}~\cite{Creutz1997,Wiese2013}
\begin{equation}
\label{Gauss_law}
\tilde G_i \ket{\rm phys} = 0\quad\forall\,i\,.
\end{equation}
In the continuum limit, this condition reduces to the usual Gauss law of quantum electrodynamics, $\vec{\nabla} \cdot \vec{E} = \rho$,  where $\rho$ is the density of charged matter.
We call states fulfilling the Gauss law the `physical' states of our problem.
The challenge for any quantum simulation of a gauge theory is to engineer the dynamics given by $\tilde  H$, which connects only physical states, while avoiding gauge-variant perturbations to states $\ket{\psi}$ with $\tilde G_i \ket{\rm \psi} \neq 0$. 
(Such perturbations correspond to, e.g., the spontaneous creation of an imbalance between quarks and anti-quarks, and are therefore `unphysical'.) 
Before explaining how this challenge can be met in the trapped-ion chain, we first discuss the basic physics of the QLM of Eq.~\eqref{eq:QLM_fermionic}.

\subsection{Charge-conjugation and parity symmetry breaking\label{cha:CPsymmetryBreaking}}

Despite its simplicity, the model of Eq.~(\ref{eq:QLM_fermionic}) captures the physics of interesting phenomena associated to gauge theories.
When the quantum links are represented by spins $S\geqslant 1$, the competition between the electric-field term $\tH_g$ and the mass term $\tH_m$, leads, for $J\ll m, g^2$, to a crossover between a `string' state characterized by $\sum\langle \tS_{i,i+1}^z \rangle \neq 0$ (for $m\gg g^2$) and a `meson' state featuring $\langle \tS_{i,i+1}^z \rangle = 0$ (for $m\ll g^2$) \cite{Banerjee2012,Marcos2013}. 

In the $S=1/2$ case that we are going to discuss in the rest of this article~\footnote{Higher spins are possible, but require considerable fine tuning of laser strengths.},
the electric-field term $\tH_g$ is constant, and the system dynamics is described by the 
competition between the kinetic and the mass term.
This competition leads to the spontaneous breaking of the parity (P) and charge-conjugation (C) symmetries, defined as
\begin{subequations}
\begin{align}
& ^{\rm P}\psi_{\frac \Nm 2+i}=\psi_{\frac \Nm 2-i}\,, \;\;\;\;\;\;\;\, ^{\rm P}\psi_{\frac \Nm 2+i}^\dagger=\psi_{\frac \Nm 2-i}^\dagger\,; \\
& ^{\rm P} U_{\frac \Nm 2+i,\frac \Nm 2+i+1}= U_{\frac \Nm 2-i-1,\frac \Nm 2-i}^\dagger\,, \\
& ^{\rm P} E_{\frac \Nm 2+i,\frac \Nm 2 +i+1}= E_{\frac \Nm 2-i-1,\frac \Nm 2-i}\,; \\
& ^{\rm C}\psi_i=(-1)^{i+1}\psi_{i+1}^{\dagger}\,, \;\;\;\;\;\; ^{\rm C}\psi_i^\dagger=(-1)^{i+1}\psi_{i+1}\,;\\
& ^{\rm C} U_{i,i+1}= U_{i+1,i+2}^\dagger\,, \;\;\;\;\;\;\; ^{\rm C} E_{i,i+1}=- E_{i+1,i+2}\,.
\end{align}
\end{subequations}
For sufficiently large positive values of $m/J$, the gauge fields are polarized, $\sum_i\langle \tS_{i,i+1}^z \rangle \neq 0$ (Fig.~\ref{fig:systems}a, right side). In the corresponding state, the parity and charge conjugation symmetries are spontaneously broken. 
For large negative values of $m/J$, parity and charge conjugation are restored, yielding $\sum_i\langle \tS_{i,i+1}^z \rangle = 0$ (Fig.~\ref{fig:systems}a, left side).

To illustrate the physical meaning of this symmetry breaking, consider the mapping to the associated quark picture (Fig.~\ref{fig:systems}b). 
In this picture, C transforms a quark on site $i$ to an anti-quark at site $i+1$, etc. 
Now, for a pictorial visualization, consider a simplified scenario where $J=0$, under periodic boundary conditions (for a discussion of various boundary conditions, see Ref.~\cite{Banerjee2012}). In that picture, at large negative $m$ the ground state consists of alternating sources and drains of flux, and preserves C and P. 
The ground state at large positive $m$, however, displays a gauge-field string that threads across the system. 
This state is doubly degenerate, corresponding to the two polarization directions, and breaks C and P. 

The transition from the former to the latter scenario is known as {\it false vacuum decay}.
In a system with open boundary conditions, it leads to the formation of meson bound states at the boundaries. In the fermionic picture, these are represented by a quark--antiquark pair in the first two and last two sites. These meson bound states are connected by a {\it polarized} bulk with a string of spins oriented along one direction. Since the true vacuum has in the average a vanishing electric flux, the C and P breaking string for $m/J\rightarrow \infty$ corresponds to a false vacuum.
The transition between these two situations in a trapped-ion setup will be the main focus of this article.
Notably, the phase diagram of the model Hamiltonian \eqref{eq:QLM_fermionic} is only qualitatively known and subject of ongoing research \cite{QLMTheoryInPreparation2013}; its experimental implementation would constitute a relevant instance of a quantum simulation, especially when considering dynamics, such as associated with the false-vacuum decay and string breaking \cite{Hebenstreit2013}.

\subsection{Mapping to a spin model \label{cha:IdealSpinModel}}

To make the QLM of Eq.~\eqref{eq:QLM_fermionic} accessible to a trapped-ion setup, we first map the fermionic fields to spin-1/2 degrees of freedom, using a Jordan--Wigner transformation,  
\begin{align}
\label{eq:}
&\psi_i^\dagger=\ue^{i\pi\sum_{k<i}(\tsigma_k^z+1)/2}\tsigma_i^{+}\,, \quad
\psi_i=\ue^{-i\pi\sum_{k<i}(\tsigma_k^z+1)/2}\tsigma_i^{-}\,, \nonumber \\
&\psi_i^\dagger \psi_i=\frac{\tsigma_i^{z} + 1}{2}\,.
\end{align}
Then, the Hamiltonian that we aim at simulating reads
\eq{
\label{eq:originalspinQLM}
\tH_{\mathrm{QLM}} = -\frac{J}{2} \sum_i \left( \tsigma_i^+ \tS_{i,i+1}^+ \tsigma_{i+1}^- + \mathrm{h.c.}\right)  +\frac m 2 \sum_i \left(-1\right)^i \tsigma_i^z \,.
}
Here, both types of spin operators denote the usual Pauli matrices, with $\tsigma_i$ associated to the matter fields and $\tS_{i,i+1}$ to the link operators. 
Translated to spin language, the tunneling process $J$ sketched in Fig.\ \ref{fig:tunneling_process}b becomes an assisted flip-flop process (Fig.\ \ref{fig:tunneling_process}d). This is similar to a ferromagnetic XY interaction, but accompanied by a flip of the additional spin degree of freedom $\tS$. The second part in $\tH_{\mathrm{QLM}}$ is simply a staggered, transverse magnetic field.
The QLM model is now expressed purely by spin operators, which can be represented in the ions by pseudo-spins consisting of two internal states. In the implementation discussed below, the two types of spins are distinguished by choosing different internal levels to represent $\sigma$ and $S$ spins, respectively. 

The gauge-invariant subspace that we are interested in fulfills the following constraint given by the spin version of the Gauss law \eqref{eq:generator_fermionic}-\eqref{Gauss_law}, 
\eq{
\tG_i\ket{\psi}=0\,,\quad\quad 
\tG_i=\frac 1 2 \left[ \tS_{i-1,i}^z - \tS_{i,i+1}^z + \tsigma_i^z + (-1)^i \right]\,.
}
Here, we represented the electric field operator by the Pauli $z$ matrix $\tS^{z}_{i,i+1}$, $E_{i,i+1}\equiv\tS^{z}_{i,i+1}/2$. 
The product states at $m/J=\pm\infty$ in this spin picture are sketched in Fig.~\ref{fig:systems}c. 

In the microscopic model, where gauge invariance is not built in \emph{a priori}, we will enforce it by adding the term 
\eqa{
\label{eq:tHG}
\tH_G &=& 2 V \sum_i (\tG_i)^2\\
	&=& V \sum_i \left[ \tsigma_i^z \tS_{i-1,i}^z - \tsigma_i^z \tS_{i,i+1}^z - \tS_{i-1,i}^z \tS_{i,i+1}^z \right. \nonumber \\
	& & \quad\quad + \left.  (-1)^i (\tsigma_i^z +   \tS_{i-1,i}^z -  \tS_{i,i+1}^z) + 2 \right] \,. 
}
For $V\gg\left| J\right|,\left| m\right|$, this energetically suppresses unphysical, gauge-variant transitions to states with $\tG_i\ket{\psi}\neq 0$. 

In one dimension, we can remove the various alternating signs by the basis transformation (corresponding to a staggered rotation about the $x$ axis)
\begin{subequations}
\label{eq:rotation_of_spins}
\begin{align}
\tsigma_{i}^z &\to (-1)^i \sigma_{i}^z\,,\quad\,\, &\tsigma_{i}^y &\to (-1)^i \sigma_{i}^y \,, \\
\tS_{i-1,i}^z &\to (-1)^i S_{i-1,i}^z\,, \quad\,\, &
\tS_{i-1,i}^y &\to (-1)^i S_{i-1,i}^y\,, 
\end{align}
\end{subequations}
while the $x$ components remain unchanged.  The product states at $m/J=\pm\infty$ are then rotated as sketched in Fig.~\ref{fig:systems}d. 
In this basis, the model Hamiltonian becomes
\eqa{
\label{eq:H0}
H_{\mathrm{QLM}} 
&=& -\frac{J}{2}\sum_i \left( \sigma_i^- S_{i,i+1}^+\sigma_{i+1}^- + \mathrm{h.c.}\right) + \frac m 2 \sum_i \sigma_i^z  \nonumber  \\
&\equiv& H_J + H_m  \,,
}
and the Gauss law reads
\eq{
\label{eq:Gausslaw}
G_i\ket{\psi}=0\,,\quad\quad 
G_i=(-1)^i \frac 1 2 \left( S_{i-1,i}^z + S_{i,i+1}^z + \sigma_i^z + 1 \right)\,. 
}
Therefore, the energetic restriction to the physical subspace is achieved by the Hamiltonian 
\eqa{
\label{eq:HG}
H_G &=& V \sum_i \left( \sigma_i^z S_{i-1,i}^z + \sigma_i^z S_{i,i+1}^z +S_{i-1,i}^z S_{i,i+1}^z \right. \nonumber \\
	& & \quad\quad + \left. \sigma_i^z + S_{i-1,i}^z + S_{i,i+1}^z +2 \right) \,. 
}
In view of experimental implementation, the basis transformation~\eqref{eq:rotation_of_spins} constitutes a major simplification, as it replaces interactions with alternating sign with uniform ones.
As it becomes apparent in the first three terms of $H_G$, the way gauge invariance is recovered in this spin basis is reminiscent of the basic frustration induced by Ising interactions in triangular (ladder) geometries at fixed magnetization \cite{Moessner2001}.  
The gauge-invariant subspace can therefore be seen as the ground-state manifold of an associated frustrated Ising model with longitudinal field. Out of this manifold (degenerate with respect to $H_G$), the dynamics generated by $H_J$ and $H_m$ selects one state as the ground state of the QLM.

\begin{figure*}
\centering
\includegraphics[width=2\columnwidth]{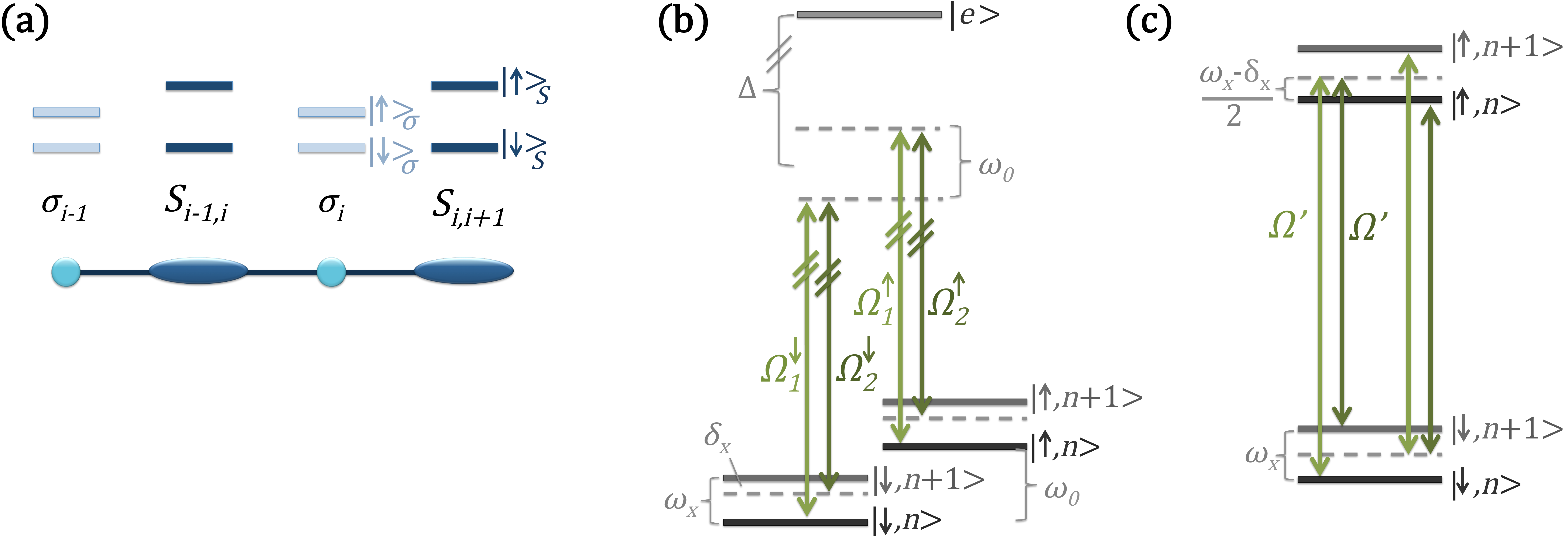}
\caption{ 
{\bf Proposed experimental scheme.}
{\bf (a) Two kinds of pseudo-spins} encode in an alternating fashion matter and gauge fields, with associated spin matrices $\sigma_i$ and $S_{i,i+1}$. The corresponding internal states are labeled as $\ket{\downarrow}_\sigma$,$\ket{\uparrow}_\sigma$ and $\ket{\downarrow}_S$,$\ket{\uparrow}_S$, respectively. 
{\bf (b-c) Spin interactions} of $zz$ type are transmitted by coupling the pseudo-spin levels (sketched for a single ion) to phonons. 
{\bf (b)} For hyperfine qubits, one can engineer these couplings via a differential light shift induced by two pairs of Raman beams (with Rabi frequencies $\Omega_{1,2}^{\uparrow,\downarrow}$ and detuning $\left|\Delta\right|\gg\omega_{\alpha}$; here $\omega_{\alpha}$ is the qubit energy difference, with $\alpha=\sigma,S$). 
The Raman transition is detuned by $\delta_x$ from the phonon frequency $\omega_x$. $\ket{n}$ denotes the phonon number Fock state. 
{\bf (c)} For optical qubits, one can use two beams with Rabi frequencies $\Omega_1=\Omega_2=\Omega^\prime$ that are tuned (with a small detuning $\delta_x$) close to half the phonon frequency $\omega_x$. 
}
\label{fig:setup}
\end{figure*}

\section{Microscopic model\label{cha:MicroscopicModel}}

In this section, we discuss how $H_{\mathrm{QLM}}$ and $H_G$ can be engineered in a trapped-ion experiment by employing effective spin--spin interactions \cite{Porras2004a,Porras2006c,Kim2008} mediated by the ion vibrations. 
Our system consists of a linear chain of ions, oriented along the $z$ direction. For simplicity, we assume equal spacings between ions, as can be achieved around the center of a long ion chain or by employing individual microtraps \cite{Chiaverini2008,Schmied2009b} (see Sec.~\ref{sec:errors} for the consequences of relaxing this conditions). 
As we will see later, to create the desired couplings $H_G$ with correct weights, we need to distinguish the $\sigma$- from the $S$-spins. 
This can be done by encoding the two types of fields in pseudo-spins with different resonance frequencies (Fig.~\ref{fig:setup}a).

Next, we describe how to engineer $H_G$, as this will enforce gauge invariance. The associated high-energy scale will also be used to implement perturbatively the effective spin coupling $\sim J$ characteristic of $H_{\mathrm{QLM}}$. 
A discussion of possible experimental errors will be given in Sec.~\ref{sec:errors}.

\subsection{Effective spin--spin interactions\label{cha:effectiveSpinSpinInteraction}}

Engineering $H_G$ in an ion setup requires the realization of single-spin operators proportional to $\sigma_i^z$ and of spin--spin interactions of $zz$ type. 
The implementation of $\sigma_i^z$ operators is straightforward, e.g., by state-selective AC-Stark shifts due to off-resonant Raman transitions. 
The $zz$ interactions can be engineered by a variety of schemes, e.g., by laser addressing of optical quadrupole qubits \cite{Kim2008,Monz2009} or by driving rf transitions in ions in spatially varying magnetic fields \cite{Mintert2001,Khromova2012}. 
The currently best-established technique for engineering $zz$ interactions \cite{Friedenauer2008,Britton2012} encodes the pseudo-spins in two hyperfine states, 
and uses Raman lasers in ``moving standing wave'' configurations to generate interactions between the spins with the help of state-dependent AC-Stark shifts \cite{Porras2004a,Porras2006c}. 
In the following, we describe the main steps for obtaining effective spin--spin interactions with the last technique. 

As sketched in Fig.~\ref{fig:setup}b, in this scheme, two lasers with optical frequencies $\omega_{1}$ and $\omega_{2}$ are far off-resonant with respect to a dipole-allowed transition to an auxiliary excited state $\ket{e}$.
To avoid spurious population of $\ket{e}$, the detuning $\Delta$ is much larger than the spontaneous decay rate of the excited state and the
Rabi frequencies $\Omega_{i}^{\mu}$, where $i=1,2$ numbers the laser beam and $\mu=\downarrow,\uparrow$ the internal state which is acted on. 
Under these conditions, one can eliminate the excited state, thus obtaining an effective Hamiltonian involving only the pseudo-spin states and collective vibrational modes. 

The resulting atom--light interaction can be written as an effective Hamiltonian for the pseudo-spins \cite{Porras2004a,Porras2006c,Bermudez2012b}
\eq{
\label{eq:Hd}
H_d=\sum_{n=1}^{N} s_n^z \frac{\hbar\Omega_n}{2}\ue^{i \vect{k}_{\mathrm{L}} \Delta \vect{r}_n - i\omega_{\mathrm{L}} t} + \mathrm{h.c.}\,. 
}
Here, $N$ is the number of ions, and the pseudo-spins are represented by operators $s_n$, which denote $\sigma$ or $S$ spins, depending on whether the $n$-th ion encodes a matter or a gauge field. 
Further, $\Delta \vect{r}_n$ is the displacement of ion $n$ from its equilibrium position; $\vect{k}_{\mathrm{L}}=\vect{k}_{1}-\vect{k}_{2}$ is the difference of the laser wave-vectors; $\omega_L=\omega_1-\omega_2$ is the difference of the laser frequencies; and $\Omega_n=(\Omega_{1,\downarrow}\Omega_{2,\downarrow}^{\star}-\Omega_{1,\uparrow}\Omega_{2,\uparrow}^{\star})_n/(2\Delta)$ is the two-photon differential Rabi frequency.
Here, we assumed that the pseudo-spin energy-splitting $\omega_{\alpha}$ is $\ll\left| \Delta\right|$ ($\alpha=\sigma,S$), meaning that the detuning to the excited level is approximately equal for the upper and the lower level. 
Additionally, in writing Eq.~\eqref{eq:Hd} we exploited the fact that unwanted transitions between the pseudo-spin levels are avoided as long as the condition $\omega_L\ll\omega_{\alpha}$ is ensured. 

For concreteness, we consider the situation where the pseudo-spin is coupled to the radial phonon modes in the $x$ direction. 
To enable this coupling, the laser beatnote $\omega_{\rm L}$ has to be tuned close to the corresponding phonon frequencies $\omega_{x}^q$.
In a linear chain of ions in a Paul trap, the radial modes are `stiff,' meaning that the energy scale associated to a local ion vibration (given by the transverse trapping potential) outweighs the coupling of vibrations via the Coulomb repulsion, i.e., $\beta_x\equiv e^2/ (M \omega_x^2 d_0^3)\ll1$, where $M$ is the ion mass, $\omega_x$ the trap frequency in $x$ direction, and $d_0$ the inter-ion distance. 
In such a situation, the radial phonon modes are only slightly dispersed and we can approximate $\omega_x^q\approx\omega_x$. 
Thus, the lasers couple the internal states almost equally to all modes.
The condition of tuning close to the phonon frequencies then becomes $\omega_L=\omega_x-\delta_x$ with $\delta_x\ll\omega_x$, as sketched in Fig.~\ref{fig:setup}b. 

Additionally, to address the vibrational modes in the $x$ direction, the difference of the laser wave-vectors $\vect{k}_{\rm L}$ has to point along the $x$ axis. Then, we can use the corresponding phonon operators to replace 
\eq{
\vect{k}_{\mathrm{L}}\cdot \Delta \vect{r}_n=\sum_{q=1}^N \eta_{nq}  {a}_q^\dagger +\mathrm{h.c.}\,.
}
Here, $\eta_{nq}={\mathcal M_{nq}} k_{\rm L}/\sqrt{2 M \omega_x^q/\hbar}$ is the Lamb--Dicke parameter, with $k_{\mathrm{L}}=\left|\vect{k}_{\mathrm{L}}\right|$. The matrices $\mathcal{M}_{n q}$, obtained by diagonalizing the elasticity matrix of the ion crystal, transform the localized vibrations of the $n$'th ion into normal modes with creation operator ${a}_q^\dagger$. 

Typically, $\eta_{nq}\ll1$, so that we can expand the exponential in Eq.~\eqref{eq:Hd} up to first order. 
Then, including the eigenenergies of the collective vibrational modes, we obtain a spin--phonon Hamiltonian of the form \cite{Porras2006c,Bermudez2012b}
\eq{
\label{eq:Hsp}
H_{\mathrm{sp}}=\sum_{q=1}^{N} \hbar \delta_x^q {a}_q^\dagger {a}_q +
i \sum_{n=1}^{N} s_n^z \frac{\hbar\Omega_{n}}{2} \sum_{q=1}^{N} \eta_{nq}{a}_q^\dagger + \mathrm{h.c.}\,.
}
Here, we transformed into an interaction picture with respect to the laser beatnote $\omega_{\mathrm{L}}$, in which the eigenfrequencies of the phonon operators are shifted to $\delta_x^q\equiv \omega_x^q-\omega_{\rm L}$. 
Moreover, we employed a rotating-wave approximation neglecting terms rotating at frequencies $\omega_x^q+\omega_L$, which is valid for $\Omega_{n}\ll \omega_{\rm L}$. Since $\omega_{\rm L}\approx \omega_{x}$, for realistic trap frequencies lying in the range $\omega_x=1-10\,$MHz, it suffices to consider $\Omega_n=0.1-1\,$MHz. 

For pseudo-spins encoded in an optical quadrupole transition, one can derive an effective Hamiltonian equivalent to Eq.~\eqref{eq:Hsp} by using two laser beams, one tuned close to the middle between carrier and red sideband, and the other between carrier and blue sideband, as depicted in Fig.~\ref{fig:setup}c. 
If these beams have the same Rabi frequency $\Omega^{\prime}_n$, one obtains the spin--phonon coupling as in Hamiltonian~\eqref{eq:Hsp}, with $\Omega_n=2\left|\Omega_n^{\prime}\right|^2/\omega_x$ \cite{Kim2008}. 
As a major advantage of this scheme, the two beams co-propagate, so that $k_{\rm L}$ is directly the laser wavenumber instead of the difference between two beams, allowing larger coupling strengths.

In Hamiltonian~\eqref{eq:Hsp}, the local spin degree of freedom couples to delocalized phonon modes, which perturbatively transmit the desired interaction between different spins. 
To see this, it is useful to separate spins and phonons by the canonical transformation $H\to \ue^{\mathcal{S}} H \ue^{-{\mathcal{S}}}$ with
\eq{
\label{eq:canonicalTransformation}
{\mathcal{S}}= i \sum_n s_n^z \frac{\Omega_n}{2} \sum_q \frac{\eta_{nq}}{\delta_x^q} {a}_q^\dagger - \mathrm{h.c.}\,.
}
In addition to linear terms $\propto\sigma_i^z$, this gives the effective spin--spin interaction \cite{Porras2004a,Deng2005,Porras2006c,Bermudez2012b}
\eq{
\label{eq:HVgeneral}
H_V=\sum_{m, n} \mathcal{V}_{mn} s_m^z s_n^z\,,
}
where the sum is taken over all pairs of ions, and where the interaction strength is
\eq{
\mathcal{V}_{mn} = - \sum_q \frac{\Re({\Omega_m}^\star \Omega_n)
\hbar^2 k_{\rm L}^2} {8 M} \frac{\mathcal{M}_{n q}\mathcal{M}_{m q}}{ \omega_{x}^q\delta_x^{q}}\,. 
\label{eq:Vgeneral}
}
For the moment, we assume that Eq.~\ref{eq:HVgeneral} is exact. We will address the errors arising in its derivation in Sec.~\ref{sec:errors}.

\subsection{Engineering the Gauss law}

When constructing $H_G$ via the spin--spin interactions \eqref{eq:HVgeneral}, we have to consider an additional subtlety: the distance $d(S_{i-1,i},S_{i,i+1})$ is twice as large as $d(S_{i-1,i},\sigma_{i})$ or $d(\sigma_{i},S_{i,i+1})$, but, for $\beta_x\ll 1$, ${\mathcal V}_{mn}$ approximately decays as a dipolar power law with distance between ions $m$ and $n$. Therefore, to obtain $H_G$ with equal couplings as given in Eq.~\eqref{eq:HG}, we need to strengthen the $S_{i-1,i}^z S_{i,i+1}^z$ interactions with respect to the terms $\sigma_{i}^z S_{i-1,i}^z$ and $\sigma_{i}^zS_{i,i+1}^z$. 
Since $\sigma$ and $S$ spins have different resonance frequencies, one can achieve different coupling strengths via global laser beams, where the beam that is close to resonance for the $S$-spin transition has a larger intensity than the one for the $\sigma$-spin transition. 
This possibility of strengthening the interaction $S_{i-1,i}^z S_{i,i+1}^z$ is the main reason why one needs two different types of pseudo-spins. 
Using $\Omega_n=\Omega_\sigma$ if $n$ is a $\sigma$ spin and $\left|\Omega_n\right|=\left|\Omega_S\right|= 8 \left|\Omega_\sigma\right|$ if $n$ denotes a $S$ spin, we obtain the Hamiltonian
\eqa{
\label{eq:HVall}
H_{V} &=& 
	    V\Bigl[\sum_{i,j} \frac{1} {D_{ij}^{\sigma S}} \sigma_i^z S_{j,j+1}^z  \\
& &+ \sum_{i< j} \frac{1}   {D_{ij}}\left( S_{i,i+1}^z  S_{j,j+1}^z + \frac{1}{2^6} \sigma_i^z \sigma_j^z \right)\Bigr]\,, \nonumber
}
with interaction energy scale 
\eq{
V =\hbar \frac{{\Re}(\Omega_S^\star\Omega_\sigma) \eta_x^2} {8\delta_{x}^2} \beta_x \omega_x\,, 
\label{eq:V}
}
where $\eta_x\equiv k_{\rm L}/\sqrt{2 M \omega_x/\hbar}$.  
In Eq.~\eqref{eq:HVall}, $D_{ij}=\left|i-j\right|^3$ and $D_{ij}^{\sigma S}=[2 \left|i-j\right|+\sgn(i-j)]^{3}$ encode the dipolar distance dependence, where we defined $\sgn(X)=+1$ if $X\geq 0$ and $\sgn(X)=-1$ otherwise. Due to the choice of $\left|\Omega_S\right|=8\left|\Omega_\sigma\right|$, the interactions for the pairs $\sigma_i^z S_{i-1,i}^z$, $\sigma_i^z S_{i,i+1}^z$, $S_{i-1,i}^z S_{i,i+1}^z$, required for $H_G$, have equal strengths, $D_{i-1,i}=D_{i,i+1}^{\sigma S}=D_{i,i-1}^{\sigma S}=1$.

Collecting the largest contributions $\sigma_i^z S_{i,i+1}^z$, $\sigma_i^z S_{i-1,i}^z$, and $S_{i-1,i}^z  S_{i,i+1}^z$, and adding suitable single-spin operators $V \sum_i (\sigma_i^z + 2 S_{i,i+1}^z)$, we obtain $H_{G}$, given by Eq.~\eqref{eq:HG}, as desired \footnote{The linear terms neglected in the transformation leading to Eq.~(\ref{eq:HVgeneral}) 
can be absorbed here.
}. 
If $V$ is stronger than all other energy scales, the system is now energetically constrained to stay in the gauge-invariant subspace. 

As an undesired byproduct, Eq.~\eqref{eq:HVall} leaves us with the interactions 
\eqa{
\label{eq:HVerr}
H_{V}^{\mathrm{err}} &=& 
	    V\sum_{i} \Bigl[ \sum_{j\neq i,i-1} \frac{1  } {D_{ij}^{\sigma S}} \sigma_i^z S_{j,j+1}^z \\
	& &\quad\,\,\,\,\,\,\,\,\,\,  +  \sum_{i<j-1 } \frac{1}   {D_{ij}} S_{i,i+1}^z  S_{j,j+1}^z + \sum_{i<j} \frac{1}   {D_{ij}} \frac{1}{2^6} \sigma_i^z \sigma_j^z  \Bigr]\,. \nonumber
}
We identify three types of (gauge invariant) imperfections, all of which follow dipolar power laws, but with different strengths due to $\left|\Omega_S\right|\neq\left|\Omega_\sigma\right|$: 
\textsl{(i)} An interaction between matter and gauge fields, with the strongest contribution having the weight $V/27$. 
\textsl{(ii)} An interaction among matter fields, with the strongest contribution being $V/64$. 
\textsl{(iii)} An interaction among gauge fields, with the strongest contribution $V/8$.  
These interactions are much smaller than $V$. They can be treated as small perturbations as long as they decay sufficiently fast to be irrelevant in the renormalization-group sense, as is the case for dipolar interactions in 1D \cite{Lahaye2009,Baranov2012}. 
An improved suppression of undesired longer-range interactions may be achievable by a suitable choice of detunings and trapping frequencies \cite{Porras2004a}.
Alternatively, but more demanding, one could engineer the spin--phonon couplings with a larger number of laser frequencies, which allows to generate only the desired interaction pattern \cite{Korenblit2012}. 

In Sec.~\ref{cha:numericsShortChains}, we will address the quantitative effects of the gauge-invariant contributions given by $H_{V}^{\mathrm{err}}$. 
Before that, we will show how one can generate the desired dynamics $H_{\mathrm{QLM}}$ within a perturbative framework, provided the system dynamics is energetically constrained onto the gauge-invariant subspace by means of a strong $H_G$.

\subsection{Engineering the system dynamics \label{cha:Hsystemdynamics}}

The desired dynamics of the QLM given by Eq.~\eqref{eq:H0} consists of the mass term $H_m$ and the matter--gauge-field interaction $H_J$. 
The single-particle terms generating $H_m$ are simple to implement, e.g., via laser-induced AC-Stark shifts. 
On the other hand, the three-body terms appearing in $H_J$ present a major challenge, 
as they are usually not generated as direct interaction processes. Nevertheless, 
taking advantage of the large energy scale $V$ imposing gauge symmetry,  they can be  obtained in second-order perturbation theory. 

Our starting point are conventional two- and one-body terms,  
\eq{
\label{eq:H1}
H_1=H_K + H_{B}= \sum_{i<j} K_{ij} \sigma_i^x\sigma_j^x + \sum_i B S_{i,i+1}^x\,, 
}
where we consider the regime $|K_{ij}|, |B|\ll V$. 
The single-spin terms of $H_{B}$ can be created in a straightforward way by radio-frequency or Raman beams driving the internal transition. 
The term $H_K$ can be implemented as M{\o}lmer--S{\o}rensen-type interactions \cite{Sorensen1999a}.
Such interactions are generated by two lasers with equal Rabi frequencies $\Omega_{\sigma,xx}$, and with optical frequencies $\omega_{\sigma}\pm \delta_\sigma$. Here, $\omega_\sigma$ is the energy difference of the $\sigma$-pseudo-spin levels and the detuning $\delta_\sigma$ lies close to vibrational frequencies $\omega_q$. 
Due to the different energy splittings of the $\sigma$ and $S$ qubits, these lasers do not couple to the $S$ spins. 
The resulting spin--spin interactions act only among $\sigma$ spins, with strength \cite{Kim2009,Kim2010,Kim2011,Islam2011,Islam2012}
\eq{
K_{ij}= -\sum_q \frac{\left|\Omega_{\sigma,xx}\right|^2\eta_{iq}\eta_{jq}\, \hbar \omega_q}{\delta_\sigma^2 - \omega_q^2}\,.
}
To avoid cross terms of the type $\sigma_i^x \sigma_j^z$, it is convenient if the phonon modes addressed here are different from those used for the $zz$ couplings, for example by choosing vibrations in a different direction. 
Alternatively, such cross terms will be energetically suppressed by the Gauss law.

Taking $H_1$ as a perturbation to $H_G$, we can obtain an effective Hamiltonian for the low-energy sector of $H_G$, given by $E_{\psi}\equiv\bra{\psi} H_G \ket{\psi} = 0$. 
This sector is equivalent to the gauge-invariant subspace defined by $G_i\ket{\psi}=0$ $\forall i$. 
In second-order perturbation theory, we obtain the effective interaction in the gauge-invariant subspace
\eq{
\label{eq:HJ}
H_J= - \frac{J}{2} \sum_i \left( \sigma_i^- S_{i,i+1}^+ \sigma_{i+1}^- + \mathrm{h.c.} \right)\,, 
}
where we defined $J\equiv\frac{K B}{V}$, assuming a homogeneous $K_{i,i+1}=K$. 
Similar hopping terms, but realized within four hyperfine states of a single ion and for a single-mode coupling, have been envisioned in Ref.~\cite{Casanova2011}.
Notably, Eq.~\eqref{eq:HJ} is a nearest-neighbor Hamiltonian; longer-range interactions and terms such as $\sigma_i^+ S_{i,i+1}^+ \sigma_{i+1}^+$ are suppressed by the energy penalty $H_G$ \footnote{This allows to use axial modes to generate $H_K$. In that case, the resulting interactions $K_{ij}$ will decay slower than dipolar, but undesired interactions beyond nearest neighbors are suppressed energetically by $H_G$}. 

\begin{figure}
\centering
\includegraphics[width=\columnwidth]{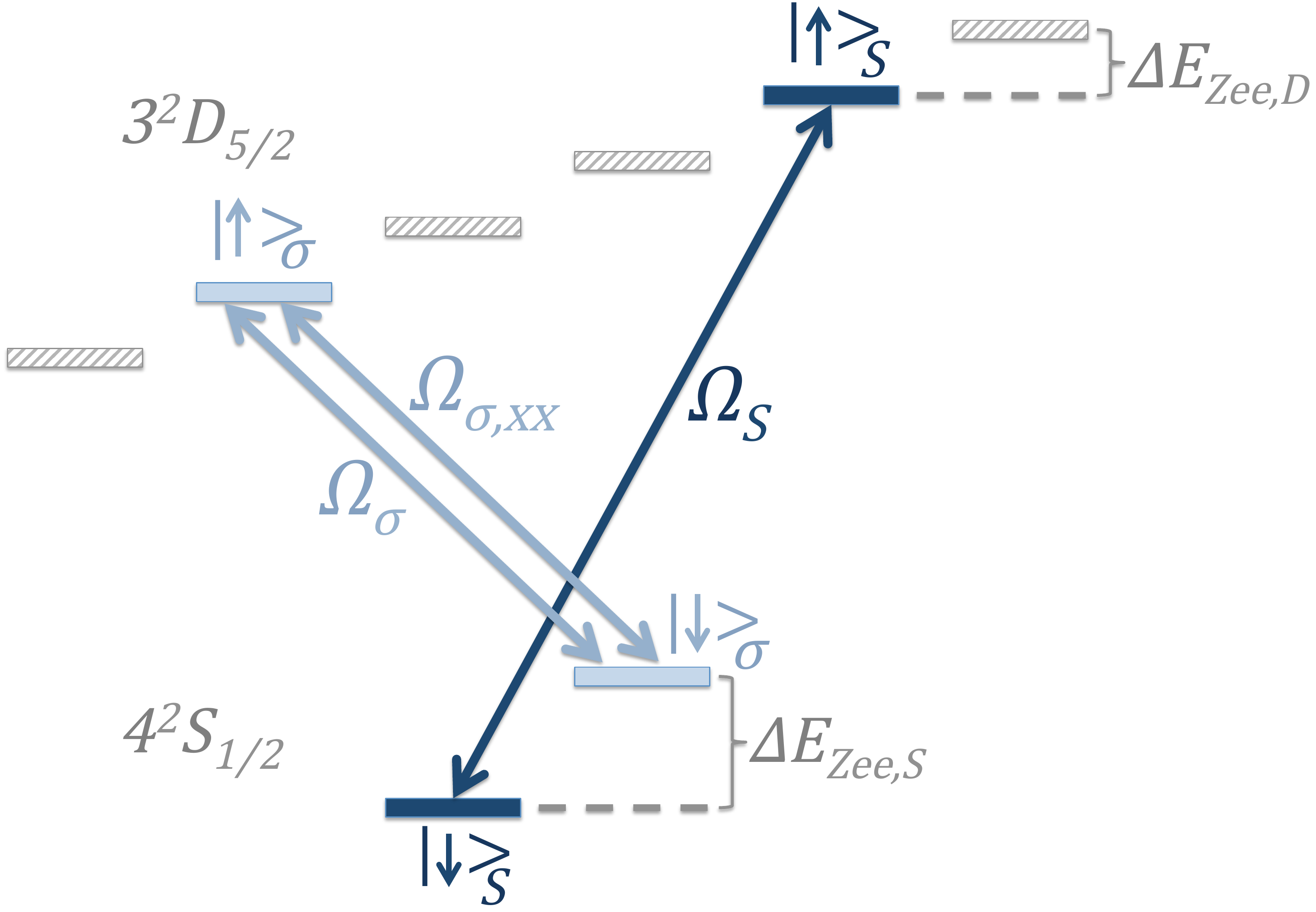}
\caption{ 
{\bf Schematical level diagram of a single ion} summarizing the applied spin--phonon couplings (sketched for optical quadrupole qubits as in $^{40}$Ca$^+$).
The processes in Fig.~\ref{fig:setup}b-c couple $\sigma_i^z$ and $S_{i,i+1}^z$ to radial phonon modes. The correct weight of $zz$ interactions for the Gauss law requires effective coupling strengths $\left|\Omega_{S}\right|=8\left|\Omega_{\sigma}\right|$. 
The $xx$ interactions in $H_K$ may be generated by M{\o}lmer--S{\o}rensen-type couplings of strength $\Omega_{\sigma,xx}$. 
Coupling to undesired internal levels is avoided by selection rules and Zeeman splittings $\Delta E_{\mathrm{Zee},D/S}$. 
Whether the ion represents a $S$ or $\sigma$ spin is determined by its initialization in the state $\ket{\downarrow}_S$ or $\ket{\downarrow}_\sigma$. 
}
\label{fig:spinphononcoupling}
\end{figure}
Summarizing, the microscopic model reads 
\eqa{
\label{eq:Hmicro}
H_{\rm micro}&=&H_K+H_B+H_m+H_G+H_V^{\rm err} \\
&=& \sum_{i< j} K_{ij} \sigma_i^x\sigma_j^x + \sum_i B S_{i,i+1}^x 
+ \frac m 2 \sum_i \sigma_i^z  \nonumber \\
&+& 2V \sum_i (G_i)^2  +H_V^{\rm err} \nonumber \,. 
}
Here, $G_i$ is given by Eq.~\eqref{eq:Gausslaw} and $H_V^{\rm err}$ by Eq.~\eqref{eq:HVerr}. 
The spin--phonon couplings required to generate $H_G$ and $H_K$ are summarized in Fig.~\ref{fig:spinphononcoupling}. 

In the limit $|K|,|B|\ll V$, the dynamics of $H_{\rm micro}$ is confined to the gauge-invariant subspace and effectively reduces to $H_{\rm QLM}+H_G$, plus the undesired dipolar $zz$ interactions $H_V^{\rm err}$. 
The accuracy of the quantum simulation will, therefore, be dominated by two error sources. 
On the one hand, for the desired dynamics $H_{\rm QLM}$ to dominate over $H_V^{\rm err}$, one needs $\left|m\right|,\left|J\right|\gtrsim V/8$. 
On the other hand, one requires $\left|J\right|\ll V$ in order to neglect contributions to the effective Hamiltonian \eqref{eq:HJ} arising in higher orders of $J/V=KB/V^2$. 
Such contributions can modify the system dynamics and compromise gauge invariance. 
We will show next that the parameter window left open by these two error sources allows to retain approximate gauge invariance while at the same time observing the correct dynamics.

\section{Numerical analysis\label{cha:numericsShortChains}}
As explained in the previous section, at small $J/V$ the microscopic model $H_{\rm micro}$, Eq.~\eqref{eq:Hmicro}, reproduces the physics of the lattice gauge theory $H_{\rm QLM}$, Eq.~\eqref{eq:H0}. 
In this section, we test these perturbation-theoretical considerations via exact diagonalizations of finite chains. 
We consider 6 matter-field plus 6 gauge-field spins, under periodic boundary conditions to avoid boundary effects (see Sec.~\ref{sec:errors} for a discussion of possible effects of non-homogeneous ion distances and open boundary conditions). 
For simplicity, we set $B=K$, and we assume a dipolar decay of the interactions in $H_V$ and $H_K$. 
In particular, we study the influence of these error terms on the system under changing the sign of $m$ (with $J>0$). 
Below, we first consider ground-state properties. Afterwards, we estimate the optimal parameter values to see the false-vacuum decay in an adiabatic sweep from the C-and-P-invariant to the C-and-P-breaking situation. 

\subsection{Ground state of the microscopic model}

\begin{figure}
\centering
\includegraphics[width=\columnwidth]{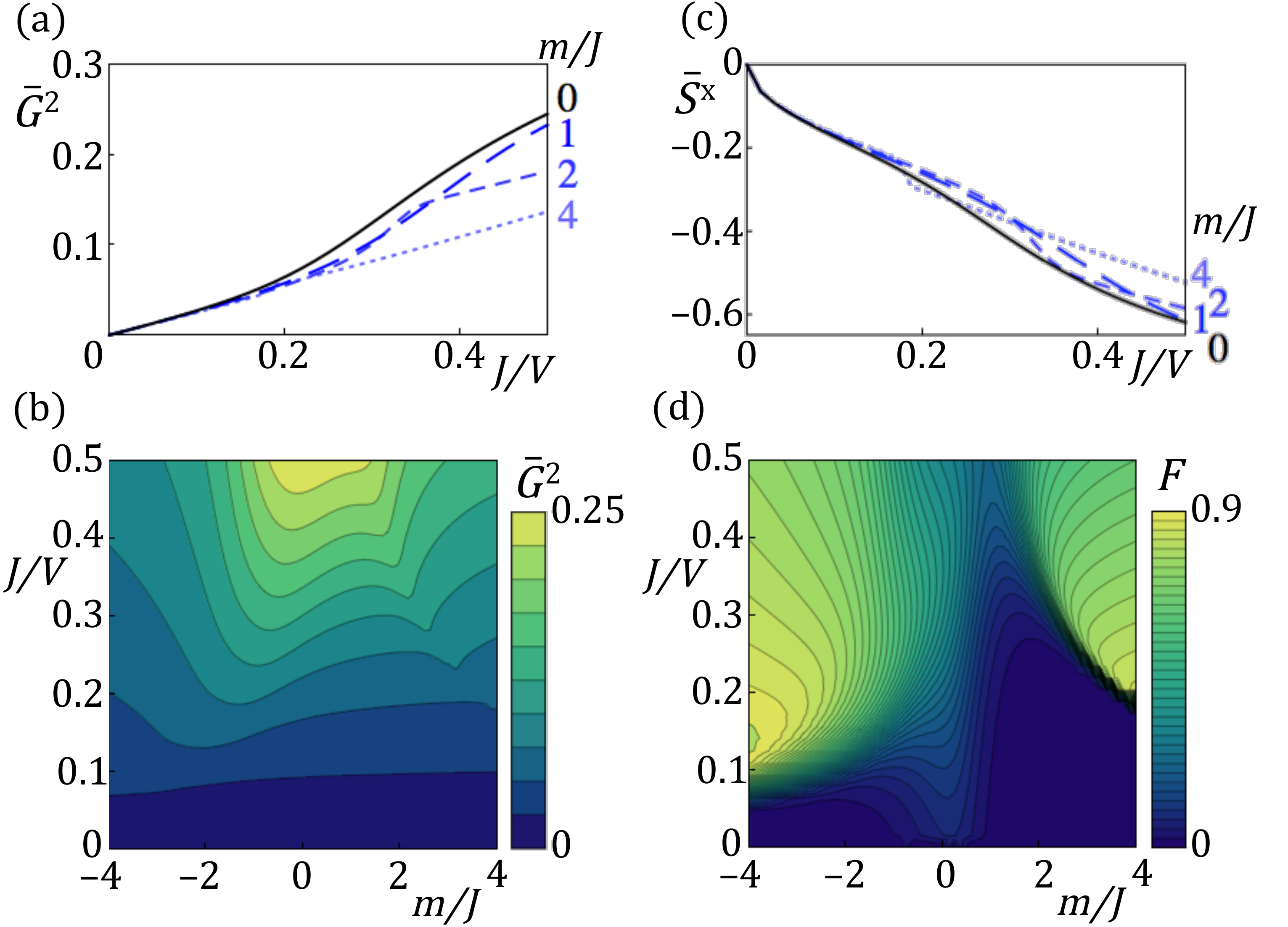}
\caption{ 
{\bf Validity of the microscopic model.} 
{\bf (a-b) Gauge-invariance,} quantified by $\widebar{G^2}\equiv\sum_i\braket{G_i^2}/N_{\rm m}$, is only violated substantially when $V/J$ is not a strong energy scale. 
{\bf (c)} At large $J/V=K B/ V^2$, the gauge fields acquire a finite $\widebar{S^x}\equiv\sum_i\braket{S_{i,i+1}^x}/N_{\rm m}$. 
The detrimental effects encountered in panels a to c are enhanced around $m=0$, i.e., close to the quantum phase transition of the QLM. 
Curves in panels a and c are for $m/J\geq0$ as given in the figures. The behavior for $m/J<0$ is very similar. 
{\bf (d)} The overlap $F=\braket{\psi(m,J,V)|\psi_{\mathrm{ideal}}(m,J)}$ between the ground states of the microscopic model and the ideal QLM, Eq.~\eqref{eq:H0}, is 
largest at intermediate $J/V$ and increases with increasing $\left|m/J\right|$. 
}
\label{fig:checks_microscopic_model}
\end{figure}

The degree of gauge invariance achieved by the microscopic model is measured by violations to the Gauss law \eqref{eq:Gausslaw}, which can be quantified by $\widebar{G^2}\equiv\sum_i\braket{G_i^2}/N_{\rm m}$ (Fig.~\ref{fig:checks_microscopic_model}a-b). 
As expected, when $KB/V^2$ is not a small energy scale, second-order perturbation theory breaks down, and $H_G$ is no longer efficient in energetically suppressing gauge-variant terms. 
The onset of this effect is linear in $J/V$ and quadratic when expressed in $K/V$, as expected from second-order perturbation theory. 
Up to the range of $J/V\approx 0.3$, however, the system remains approximately gauge invariant. 

As $J/V$ is increased, due to $H_B$ the gauge fields acquire a finite polarization $\widebar{S^x}\equiv\sum_i\braket{S_{i,i+1}^x}/N_{\rm m}$  (Fig.~\ref{fig:checks_microscopic_model}c). 
Simultaneously, due to $H_K$ the matter fields display an antiferromagnetic ordering in the $x$ direction. 
Both effects are not foreseen in the ideal model \eqref{eq:H0}. 
These effects are more pronounced close to $m=0$, which corresponds to a potential critical region in the QLM.

A quantitative indicator for deviations resulting from these perturbations is provided by the overlap of the exact microscopic ground-state to the ideal one (Fig.~\ref{fig:checks_microscopic_model}d). 
In general, this overlap is a very strict measure, and we can understand it as a loose lower bound to the quality of the ground state. 
As expected from the analysis of $\widebar{S^x}$ and the gauge invariance, it decreases at large $J/V$ and in a relatively broad region around $m=0$. 
Additionally, the overlap is strongly suppressed at small $J/V$, due to the undesired dipolar terms $H_{V}^{\mathrm{err}}$. These terms form a relevant contribution when $J/V\lesssim 1/8$ or $\left|m\right|/V\lesssim 1/8$. 
Being gauge invariant, they do not affect the Gauss law, but they do change the dynamics of the system.

\begin{figure}
\centering
\includegraphics[width=1\columnwidth]{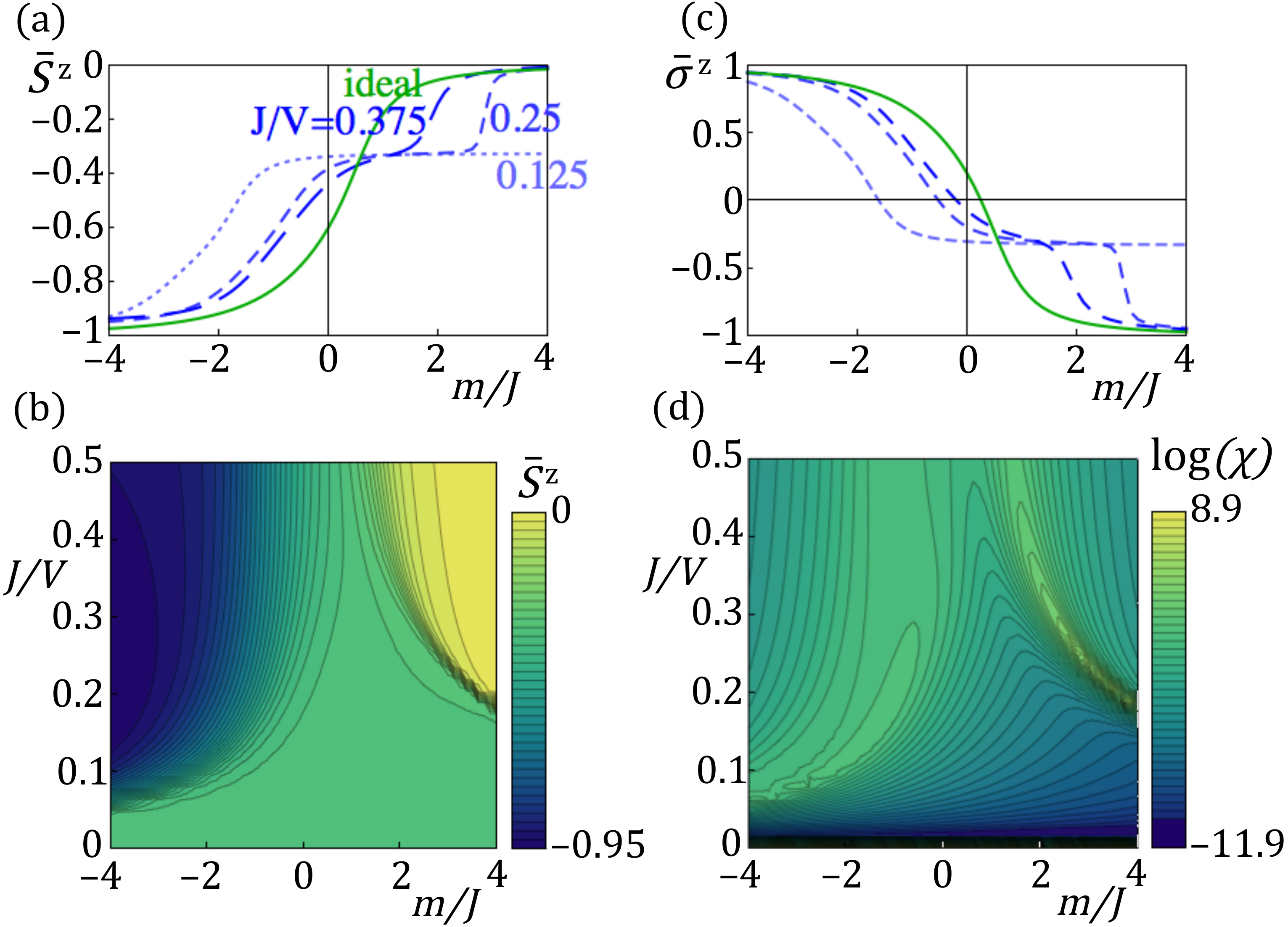}
\caption{ 
{\bf Ground-state behavior of the microscopic model.} 
{\bf(a-c)} Similar to the ideal QLM, the system reaches $\widebar{S^z}\approx-1$ and  $\widebar{\sigma^z}\approx 1$ for $m/J\to-\infty$, and $\widebar{S^z}\approx0$ and $\widebar{\sigma^z}\approx-1$ for $m/J\to\infty$. 
This denotes a transition from a parity- and charge-symmetric to a parity- and charge-symmetry breaking state. 
Curves from lighter to darker and shorter to longer dashes: $J/V=0.125,0.25,0.275$. Solid line: ideal QLM. 
The curves change their sign of curvature twice, indicating two phase transitions. 
{\bf(d)} 
Similarly, the fidelity susceptibility $\chi(m,J)$ has two peaks, suggesting two phase transitions. 
This behavior is in contrast to the ideal model, where $\chi(m,J)$ has a single peak at $m=0$. 
 }
\label{fig:phd_microscopic_model}
\end{figure}

To analyze in how far these deviations affect the ground-state phase diagram, we now study the order parameters $\widebar{S^z}\equiv \sum_i \braket{S_{i,i+1}^z}/N_{\rm m}$ (Fig.~\ref{fig:phd_microscopic_model}a-b)  and $\widebar{\sigma^z}\equiv\sum_i\braket{\sigma_{i}^z}/N_{\rm m}$ (Fig.~\ref{fig:phd_microscopic_model}c). 
As discussed in Fig.~\ref{fig:systems}d, 
at $m/J=-\infty$ we expect for the ideal model $\widebar{\sigma^z}=1$ and $\widebar{S^z}=-1$. 
At $m/J=\infty$, we expect $\widebar{\sigma^z}=-1$ and $\widebar{S^z}=0$. The last value is realized by a superposition of the two parity-breaking states with fluxes in negative and positive direction (see Fig.~\ref{fig:systems}a), i.e., by the two antiferromagnetic configurations of $S_{i,i+1}^z$ that are shifted by one lattice spacing (Fig.~\ref{fig:systems}d). 
The expected behavior at large $\left|m\right|$ is indeed what is observed in the microscopic model. 

Deviating from the QLM, around $m=0$, we find the precursor of a plateau 
where the order parameters remain constant at $\widebar{S^z}\approx\widebar{\sigma^z}\approx -1/3$ over an entire range of $m/J$. 
This plateau becomes more pronounced at smaller $J/V$ (for the curve at $J/V=0.125$, it stretches beyond the plotted range).  
As a result, the corresponding curves change the sign of their curvature twice. 
Since peaks in $\partial\, \widebar{\sigma^z}/\partial (m/J)$ or $\partial\,\widebar{S^z}/\partial (m/J)$ are good indicators for quantum phase transitions in the thermodynamic limit, this suggests two distinct transition points in the microscopic model. This is in contrast to the ideal QLM, where we expect only a single quantum phase transition \cite{QLMTheoryInPreparation2013}. 

The assumption of two transitions is confirmed by the fidelity susceptibility, which is a reliable tool to detect conventional quantum phase transitions \cite{Zanardi2006,Gu2010}. 
It is defined as 
\begin{equation}
\chi(m,J)= \lim_{\delta m \to 0} -\frac{\log\left|\braket{\psi(m,J)|\psi(m+\delta m,J)}\right|^2}{\delta m^2}\,,
\end{equation}
where $\ket{\psi(m,J)}$ is the ground state of the microscopic model at fixed $m$ and $J$. 
In analogy to other, long-established susceptibilities, the fidelity susceptibility measures the response of the ground state towards an external perturbation, in this case a change of the Hamiltonian. 
Close to a quantum critical point, the wave function changes its behavior drastically, yielding peaks in $\chi(m,J)$. 
For the microscopic model, we find two such peaks (see Fig.~\ref{fig:phd_microscopic_model}d). 
As it can be expected, at the position of these peaks the Gauss law is more strongly violated (see Fig.~\ref{fig:checks_microscopic_model}b). 

The presence of two peaks in the fidelity susceptibility indicates the appearance of an intermediate phase around $m=0$, which is absent in the ideal QLM. 
In this central region, the physics of the microscopic model is governed by a complex interplay between $H_m$ and $H_V^{\mathrm{err}}$ --- while the mass term $H_m$ favors the states sketched in Fig.~\ref{fig:systems}d, with a finite $z$ polarization of the matter fields, the frustration arising from $H_V^{\mathrm{err}}$ tends to suppress such polarizations. 
Due to this interplay, transitions away from this region occur roughly when $H_m$ dominates over $H_V^{\mathrm{err}}$, i.e., when $\left|m\right|/V\gtrsim 1/8$ or, equivalently, $\left|m\right|/J\gtrsim V/(8J)$. 
For this reason, the extent of the intermediate region increases at smaller $J/V$. 
However, due to the dipolar nature of $H_V^{\mathrm{err}}$, this region underlies appreciable finite-size effects at the considered system sizes.  
Additionally, it presents a considerable asymmetry upon $m\to-m$. 
This asymmetry derives from the $\sigma_i^zS_{i,i+1}^z$ interactions, which favor the state achieved at $m/J\to-\infty$ over the one reached at $m/J\to+\infty$. 

To summarize this analysis, while there seems to appear a novel phase due to $H_V^{\mathrm{err}}$ in the region around $m=0$, the transition from a parity-conserving to a parity-broken state can be clearly observed in the microscopic model, even at small chain lengths. 
Further, by restricting oneself to the regime of $J/V\approx 0.2$, one can realize a situation where $J$ dominates over the error terms $H_V^{\rm err}$ -- and thus induces the desired dynamics -- while retaining approximate gauge invariance.

\subsection{Sweep across the transition\label{cha:sweep}}

We now want to check the possibility of observing the false-vacuum decay and the associated spontaneous C and P breaking in experiments along the lines of Refs.~\cite{Friedenauer2008,Kim2010,Kim2011,Islam2011,Islam2012}. 
For this end, we consider an adiabatic sweep from the C-and-P-invariant to the C-and-P-broken region. 
We keep $V$ and $J$ positive and constant and apply a linear quench of the mass term, $m(t)=m_{\mathrm{init}} + t\, {\delta m}/{\delta t}$, with $m_{\mathrm{init}}\equiv m(t=0)<0$.
We finish the sweep at $m_{\mathrm{fin}}\equiv m(t_{\mathrm{fin}})=-m_{\mathrm{init}}$. 
The duration of the quench, $t_{\mathrm{fin}}$, depends on quench range and speed. 
In our analysis, we consider a system of 4 matter plus 4 gauge-field spins for similar parameters as in the preceding section. 

\begin{figure}
\centering
\includegraphics[width=1\columnwidth]{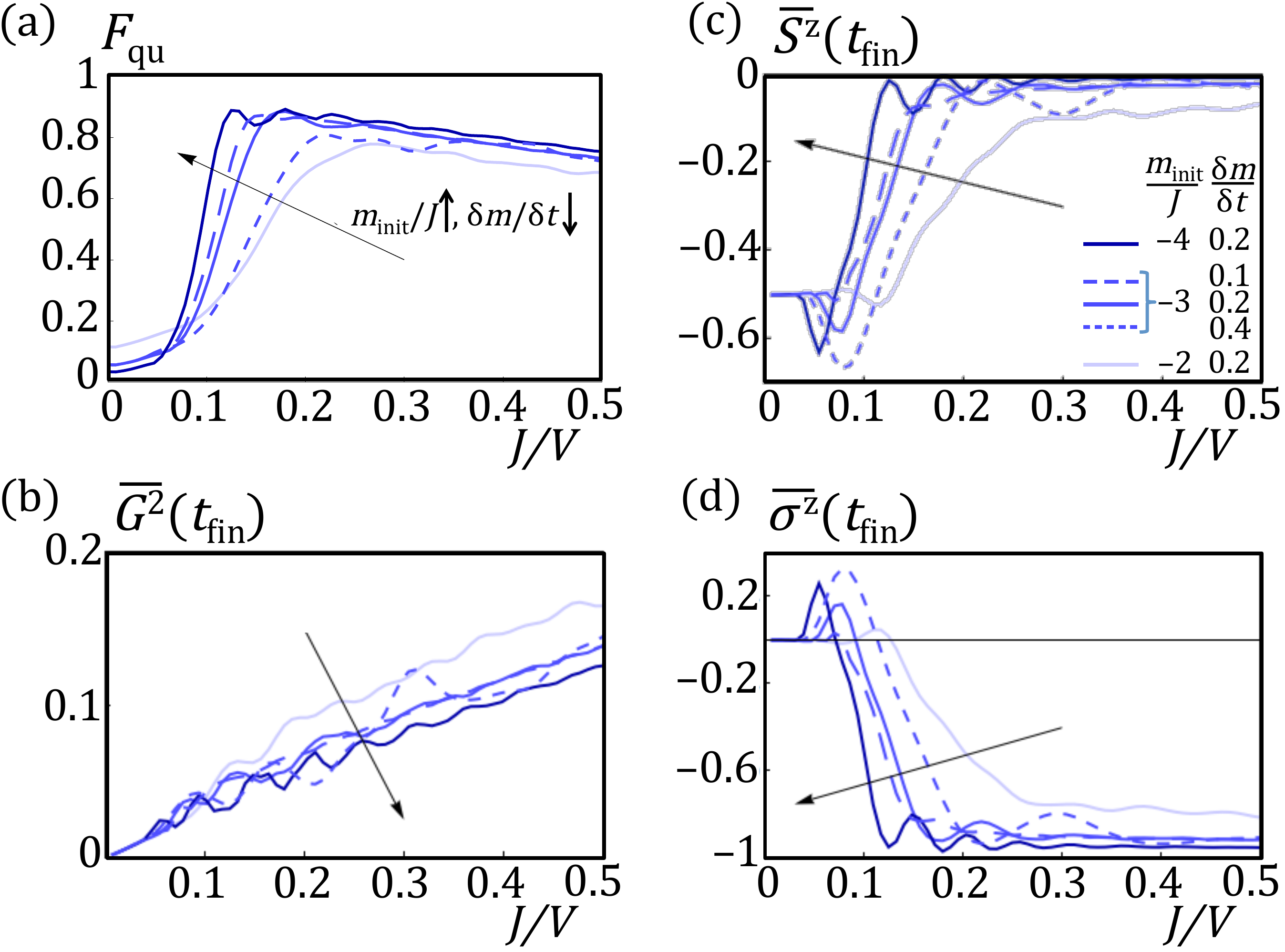}
\caption{ 
{\bf Final state after a linear sweep,} $m(t)=m_{\mathrm{init}} + t\,\delta{m}/\delta{t}$, with $m_{\mathrm{fin}}=-m_{\mathrm{init}}$ and fixed $V$ and $J$. 
Solid lines from lighter to darker shades (bottom to top in panel a): $m_{\mathrm{init}}/J=-2,-3,-4$ for $\delta{m}/\delta{t}=0.2\, V^2/\hbar$. Short (long) dashed lines:  $m_{\mathrm{init}}/J=-3$ and $\delta{m}/\delta{t}=0.4\, (0.1)\,  V^2/\hbar$. 
The arrows show how the precision of the quench increases for larger $m_{\mathrm{init}}$ and smaller quench velocities $\delta{m}/\delta{t}$.
The total times considered range from $t_{\mathrm{fin}}=15 \hbar/V$ (for $m_{\mathrm{init}}/J=-3$, $\delta{m}/\delta{t}=0.4 V^2/\hbar$) to $60 \hbar/V$ ($m_{\mathrm{init}}/J=-3$, $\delta{m}/\delta{t}=0.1$). For $V/\hbar= 2\pi\, 1\,$kHz \cite{Britton2012,Schneider2012}, these correspond to $3.2-9.5\,$ms. 
{\bf (a-d)}
For a value of $J/V\approx 0.15-0.2$, the final, C-and-P-breaking state is reached with (a) high fidelity, (b) small gauge variance, and the desired behavior for (c) gauge field and (d) matter field. 
 }
\label{fig:quench}
\end{figure}

As a strict measure for how well this sweep reproduces the false-vacuum decay, we study the quench fidelity. 
It is defined as the overlap of the microscopic state after the sweep, $\psi_{\mathrm{fin}}$, with the ground state of the ideal QLM at the final parameter values, $\psi_{\mathrm{ideal}}(m_{\mathrm{fin}}/J)$, i.e., ${F}_{\mathrm{qu}}\equiv\left|\braket{\psi_{\mathrm{fin}}| \psi_{\mathrm{ideal}}(m_{\mathrm{fin}}/J)}\right|^2$. 
As seen in Fig.~\ref{fig:quench}a, with increasing $J/V$ the errors committed in the perturbative treatment of Sec.~\ref{cha:Hsystemdynamics} reduce ${F}_{\mathrm{qu}}$. 
More drastic, however, is the effect of the dipolar error terms $H_V^{\mathrm{err}}$: In order to start and finish in the correct states of Fig.~\ref{fig:systems}d, we require $\left|m_{\mathrm{init}}/J\right| \gg V/(8J)$. 
Otherwise, the sweep happens within the central phase that is not present in the ideal QLM, and the overlap with the ideal final state vanishes. 

Furthermore, as usual, a small $\delta{m}/\delta{t}$ is desirable to assure the adiabicity of the process. The parameter region that is most relevant for our purposes, $1/8\lesssim J/V \ll 1$, is especially sensitive to a change of sweep speeds. 
For a good fidelity, one will thus want to choose not only a large $\left|m_{\mathrm{init}}/J\right|$ but also a small $\delta{m}/\delta{t}$, both of which increase the total quench time $t_{\mathrm{fin}}$. 
In experiment, where available time scales are restricted (e.g., by the life-time of the upper pseudo-spin level and spontaneous Raman scattering from the optical beams \cite{Uys2010}), 
one will therefore have to find a compromise. 

In contrast to the quench fidelity, the gauge invariance is not decreased by a possible non-adiabicity of the sweep. Instead, it remains on the level of the ground-state values at $m_{\mathrm{init}}/J$, as can be appreciated by comparing Fig.~\ref{fig:quench}b to Fig.~\ref{fig:checks_microscopic_model}a. 
For this reason, choosing a large $\left|(m/J)_{\mathrm{init}}\right|$ appears more important than a small $\delta{m}/\delta{t}$. 

For the parameter regimes where the quench fidelity is large, the order parameters 
$\widebar{S^z}$ (Fig.~\ref{fig:quench}c)  and $\widebar{\sigma^z}$ (Fig.~\ref{fig:quench}d) reach the final value expected for the C-and-P breaking state ($0$ and $-1$, respectively). 
For $J/V\lesssim0.05$, however, the quench happens entirely in the central region and $\widebar{S^z}$ and $\widebar{\sigma^z}$ practically remain at their initial values; for a system of 4 matter-field and 4 gauge-field spins, these are $\widebar{S^z}=-0.5$ and $\widebar{\sigma^z}=0$. 
In the range $0.05\lesssim J/V\lesssim0.1$, the quench still ends in the central region, but now starts close to the C-and-P retaining state sketched in Fig.~\ref{fig:systems}d, left side, where $\widebar{S^z}=-1$ and $\widebar{\sigma^z}=1$. 
The starting values of the order parameters lie now farther away from the ones of the C-and-P breaking state. Therefore, in this parameter region of $0.05\lesssim J/V\lesssim0.1$, the final values of the order parameters seem actually worse than for $J/V\lesssim0.05$. 
Only when the sweep starts in a C-and-P retaining state and ends in the C-and-P breaking region (or vice versa), do we obtain the expected final values of the order parameters $\widebar{S^z}$ and $\widebar{\sigma^z}$. Depending on $\left|(m/J)_{\mathrm{init}}\right|$, this is achieved for $J/V\gtrsim 0.2$. 

Besides this overall behavior, all considered quantities display some oscillations as a function of $J/V$. These derive from oscillations during the time evolution, the frequency of which changes as a function of the coupling parameters $J/V$. 
Since we take a snapshot at the fixed time $t_{\mathrm{fin}}$, these oscillations are imprinted on the final state. 

To summarize this part, the fidelity of the process can reach large values $\approx 90\%$ while retaining approximate gauge invariance. 
A reasonable choice of parameters could be $m_{\mathrm{init}}/J=-4$ and $\delta{m}/\delta{t}=0.2 V^2/\hbar$. 
For a realistic value of $V/\hbar= 2\pi\, 1\,$kHz \cite{Britton2012,Schneider2012} (see Section \ref{sec:errors} below), this corresponds to a sweep time $t_{\mathrm{fin}}=40 \hbar/V=6.5\,$ms. 
This is a reasonable value compared to experimental time scales. The latter are limited by dephasing times, which typically lie in the range $5-10\,$ms and can reach up to $100\,$ms \cite{Britton2012}. 
The quality of the final state could be further increased by optimizing the quench protocol, for example, by choosing faster sweep velocities in regions where the gap to the first excited state is large \cite{Richerme2013}.

\section{Minimal implementation using four ions\label{cha:minimalModel}}

Up to now, we have considered an implementation of the QLM \eqref{eq:H0} that is suitable also for large ion chains. However, current experiments are typically restricted to a few ions. 
In the following, we discuss a minimal experiment that can be realized with four ions and by only addressing axial modes, but which already contains the physics of false vacuum decay explained above.
\begin{figure}
\centering
\includegraphics[width=0.75\columnwidth]{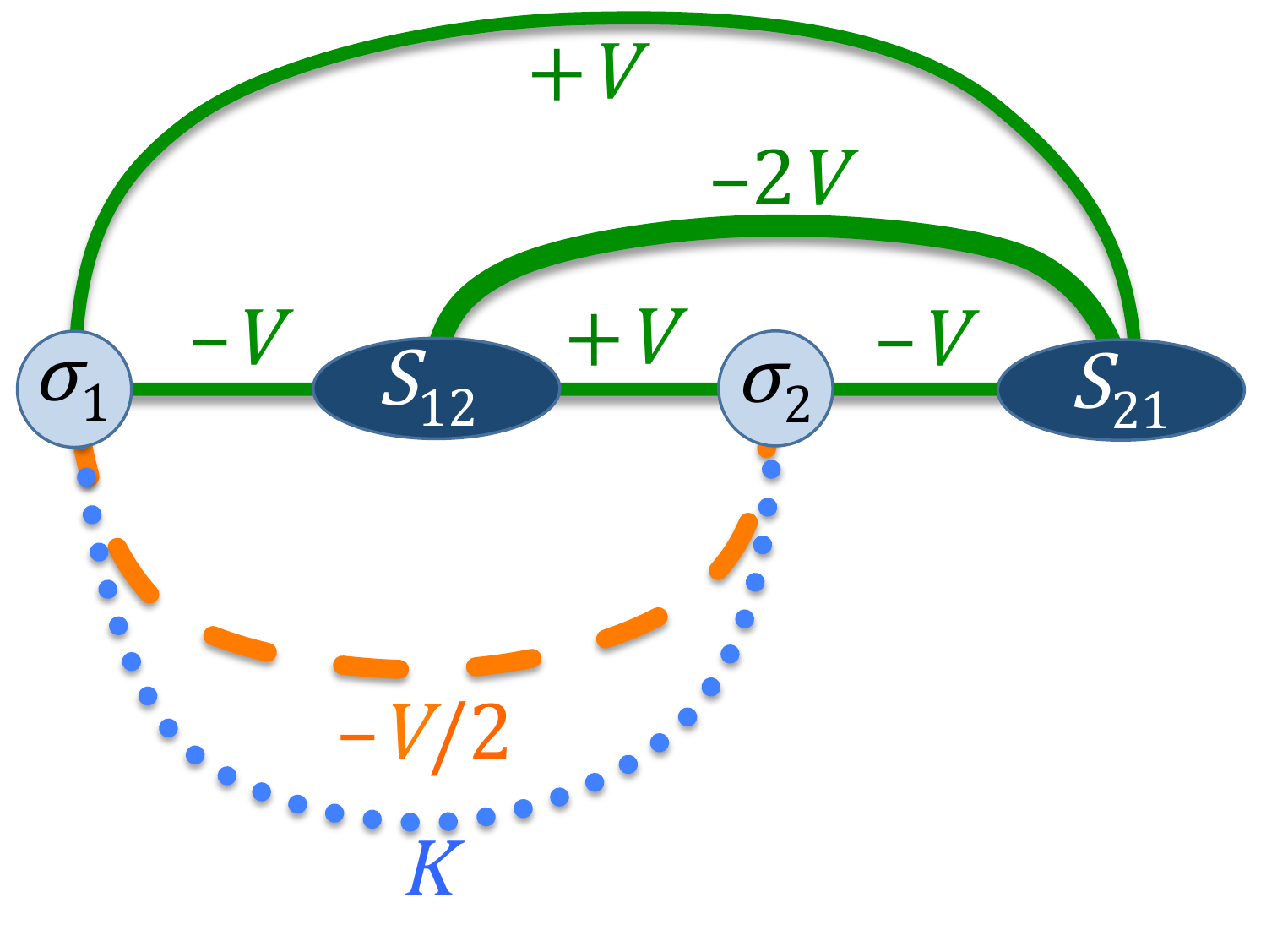}
\caption{ 
{\bf Setup for the minimal implementation using four ions.} 
Spin--spin interactions are created by pairs of Raman beams as explained in Fig.~\ref{fig:setup}, with strengths $\Omega_S=2\Omega_\sigma$.  By tuning close to the vibrational mode $q_0$ characterized by amplitudes $\mathcal M_{n q_0}=\left(1,-1,-1,1\right)_n/2$, one obtains $zz$ spin--spin couplings as desired for $\tH_G$, Eq.~\eqref{eq:tHG} (green solid lines). 
An additional undesired, weaker interaction (orange dashed line) does not affect the system dynamics. 
Mediated by a different vibrational mode, an additional laser pair generates the $xx$ interaction as required for $H_K$ (blue dotted line). 
 }
\label{fig:minimal_model_setup}
\end{figure}
We focus on the unrotated QLM~\eqref{eq:originalspinQLM} with $S=1/2$ and two sites plus two links, under periodic boundary conditions. 
This system is sketched in Fig.~\ref{fig:minimal_model_setup}. 

\subsection{Possible implementation}
The implementation of this two-site QLM is straightforward: to induce the interactions for $\tH_G$, we exploit that in a chain of four ions, there is a vibrational mode $q_0$ with amplitudes ${\mathcal M}_{nq_0}=\left(1,-1,-1,1\right)_n/2$. The index $n$ labels the ions from 1 to 4, where $n=1,3$ denotes the matter fields $\sigma_1$ and $\sigma_2$, and $n=2,4$ the gauge fields $S_{12}$ and $S_{21}$, respectively. 
The frequency spacing of the phonon modes is $\propto \sqrt{\beta_\alpha} \omega_\alpha$, where, as before, $\omega_{\alpha}$ denotes the trap frequency in direction $\alpha$, and $\beta_{\alpha}= e^2/ (M \omega_\alpha^2 d_0^3)$ is the associated stiffness parameter. 
For the axial direction, this energy spacing is sufficiently large to tune the beatnote $\omega_{\rm L}$ of a laser pair close to mode $q_0$, with detuning $\delta_{q_0}=\omega_{q_0}-\omega_{\mathrm L}$, while neglibly driving the other modes. 

To obtain the correct weights of the interactions, we assume that the effective Rabi frequencies $\Omega_n$ coupling the $S$ spins to the phonons is twice as large as for the $\sigma$ ions, $\Omega_S=2\Omega_{\sigma}$. 
Applying Eq.~\eqref{eq:Vgeneral} to this case gives $\mathcal{V}_{mn}\equiv- {{\mathcal R}(\Omega_m^\star \Omega_n) \hbar^2 k_{\rm L}^2
{\mathcal M}_{mq_0} {\mathcal M}_{nq_0}}/ \left( {8 M \omega_{q_0} \delta_{q_0} } \right)$. 
We thus obtain the interaction strength between ion $m$ and $n$
\eq{
\mathcal{V}_{mn} = - V
\left( \begin{array}{cccc}
0 & -1 & -\frac{1}{2} & 1 \\
 -1 & 0 & 1 & -2 \\
 -\frac 1 2 &  1  & 0 & -1\\
 1 & -2 & -1 & 0
 \end{array} \right)_{mn}\,, 
}
with $V={\left|\Omega_\sigma\right|^2 \hbar^2 k_{\rm L}^2} / ( {16 M \omega_{q_0} \delta_{q_0} } )$. 
For negative detuning, these interactions constitute exactly the $zz$ interaction terms of the  Gauss-law constraint \eqref{eq:tHG} [i.e., the one before the rotation \eqref{eq:rotation_of_spins} in spin space]. 
Additionally, we obtain the undesired, but weaker interaction $\sigma_1^z \sigma_2^z V /2 $. This interaction is gauge invariant, and we find that it does not alter the system dynamics (see Sec.~\ref{cha:numerics_minimalmodel}). 

The $\sigma_1^x \sigma_2^x$ interactions that generate $H_K$ can be mediated via a different phonon mode, with lasers that address only the $\sigma$ pseudo-spins. 
Notably, due to the large mode spacing, all interactions in this minimal implementation can be generated by using only axial modes. 
For the architecture presented in Secs.~\ref{cha:MicroscopicModel} and~\ref{cha:numericsShortChains}, this was not possible, since a sufficiently fast decay of $H_V^{\rm err}$ required the use of radial modes.

An alternative four-ion implementation that avoids the use of two types of pseudo-spins can be obtained by grouping the gauge-field spins in the center. 
The order of the ions in the trap does then not correspond to their position in the lattice, but an adequate addressing of the vibrational modes can engineer the desired interactions. 
For example, the correct weight of the interactions in $H_G$ can be generated simply by off-resonantly driving the center-of-mass mode with a laser that is focused more strongly on the center of the chain. 
The $xx$ interactions between matter-field spins 1 and 2 (now at position 1 and 4 of the ion chain) can be transmitted by a mode that has amplitudes ${\mathcal M}_{nq}\approx\left(-0.66,-0.25,0.25,0.66\right)_n$. 
We checked numerically that the additional, unwanted $xx$ interactions induced by that mode do not compromise the ground-state behavior. 
In the following we focus on the cleaner case described in Fig.~\ref{fig:minimal_model_setup}.

\subsection{Numerical analysis of the ground-state behavior \label{cha:numerics_minimalmodel}}

\begin{figure}
\centering
\includegraphics[width=1\columnwidth]{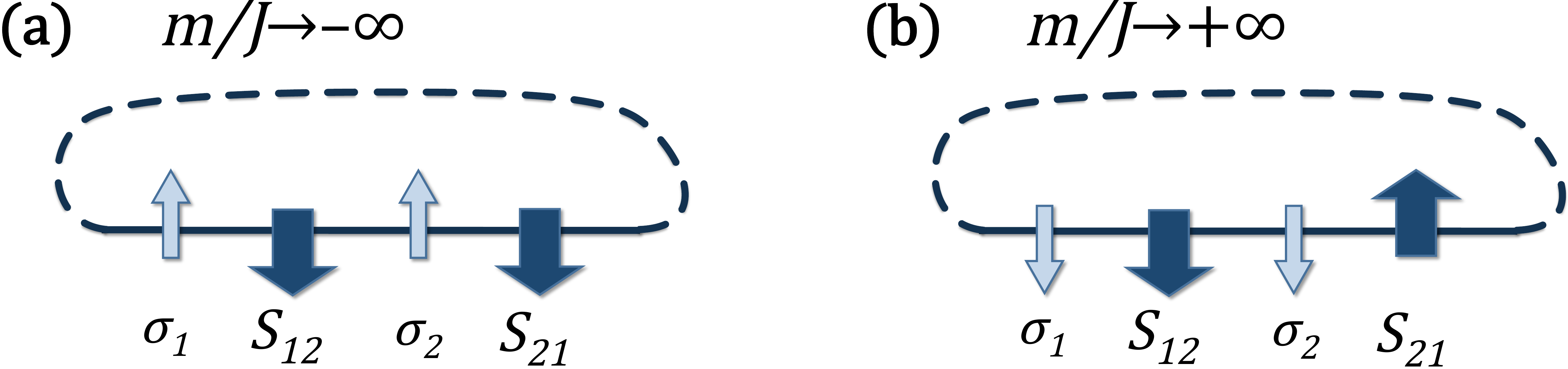}
\caption{ 
{\bf Product states at $m/J=\pm\infty$ for the minimal implementation of four ions.}
{\bf (a)} State at $m/J\to-\infty$, conserving C and P invariance. 
{\bf (b)} State at $m/J\to+\infty$, breaking C and P invariance. This state is degenerate with the opposite configuration of gauge-field spins 
(i.e., $\ket{S_{12}}=\ket{\uparrow}$, $\ket{S_{21}}=\ket{\downarrow}$). 
 }
\label{fig:productstates_minimal_model}
\end{figure}

\begin{figure}
\centering
\includegraphics[width=1\columnwidth]{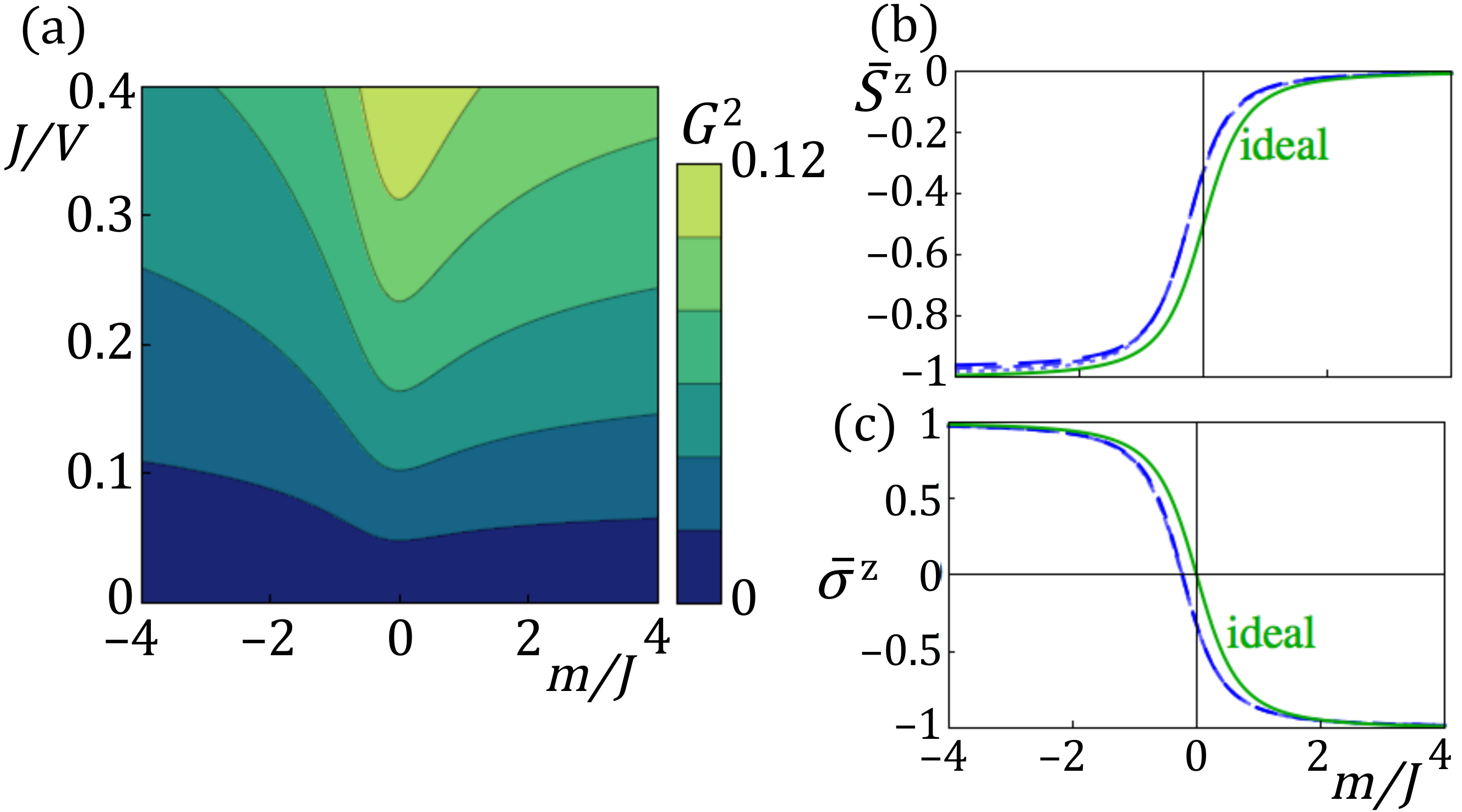}
\caption{ 
{\bf Observables in the minimal implementation.} 
{\bf(a)} The gauge-invariance is only violated substantially where $V/J$ is not a strong energy scale. 
The order parameters 
{\bf(b)} $\widebar{S^z}\equiv (\braket{S_{12}^z} + \braket{S_{21}^z})/2$  and 
{\bf(c)} $\widebar{\sigma^z}\equiv (\braket{\sigma_{1}^z} + \braket{\sigma_{2}^z})/2$
behave similar to the ideal QLM (solid line). 
The transition from a C and P invariant to a C and P breaking state can be clearly seen. 
The blue curves from lighter to darker and shorter to longer dashes denote $J/V=0.125,0.25,0.275$, but practically coincide. 
 }
\label{fig:phd_minimal_model}
\end{figure}

We now illustrate that the minimal implementation of Fig.~\ref{fig:minimal_model_setup} already contains much of the physics of the full-fledged many-body problem. 
Using again exact diagonalizations, we analyze  in this section the ground state behavior. In the next section, we study the sweep dynamics. 
For better comparison with the implementation discussed in Sec.~\ref{cha:numericsShortChains}, we interpret our results, presented in Fig.~\ref{fig:phd_minimal_model},  in the rotated basis \eqref{eq:rotation_of_spins}. 
For convenience, we reproduce in Fig.~\ref{fig:productstates_minimal_model} the product states at $m/J\pm\infty$ for the present case of two sites under periodic boundary conditions. 

As seen in Fig.~\ref{fig:phd_minimal_model}a, for this small system, gauge invariance is even better retained than what we saw in Sec.~\ref{cha:numericsShortChains}. 
Again, deviations are stronger at large $J/V$ and around $m=0$, i.e., close to the expected phase transition of the ideal QLM \eqref{eq:H0}.
The order parameters $\widebar{S^z}$ and $\widebar{\sigma^z}$ behave very similarly to an ideal QLM of the same size (Fig.~\ref{fig:phd_minimal_model}b-c).
In particular, at large $\left|m\right|$, the polarization of the product states sketched in Fig.~\ref{fig:productstates_minimal_model} is attained. 
The behavior of the order parameters is practically independent of $J/V$. 
This finding suggests that in the considered parameter range the errors are negligible that are introduced by generating the system dynamics perturbatively (see Sec.~\ref{cha:Hsystemdynamics}). 
Moreover, the interactions $V \sigma_1^z \sigma_2^z/2$ do not alter the ground state. Such interactions would favor an antiferromagnetic correlation between $\sigma_1^z$ and $\sigma_2^z$, a behavior which violates the Gauss law \eqref{eq:Gausslaw}. The influence of this interaction is therefore cancelled by the stronger $H_G$. 
As a result, the region at small $m/V$ that is not contained in the ideal QLM, and which appeared in the large-chain scheme due to the dipolar interactions $H_V^{\mathrm{err}}$, is absent in this minimal model. 
Consequently, the fidelity susceptibility (not shown) has only a single peak at $m=0$, in accordance with the ideal QLM.

\subsection{Numerical analysis of adiabatic sweeps}

In order to provide a better connection to experimental realizations, we now study adiabatic sweeps through the C and P breaking transition. 
For ease of comparison, we consider the same linear quench protocol as in Sec.~\ref{cha:sweep}, namely, a sweep from $m/J\ll 0$ to $m/J\gg 0$, with the same sweep speeds and ranges.

\begin{figure}
\centering
\includegraphics[width=1\columnwidth]{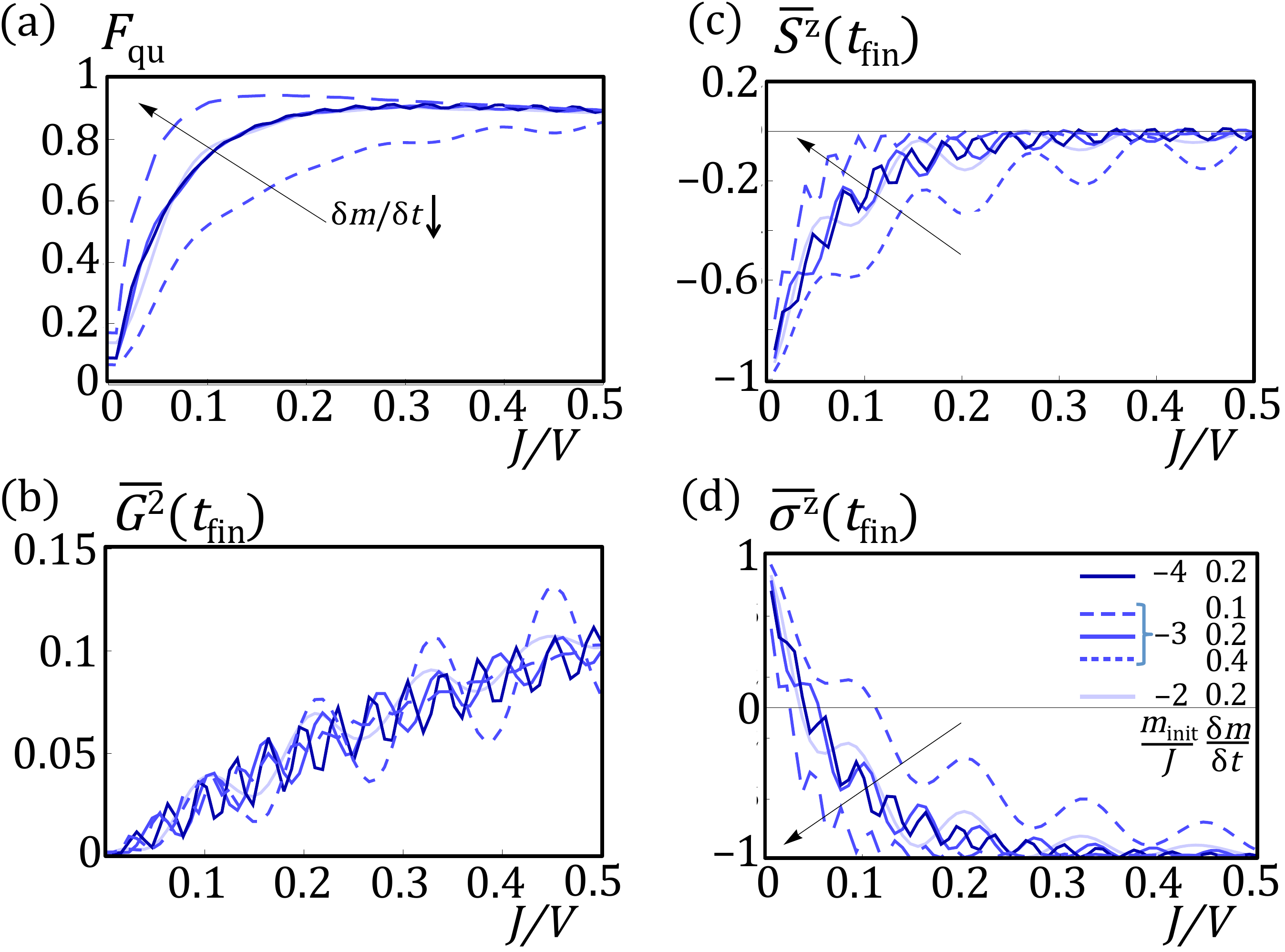}
\caption{ 
{\bf Final state after a linear sweep for the minimal implementation.} 
We change $m(t)=m_{\mathrm{init}} + t\,\delta{m}/\delta{t}$, with $m_{\mathrm{fin}}=-m_{\mathrm{init}}$ and keep $V$ and $J$ fixed. 
Solid lines from lighter to darker shades: $m_{\mathrm{init}}/J=-2,-3,-4$ for $\delta{m}/\delta{t}=0.2\, V^2/\hbar$. Short (long) dashed lines:  $m_{\mathrm{init}}/J=-3$ and $\delta{m}/\delta{t}=0.4\, (0.1)\,  V^2/\hbar$. 
The arrows show how the precision of the quench increases for smaller quench velocities $\delta{m}/\delta{t}$.
The total times considered range from $t_{\mathrm{fin}}=15 \hbar/V$ (for $m_{\mathrm{init}}/J=-3$, $\delta{m}/\delta{t}=0.4 V^2/\hbar$) to $60 \hbar/V$ ($m_{\mathrm{init}}/J=-3$, $\delta{m}/\delta{t}=0.1$). For $V/\hbar= 2\pi\, 1\,$kHz \cite{Britton2012,Schneider2012}, these correspond to $3.2-9.5\,$ms. 
{\bf (a-d)}
For a value of $J/V\gtrsim 0.1$, the final, C-and-P-breaking state is reached (a) with high fidelity, (b) good gauge invariance, and the desired behavior for (c) gauge field and (d) matter field. 
 }
\label{fig:dynamics_minimal_model}
\end{figure}

As the quench fidelity shows (Fig.~\ref{fig:dynamics_minimal_model}a), for the considered parameter range, the accuracy of the sweep does practically not depend on $m_{\mathrm{init}}/J$, but only on the chosen quench speed. 
As in the scheme of Sec.~\ref{cha:numericsShortChains}, we remark a considerable drop of fidelity at low $J/V$. In this case, however, the drop has its origin not in dipolar error terms, but in the fact that the intrinsic energy scale of the system dynamics decreases with $J/V$. This results in a smaller gap, thus requiring slower evolutions to retain adiabaticity. 
Therefore, slower sweeps particularly improve the quench fidelity at small $J/V$. 

This behavior is reflected in the order parameters $\widebar{S^z}$ (Fig.~\ref{fig:dynamics_minimal_model}c) and $\widebar{\sigma^z}$ (Fig.~\ref{fig:dynamics_minimal_model}d). 
For the considered quench velocities, the sweep is non-adiabatic at low $J/V$, and the order parameters almost retain their initial values $\widebar{S^z}=-1$ and $\widebar{\sigma^z}=1$ (c.f.\ Fig.~\ref{fig:productstates_minimal_model}a). 
Only when $J/V\gtrsim 0.1$, the final values of the order parameters approach the values expected for the ideal QLM at large $m$, namely,  $\widebar{S^z}=0$ and $\widebar{\sigma^z}=-1$ (c.f.\ Fig.~\ref{fig:productstates_minimal_model}b). 
We observe again an oscillatory behavior as a function of $J/V$. 
Similar to Sec.~\ref{cha:sweep}, these oscillations appear because we take a snapshot of the evolution at a fixed final time $t_{\mathrm{fin}}$. 

In contrast to quench fidelity and order parameters, the overall breaking of gauge invariance rises with $J/V$ and does not depend much on the quench speed (Fig.~\ref{fig:dynamics_minimal_model}b). 
It does, however, acquire stronger oscillations for faster sweeps. 
Thus, while adiabicity requires a sufficiently large $J/V$, to retain approximate gauge invariance, large $J/V$ should be avoided -- 
again there is an optimal parameter window. 
For the considered parameters, it lies around $J/V\approx0.1-0.2$. 
Staying within this window, the minimal model retains approximate gauge invariance while displaying the C and P breaking physics characteristic of the full many-body problem.

\section{Experimental considerations\label{cha:errors}}

In this section, we discuss some experimental issues, including an outline of a meaningful experimental run, as well as possible error sources.

\subsection{Experimental sequence\label{sec:experimental_sequence}}

A meaningful experimental sequence should typically contain \textsl{(i)} initialization in a state that can be prepared with high fidelity, \textsl{(ii)} time evolution under a desired Hamiltonian, and \textsl{(iii)} measurement of a relevant observable. 
All of these points can be implemented with high accuracy in state-of-the-art trapped-ion experiments. 

For the initialization, one can choose the ground states at $m/J= \pm\infty$. Since these are product states (as sketched in Fig.~\ref{fig:systems}d), they can be prepared with high fidelity by optically pumping each ion into the appropriate internal state (this step includes preparing the ions as $\sigma$ or $S$ spins, as appropriate). 
Then, to protect gauge invariance, one needs to induce the Gauss-law interactions $H_G$ by turning on the corresponding lasers. 
Afterwards, one can turn on the system dynamics given by $H_K+H_B+H_m$, in an adiabatic or an abrupt manner, depending on whether one wants to study ground states or quantum dynamics after a sudden quench. 
Notably, applying only $H_K+H_B$, without generating $H_m$, corresponds to a quench from $m/J=\pm\infty$ to $m=0$. This quench ends in the critical region, which is a highly interesting situation \cite{Polkovnikov2011}. 

The most important information can be read out easily by measuring the local polarization, i.e., the occupation of the pseudo-spin levels. 
From this simple measurement, one can extract the signature of the parity-breaking phase transition through $\braket{S_{i,i+1}^z}$ or $\braket{\sigma_{i}^z}$. 
From the corresponding two-point correlations, one can compute $\braket{G_{i}^2}$ and thus quantify the breaking of gauge invariance. 
Complementary observables such as the $x$ spin components can be measured by rotating the spins prior to detection.

\subsection{Error sources\label{sec:errors}}

An important restriction of the proposed scheme is the condition $J/V\ll 1$. A small $J$ signifies a slow dynamics, so that one may ask whether physical phenomena associated to the QLM \eqref{eq:H0} are observable within realistic experimental time scales. 
For approximately realizing the correct gauge-invariant dynamics, 
we require values of $J\approx V/5$ (see Sec.~\ref{cha:numericsShortChains}). 
State-of-the-art experiments reach interaction strengths as large as $V/\hbar \approx 2\pi \, 1\,$kHz \cite{Britton2012,Schneider2012} \footnote{
This spin--spin interaction strength has been achieved for lasers with equal intensities on all involved ions. Here, we require $\left|\Omega_S\right|=8\left|\Omega_\sigma\right|$. 
To retain $V/\hbar \approx 2\pi \, 1\,$kHz in the case of the hyperfine qubits sketched in Fig.~\ref{fig:setup}b, one needs an increase of 
$\left|\Omega_{1,2}^{\uparrow,\downarrow}\right|$ by a factor of $\sqrt{8}$ for both the $S$ and $\sigma$ transition (while increasing the detuning $\Delta$ by a factor of $\sqrt{8}$ to keep a constant off-resonant scattering rate). While challenging, this improvement of laser power seems technically feasible.
}. 
The corresponding time scale for the system dynamics is $\hbar/J\approx 0.8\, {\mathrm{ms}}$. 
This is well below the dephasing time, which typically is $5-10\,$ms and can reach up to $100\,$ms \cite{Britton2012}. 
It is also far below realistic relaxation times, which can reach up to $1\,$s \cite{Schneider2012,Britton2012}. 
As we have seen in Fig.~\ref{fig:quench}, an adiabatic sweep with good fidelity requires only a few ms, so that the C-and-P-breaking transition should be observable in realistic setups. 

For small $J/V$, another type of error might arise in the derivation of the effective spin--spin interactions.
Namely, the step from the spin--phonon coupling~\eqref{eq:Hsp} to the spin--spin interaction~\eqref{eq:HVgeneral} involves truncating the canonical transformation~\eqref{eq:canonicalTransformation} to leading order in $\gamma\equiv\left|\Omega_n\right|\eta_{nq}/(2\delta_x^q)$. 
One could think that higher-order terms neglected when deriving the strong interactions $V$ may compete with the first-order terms of weaker interactions such as $H_1$, Eq.~\eqref{eq:H1}. 
However, the Lamb--Dicke parameter is $\eta_{nq}\ll 1$ in typical trapped-ion quantum simulations, so that $\gamma$ can be easily on the order of $1/8$ \cite{Islam2011}.
Consequently, higher-order terms in the canonical transformation are of order $\gamma^2=0.01$ \cite{Porras2004a}. 
We found in Sec.~\ref{cha:numericsShortChains} that reasonable parameters for $H_1$ are $J=\frac{KB}{V}\approx V/5$, or equivalently $K/V,B/V\approx 0.45\gg 0.01$. 

Generally speaking, the inherent errors discussed in Sec.~\ref{cha:numericsShortChains}, namely the dipolar terms $H_V^{\mathrm{err}}$ and gauge-variant contributions to $H_1$ that are not sufficiently suppresed by $H_V$, will by far dominate over other error sources, such as experimental inaccuracies. 
For example, for typical experimental parameters, errors due to finite phonon occupation, phonon heating, and micromotion can be estimated to be at most a few percent \cite{Porras2004a,Bermudez2012b}. 
Additional errors, appearing if non-commuting spin--spin interactions are induced simultaneously, are strongly suppressed if the used phonon frequencies are sufficiently different \cite{Porras2004a}. 
Let us also remark that the low-energy, many-body properties of the system are expected to be robust in the presence of small gauge-variant terms (see Ref.~\cite{Foerster1980} and \cite{Kasamatsu2012} for a general treatment and a discussion in the context of quantum simulators). 

In our analysis for the longer chains (Sec.~\ref{cha:numericsShortChains}), we assumed equidistant ions, but in a realistic linear Paul trap this is typically not truly fulfilled, resulting in inhomogeneous coupling strengths. To achieve equal distances, one could employ unharmonic trap potentials, study only the central part of a large ion chain, or resort to individual ion traps \cite{Chiaverini2008,Schmied2009b}. 
However, even if the interactions are not homogeneous, the preservation of gauge invariance enforced by $H_G$ will remain valid, although with a position-dependent strength (the position-dependence of $V$ may possibly even be compensated by the one of $K$, so that $J=K B/V$ remains homogenous). 

Moreover, realistic ion chains are governed by open boundary conditions, while in our numerical calculations of Sec.~\ref{cha:numericsShortChains} we considered periodic boundary conditions, because these reproduce better the physics of larger systems. Nevertheless, since all involved phases are gapped, we can expect a strong influence of the boundary conditions only close to quantum phase transitions. The physics of the false-vacuum decay should depend only weakly on the choice of boundary conditions.

Finally, let us note that the interactions used in our scheme have, in a different context, all been successfully implemented in experiments. 
Namely, quantum simulations using hyperfine qubits have been realized employing both the $zz$ couplings depicted in Fig.~\ref{fig:setup}b-c \cite{Friedenauer2008,Britton2012} as well as $xx$-type spin--spin couplings generated by M{\o}lmer--S{\o}rensen gates \cite{Kim2009,Kim2010,Kim2011,Islam2011,Islam2012,Richerme2013a}. 
For optical qubits, high-fidelity gates of $xx$ type \cite{Benhelm2008,Lanyon2011} and $zz$ type have been demonstrated \cite{Monz2009}. 
An alternative implementation, based on rf-coupling of hyperfine qubits in spatially varying magnetic fields, has recently also been realized \cite{Khromova2012}. 
Rotations of the coordinate system allow to choose from these coupling schemes the most practical one for a given setup. For example, it may be simpler to implement $yy$ than $zz$ interactions, and to redefine the coordinate system accordingly.

\section{Conclusion and outlook\label{sec:outlook}}

In conclusion, we have presented a scheme to realize, with existing technology, a lattice gauge theory in a chain of trapped ions. 
Studying ground-state and dynamical behavior via exact diagonalizations, we identified the optimal parameter regime for the observation of false-vacuum decay. 
An interesting roadmap for experiments would be to demonstrate the basic building block with two or three ions. Then, it could proceed to the proof-of-principle realization of the parity-breaking transition with the use of four ions, and finally aim at a full-fledged quantum simulation on larger ion chains.

Our scheme leaves way for adjustment to the concrete experimental situation. For example, the distinction between matter and gauge-field spins  could alternatively by achieved via staggered fields that shift the resonance frequencies for every other spin, or via individual laser addressing, 
or by separating the ions spatially, e.g., by employing a narrow zig-zag chain \cite{Kaufmann2012,Ulm2013,Partner2013,Mielenz2013}. 
Moreover, if a large number of laser frequencies is available, some of the most prominent errors due to dipolar interactions could be removed by appropriately engineering the spin--phonon couplings \cite{Korenblit2012}. 
An additional advantage of our scheme is that imperfections in the realization of the Gauss law can be monitored by means of post-selection measurement of the local pseudo-spin polarization along the $z$-direction. 

Regarding future perspectives, since the Jordan--Wigner transformation is only practical in one dimension, it seems to be out of reach for trapped ions to quantum simulate lattice gauge theories of fermionic matter in higher dimensions in an analog setup. 
This limitation can be overcome in digital quantum simulations, where the Jordan--Wigner mapping is useful also in higher dimensions, resulting in a number of gates that scales polynomially with the number of involved fermionic modes 
\cite{Nielsen2000,Ortiz2001,Casanova2012}. 
While analog quantum simulations have, due to the `always-on' nature of interactions, advantages in scalability, digital approaches allow larger control and especially error correction~\cite{Hauke2011d}. 
But despite the limitations set by the Jordan--Wigner transformation, a variety of relevant lattice gauge theories should be accessible to analog quantum simulation in larger dimensions, e.g., in the two-dimensional ion crystals that can now be realized in experiment \cite{Britton2012}. 
Interesting examples may include gauge fields coupled to bosonic matter~\cite{Lacroix2010}, 
possibly allowing the observation of the Higgs mechanism according to the Fradkin--Shenker scenario in 2D systems \cite{Fradkin1979} (this phenomenon, beyond its intrinsic interest, has drawn particular attention in the context of condensed-matter theory, see, e.g., \cite{Motrunich2002}).
Another interesting future direction would be the investigation of pure gauge theories, either Abelian or non-Abelian~\cite{Lacroix2010}. The lattice versions of such gauge theories are often directly formulated in the language of spins, which makes trapped-ion setups seem rather promising. For example, a natural next step would be the quantum simulation of quantum dimer models as are usually obtained from frustrated spin Hamiltonians~\cite{Moessner2001,Lacroix2010}.

In the shorter term, a number of rather straightforward extensions to the scheme proposed here seems foreseeable. 
For example, one could use three different types of pseudo-spins to simulate two flavors of matter fields coupled to a single gauge field. 
As another conceptually simple generalization, one could mimic $S=1$ gauge fields by the use of three internal levels \cite{Ivanov2011,Grass2013}. 
Finally, it will be interesting to explore alternative possibilities to protect the gauge invariance, e.g., by the use of classical noise  \cite{Stannigel2013}.

{\bf Acknowledgments --}
We thank D.~Banerjee,  R.~Blatt, R.~Gerritsma, C.~Hempel, P.~Jurcevic, B.~Lanyon, S.~Montangero, 
M.~M\"uller, E.~Rico, Ch.~Roos, and U.-J.~Wiese for useful and fruitful discussions. 
M.D.~acknowledges support by the
European Commission via the integrated project AQUTE. We further acknowledge
support by SIQS, ERC Synergy grant UQUAM, and the Austrian Science Fund FWF (SFB FOQUS F4015-N16).


\begin{thebibliography}{97}%
\makeatletter
\providecommand \@ifxundefined [1]{%
 \@ifx{#1\undefined}
}%
\providecommand \@ifnum [1]{%
 \ifnum #1\expandafter \@firstoftwo
 \else \expandafter \@secondoftwo
 \fi
}%
\providecommand \@ifx [1]{%
 \ifx #1\expandafter \@firstoftwo
 \else \expandafter \@secondoftwo
 \fi
}%
\providecommand \natexlab [1]{#1}%
\providecommand \enquote  [1]{``#1''}%
\providecommand \bibnamefont  [1]{#1}%
\providecommand \bibfnamefont [1]{#1}%
\providecommand \citenamefont [1]{#1}%
\providecommand \href@noop [0]{\@secondoftwo}%
\providecommand \href [0]{\begingroup \@sanitize@url \@href}%
\providecommand \@href[1]{\@@startlink{#1}\@@href}%
\providecommand \@@href[1]{\endgroup#1\@@endlink}%
\providecommand \@sanitize@url [0]{\catcode `\\12\catcode `\$12\catcode
  `\&12\catcode `\#12\catcode `\^12\catcode `\_12\catcode `\%12\relax}%
\providecommand \@@startlink[1]{}%
\providecommand \@@endlink[0]{}%
\providecommand \url  [0]{\begingroup\@sanitize@url \@url }%
\providecommand \@url [1]{\endgroup\@href {#1}{\urlprefix }}%
\providecommand \urlprefix  [0]{URL }%
\providecommand \Eprint [0]{\href }%
\providecommand \doibase [0]{http://dx.doi.org/}%
\providecommand \selectlanguage [0]{\@gobble}%
\providecommand \bibinfo  [0]{\@secondoftwo}%
\providecommand \bibfield  [0]{\@secondoftwo}%
\providecommand \translation [1]{[#1]}%
\providecommand \BibitemOpen [0]{}%
\providecommand \bibitemStop [0]{}%
\providecommand \bibitemNoStop [0]{.\EOS\space}%
\providecommand \EOS [0]{\spacefactor3000\relax}%
\providecommand \BibitemShut  [1]{\csname bibitem#1\endcsname}%
\let\auto@bib@innerbib\@empty
%</preamble>
\bibitem [{\citenamefont {H{\"a}ffner}\ \emph {et~al.}(2008)\citenamefont
  {H{\"a}ffner}, \citenamefont {Roos},\ and\ \citenamefont
  {Blatt}}]{Haeffner2008}%
  \BibitemOpen
  \bibfield  {author} {\bibinfo {author} {\bibfnamefont {H.}~\bibnamefont
  {H{\"a}ffner}}, \bibinfo {author} {\bibfnamefont {C.F.}\ \bibnamefont
  {Roos}}, \ and\ \bibinfo {author} {\bibfnamefont {R.}~\bibnamefont {Blatt}},\
  }\bibfield  {title} {\enquote {\bibinfo {title} {Quantum computing with
  trapped ions},}\ }\href@noop {} {\bibfield  {journal} {\bibinfo  {journal}
  {Phys. Rep.}\ }\textbf {\bibinfo {volume} {469}},\ \bibinfo {pages} {155}
  (\bibinfo {year} {2008})}\BibitemShut {NoStop}%
\bibitem [{\citenamefont {Monroe}\ and\ \citenamefont
  {Kim}(2013)}]{Monroe2013}%
  \BibitemOpen
  \bibfield  {author} {\bibinfo {author} {\bibfnamefont {C.}~\bibnamefont
  {Monroe}}\ and\ \bibinfo {author} {\bibfnamefont {J.}~\bibnamefont {Kim}},\
  }\bibfield  {title} {\enquote {\bibinfo {title} {Scaling the ion trap quantum
  processor},}\ }\href@noop {} {\bibfield  {journal} {\bibinfo  {journal}
  {Science}\ }\textbf {\bibinfo {volume} {339}},\ \bibinfo {pages} {1164}
  (\bibinfo {year} {2013})}\BibitemShut {NoStop}%
\bibitem [{\citenamefont {Benhelm}\ \emph {et~al.}(2008)\citenamefont
  {Benhelm}, \citenamefont {Kirchmair}, \citenamefont {Roos},\ and\
  \citenamefont {Blatt}}]{Benhelm2008}%
  \BibitemOpen
  \bibfield  {author} {\bibinfo {author} {\bibfnamefont {J.}~\bibnamefont
  {Benhelm}}, \bibinfo {author} {\bibfnamefont {G.}~\bibnamefont {Kirchmair}},
  \bibinfo {author} {\bibfnamefont {C.F.}\ \bibnamefont {Roos}}, \ and\
  \bibinfo {author} {\bibfnamefont {R.}~\bibnamefont {Blatt}},\ }\bibfield
  {title} {\enquote {\bibinfo {title} {Towards fault-tolerant quantum computing
  with trapped ions},}\ }\href@noop {} {\bibfield  {journal} {\bibinfo
  {journal} {Nature Physics}\ }\textbf {\bibinfo {volume} {4}},\ \bibinfo
  {pages} {463} (\bibinfo {year} {2008})}\BibitemShut {NoStop}%
\bibitem [{\citenamefont {Monz}\ \emph {et~al.}()\citenamefont {Monz},
  \citenamefont {Kim}, \citenamefont {Villar}, \citenamefont {Schindler},
  \citenamefont {Chwalla}, \citenamefont {Riebe}, \citenamefont {Roos},
  \citenamefont {H\"affner}, \citenamefont {H\"ansel}, \citenamefont
  {Hennrich},\ and\ \citenamefont {Blatt}}]{Monz2009}%
  \BibitemOpen
  \bibfield  {author} {\bibinfo {author} {\bibfnamefont {T.}~\bibnamefont
  {Monz}}, \bibinfo {author} {\bibfnamefont {K.}~\bibnamefont {Kim}}, \bibinfo
  {author} {\bibfnamefont {A.~S.}\ \bibnamefont {Villar}}, \bibinfo {author}
  {\bibfnamefont {P.}~\bibnamefont {Schindler}}, \bibinfo {author}
  {\bibfnamefont {M.}~\bibnamefont {Chwalla}}, \bibinfo {author} {\bibfnamefont
  {M.}~\bibnamefont {Riebe}}, \bibinfo {author} {\bibfnamefont {C.~F.}\
  \bibnamefont {Roos}}, \bibinfo {author} {\bibfnamefont {H.}~\bibnamefont
  {H\"affner}}, \bibinfo {author} {\bibfnamefont {W.}~\bibnamefont {H\"ansel}},
  \bibinfo {author} {\bibfnamefont {M.}~\bibnamefont {Hennrich}}, \ and\
  \bibinfo {author} {\bibfnamefont {R.}~\bibnamefont {Blatt}},\ }\bibfield
  {title} {\enquote {\bibinfo {title} {Realization of universal ion-trap
  quantum computation with decoherence-free qubits},}\ }\href@noop {}
  {\bibfield  {journal} {\bibinfo  {journal} {Phys. Rev. Lett.}\ }\textbf
  {\bibinfo {volume} {103}},\ \bibinfo {pages} {200503} (\bibinfo {year} {2009})}\BibitemShut {NoStop}%
\bibitem [{\citenamefont {Lanyon}\ \emph {et~al.}(2011)\citenamefont {Lanyon},
  \citenamefont {Hempel}, \citenamefont {Nigg}, \citenamefont {M{\"u}ller},
  \citenamefont {Gerritsma}, \citenamefont {Z{\"a}hringer}, \citenamefont
  {Schindler}, \citenamefont {Barreiro}, \citenamefont {Rambach}, \citenamefont
  {Kirchmair}, \citenamefont {Hennrich}, \citenamefont {Zoller}, \citenamefont
  {Blatt},\ and\ \citenamefont {Roos}}]{Lanyon2011}%
  \BibitemOpen
  \bibfield  {author} {\bibinfo {author} {\bibfnamefont {B.~P.}\ \bibnamefont
  {Lanyon}}, \bibinfo {author} {\bibfnamefont {C.}~\bibnamefont {Hempel}},
  \bibinfo {author} {\bibfnamefont {D.}~\bibnamefont {Nigg}}, \bibinfo {author}
  {\bibfnamefont {M.}~\bibnamefont {M{\"u}ller}}, \bibinfo {author}
  {\bibfnamefont {R.}~\bibnamefont {Gerritsma}}, \bibinfo {author}
  {\bibfnamefont {F.}~\bibnamefont {Z{\"a}hringer}}, \bibinfo {author}
  {\bibfnamefont {P.}~\bibnamefont {Schindler}}, \bibinfo {author}
  {\bibfnamefont {J.~T.}\ \bibnamefont {Barreiro}}, \bibinfo {author}
  {\bibfnamefont {M.}~\bibnamefont {Rambach}}, \bibinfo {author} {\bibfnamefont
  {G.}~\bibnamefont {Kirchmair}}, \bibinfo {author} {\bibfnamefont
  {M.}~\bibnamefont {Hennrich}}, \bibinfo {author} {\bibfnamefont
  {P.}~\bibnamefont {Zoller}}, \bibinfo {author} {\bibfnamefont
  {R.}~\bibnamefont {Blatt}}, \ and\ \bibinfo {author} {\bibfnamefont {C.~F.}\
  \bibnamefont {Roos}},\ }\bibfield  {title} {\enquote {\bibinfo {title}
  {Universal digital quantum simulation with trapped ions},}\ }\href {\doibase
  10.1126/science.1208001} {\bibfield  {journal} {\bibinfo  {journal}
  {Science}\ }\textbf {\bibinfo {volume} {334}},\ \bibinfo {pages} {57} (\bibinfo
  {year} {2011})}\BibitemShut {NoStop}%
\bibitem [{\citenamefont {Khromova}\ \emph {et~al.}(2012)\citenamefont
  {Khromova}, \citenamefont {Piltz}, \citenamefont {Scharfenberger},
  \citenamefont {Gloger}, \citenamefont {Johanning}, \citenamefont
  {Var{\'o}n},\ and\ \citenamefont {Wunderlich}}]{Khromova2012}%
  \BibitemOpen
  \bibfield  {author} {\bibinfo {author} {\bibfnamefont {A.}~\bibnamefont
  {Khromova}}, \bibinfo {author} {\bibfnamefont {Ch.}\ \bibnamefont {Piltz}},
  \bibinfo {author} {\bibfnamefont {B.}~\bibnamefont {Scharfenberger}},
  \bibinfo {author} {\bibfnamefont {T.~F.}\ \bibnamefont {Gloger}}, \bibinfo
  {author} {\bibfnamefont {M.}~\bibnamefont {Johanning}}, \bibinfo {author}
  {\bibfnamefont {A.~F.}\ \bibnamefont {Var{\'o}n}}, \ and\ \bibinfo {author}
  {\bibfnamefont {Ch.}\ \bibnamefont {Wunderlich}},\ }\bibfield  {title}
  {\enquote {\bibinfo {title} {Designer spin pseudomolecule implemented with
  trapped ions in a magnetic gradient},}\ }\href@noop {} {\bibfield  {journal}
  {\bibinfo  {journal} {Phys. Rev. Lett.}\ }\textbf {\bibinfo {volume} {108}},\
  \bibinfo {pages} {220502} (\bibinfo {year} {2012})}\BibitemShut {NoStop}%
\bibitem [{\citenamefont {Johanning}\ \emph {et~al.}(2009)\citenamefont
  {Johanning}, \citenamefont {Var\'{o}n},\ and\ \citenamefont
  {Wunderlich}}]{Johanning2009}%
  \BibitemOpen
  \bibfield  {author} {\bibinfo {author} {\bibfnamefont {M.}~\bibnamefont
  {Johanning}}, \bibinfo {author} {\bibfnamefont {A.F.}\ \bibnamefont
  {Var\'{o}n}}, \ and\ \bibinfo {author} {\bibfnamefont {Ch.}\ \bibnamefont
  {Wunderlich}},\ }\bibfield  {title} {\enquote {\bibinfo {title} {Quantum
  simulations with cold trapped ions},}\ }\href@noop {} {\bibfield  {journal}
  {\bibinfo  {journal} {J. Phys. B}\ }\textbf {\bibinfo {volume} {42}},\
  \bibinfo {pages} {154009} (\bibinfo {year} {2009})}\BibitemShut {NoStop}%
\bibitem [{\citenamefont {Schneider}\ \emph {et~al.}(2012)\citenamefont
  {Schneider}, \citenamefont {Porras},\ and\ \citenamefont
  {Schaetz}}]{Schneider2012}%
  \BibitemOpen
  \bibfield  {author} {\bibinfo {author} {\bibfnamefont {Ch.}\ \bibnamefont
  {Schneider}}, \bibinfo {author} {\bibfnamefont {D.}~\bibnamefont {Porras}}, \
  and\ \bibinfo {author} {\bibfnamefont {T.}~\bibnamefont {Schaetz}},\
  }\bibfield  {title} {\enquote {\bibinfo {title} {Experimental quantum
  simulations of many-body physics with trapped ions},}\ }\href {\doibase
  10.1088/0034-4885/75/2/024401} {\bibfield  {journal} {\bibinfo  {journal}
  {Rep. Prog. Phys.}\ }\textbf {\bibinfo {volume} {75}},\ \bibinfo {pages}
  {024401} (\bibinfo {year} {2012})}\BibitemShut {NoStop}%
\bibitem [{\citenamefont {Blatt}\ and\ \citenamefont {Roos}(2012)}]{Blatt2012}%
  \BibitemOpen
  \bibfield  {author} {\bibinfo {author} {\bibfnamefont {R.}~\bibnamefont
  {Blatt}}\ and\ \bibinfo {author} {\bibfnamefont {C.~F.}\ \bibnamefont
  {Roos}},\ }\bibfield  {title} {\enquote {\bibinfo {title} {Quantum
  simulations with trapped ions},}\ }\href {\doibase doi:10.1038/nphys2252}
  {\bibfield  {journal} {\bibinfo  {journal} {Nat. Phys.}\ }\textbf {\bibinfo
  {volume} {8}},\ \bibinfo {pages} {277} (\bibinfo {year} {2012})}\BibitemShut
  {NoStop}%
\bibitem [{\citenamefont {Barreiro}\ \emph {et~al.}(2011)\citenamefont
  {Barreiro}, \citenamefont {M{\"u}ller}, \citenamefont {Schindler},
  \citenamefont {Nigg}, \citenamefont {Monz}, \citenamefont {Chwalla},
  \citenamefont {Hennrich}, \citenamefont {Roos}, \citenamefont {Zoller},\ and\
  \citenamefont {Blatt}}]{Barreiro2011}%
  \BibitemOpen
  \bibfield  {author} {\bibinfo {author} {\bibfnamefont {J.}~\bibnamefont
  {Barreiro}}, \bibinfo {author} {\bibfnamefont {M.}~\bibnamefont
  {M{\"u}ller}}, \bibinfo {author} {\bibfnamefont {Ph.}\ \bibnamefont
  {Schindler}}, \bibinfo {author} {\bibfnamefont {D.}~\bibnamefont {Nigg}},
  \bibinfo {author} {\bibfnamefont {Th.}\ \bibnamefont {Monz}}, \bibinfo
  {author} {\bibfnamefont {M.}~\bibnamefont {Chwalla}}, \bibinfo {author}
  {\bibfnamefont {M.}~\bibnamefont {Hennrich}}, \bibinfo {author}
  {\bibfnamefont {Ch.}\ \bibnamefont {Roos}}, \bibinfo {author} {\bibfnamefont
  {P.}~\bibnamefont {Zoller}}, \ and\ \bibinfo {author} {\bibfnamefont
  {R.}~\bibnamefont {Blatt}},\ }\bibfield  {title} {\enquote {\bibinfo {title}
  {An open-system quantum simulator with trapped ions},}\ }\href {\doibase
  doi:10.1038/nature09801} {\bibfield  {journal} {\bibinfo  {journal} {Nature}\
  }\textbf {\bibinfo {volume} {470}},\ \bibinfo {pages} {486} (\bibinfo {year}
  {2011})}\BibitemShut {NoStop}%
\bibitem [{\citenamefont {Friedenauer}\ \emph {et~al.}(2008)\citenamefont
  {Friedenauer}, \citenamefont {Schmitz}, \citenamefont {Glueckert},
  \citenamefont {Porras},\ and\ \citenamefont {Schaetz}}]{Friedenauer2008}%
  \BibitemOpen
  \bibfield  {author} {\bibinfo {author} {\bibfnamefont {A.}~\bibnamefont
  {Friedenauer}}, \bibinfo {author} {\bibfnamefont {H.}~\bibnamefont
  {Schmitz}}, \bibinfo {author} {\bibfnamefont {J.~T.}\ \bibnamefont
  {Glueckert}}, \bibinfo {author} {\bibfnamefont {D.}~\bibnamefont {Porras}}, \
  and\ \bibinfo {author} {\bibfnamefont {T.}~\bibnamefont {Schaetz}},\
  }\bibfield  {title} {\enquote {\bibinfo {title} {Simulating a quantum magnet
  with trapped ions.}}\ }\href@noop {} {\bibfield  {journal} {\bibinfo
  {journal} {Nat. Phys.}\ }\textbf {\bibinfo {volume} {4}},\ \bibinfo {pages}
  {757} (\bibinfo {year} {2008})}\BibitemShut {NoStop}%
\bibitem [{\citenamefont {Kim}\ \emph {et~al.}(2009)\citenamefont {Kim},
  \citenamefont {Chang}, \citenamefont {Islam}, \citenamefont {Korenblit},
  \citenamefont {Duan},\ and\ \citenamefont {Monroe}}]{Kim2009}%
  \BibitemOpen
  \bibfield  {author} {\bibinfo {author} {\bibfnamefont {K.}~\bibnamefont
  {Kim}}, \bibinfo {author} {\bibfnamefont {M.-S.}\ \bibnamefont {Chang}},
  \bibinfo {author} {\bibfnamefont {R.}~\bibnamefont {Islam}}, \bibinfo
  {author} {\bibfnamefont {S.}~\bibnamefont {Korenblit}}, \bibinfo {author}
  {\bibfnamefont {L.-M.}\ \bibnamefont {Duan}}, \ and\ \bibinfo {author}
  {\bibfnamefont {C.}~\bibnamefont {Monroe}},\ }\bibfield  {title} {\enquote
  {\bibinfo {title} {Entanglement and tunable spin-spin couplings between
  trapped ions using multiple transverse modes},}\ }\href@noop {} {\bibfield
  {journal} {\bibinfo  {journal} {Phys. Rev. Lett.}\ }\textbf {\bibinfo
  {volume} {103}},\ \bibinfo {pages} {120502} (\bibinfo {year}
  {2009})}\BibitemShut {NoStop}%
\bibitem [{\citenamefont {Kim}\ \emph {et~al.}(2010)\citenamefont {Kim},
  \citenamefont {Chang}, \citenamefont {Korenblit}, \citenamefont {Islam},
  \citenamefont {Edwards}, \citenamefont {Freericks}, \citenamefont {Lin},
  \citenamefont {Duan},\ and\ \citenamefont {Monroe}}]{Kim2010}%
  \BibitemOpen
  \bibfield  {author} {\bibinfo {author} {\bibfnamefont {K.}~\bibnamefont
  {Kim}}, \bibinfo {author} {\bibfnamefont {M.-S.}\ \bibnamefont {Chang}},
  \bibinfo {author} {\bibfnamefont {S.}~\bibnamefont {Korenblit}}, \bibinfo
  {author} {\bibfnamefont {R.}~\bibnamefont {Islam}}, \bibinfo {author}
  {\bibfnamefont {E.~E.}\ \bibnamefont {Edwards}}, \bibinfo {author}
  {\bibfnamefont {J.~K.}\ \bibnamefont {Freericks}}, \bibinfo {author}
  {\bibfnamefont {G.-D.}\ \bibnamefont {Lin}}, \bibinfo {author} {\bibfnamefont
  {L.-M.}\ \bibnamefont {Duan}}, \ and\ \bibinfo {author} {\bibfnamefont
  {C.}~\bibnamefont {Monroe}},\ }\bibfield  {title} {\enquote {\bibinfo {title}
  {Quantum simulation of frustrated {I}sing spins with trapped ions},}\
  }\href@noop {} {\bibfield  {journal} {\bibinfo  {journal} {Nature}\ }\textbf
  {\bibinfo {volume} {465}},\ \bibinfo {pages} {590} (\bibinfo {year}
  {2010})}\BibitemShut {NoStop}%
\bibitem [{\citenamefont {Kim}\ \emph {et~al.}(2011)\citenamefont {Kim},
  \citenamefont {Korenblit}, \citenamefont {Islam}, \citenamefont {Edwards},
  \citenamefont {Chang}, \citenamefont {Noh}, \citenamefont {Carmichael},
  \citenamefont {Lin}, \citenamefont {Duan}, \citenamefont {Wang},
  \citenamefont {Freericks},\ and\ \citenamefont {Monroe}}]{Kim2011}%
  \BibitemOpen
  \bibfield  {author} {\bibinfo {author} {\bibfnamefont {K.}~\bibnamefont
  {Kim}}, \bibinfo {author} {\bibfnamefont {S.}~\bibnamefont {Korenblit}},
  \bibinfo {author} {\bibfnamefont {R.}~\bibnamefont {Islam}}, \bibinfo
  {author} {\bibfnamefont {E.~E.}\ \bibnamefont {Edwards}}, \bibinfo {author}
  {\bibfnamefont {M.-S.}\ \bibnamefont {Chang}}, \bibinfo {author}
  {\bibfnamefont {C.}~\bibnamefont {Noh}}, \bibinfo {author} {\bibfnamefont
  {H.}~\bibnamefont {Carmichael}}, \bibinfo {author} {\bibfnamefont {G.-D.}\
  \bibnamefont {Lin}}, \bibinfo {author} {\bibfnamefont {L.-M.}\ \bibnamefont
  {Duan}}, \bibinfo {author} {\bibfnamefont {C.~C.~Joseph}\ \bibnamefont
  {Wang}}, \bibinfo {author} {\bibfnamefont {J.~K.}\ \bibnamefont {Freericks}},
  \ and\ \bibinfo {author} {\bibfnamefont {C.}~\bibnamefont {Monroe}},\
  }\bibfield  {title} {\enquote {\bibinfo {title} {Quantum simulation of the
  transverse {I}sing model with trapped ions},}\ }\href {\doibase
  doi:10.1088/1367-2630/13/10/105003} {\bibfield  {journal} {\bibinfo
  {journal} {New J. Phys.}\ }\textbf {\bibinfo {volume} {13}},\ \bibinfo
  {pages} {105003} (\bibinfo {year} {2011})}\BibitemShut {NoStop}%
\bibitem [{\citenamefont {Islam}\ \emph {et~al.}(2011)\citenamefont {Islam},
  \citenamefont {Edwards}, \citenamefont {Kim}, \citenamefont {Korenblit},
  \citenamefont {Noh}, \citenamefont {Carmichael}, \citenamefont {Lin},
  \citenamefont {Duan}, \citenamefont {Wang}, \citenamefont {Freericks},\ and\
  \citenamefont {Monroe}}]{Islam2011}%
  \BibitemOpen
  \bibfield  {author} {\bibinfo {author} {\bibfnamefont {R.}~\bibnamefont
  {Islam}}, \bibinfo {author} {\bibfnamefont {E.E.}\ \bibnamefont {Edwards}},
  \bibinfo {author} {\bibfnamefont {K.}~\bibnamefont {Kim}}, \bibinfo {author}
  {\bibfnamefont {S.}~\bibnamefont {Korenblit}}, \bibinfo {author}
  {\bibfnamefont {C.}~\bibnamefont {Noh}}, \bibinfo {author} {\bibfnamefont
  {H.}~\bibnamefont {Carmichael}}, \bibinfo {author} {\bibfnamefont {G.-D.}\
  \bibnamefont {Lin}}, \bibinfo {author} {\bibfnamefont {L.-M.}\ \bibnamefont
  {Duan}}, \bibinfo {author} {\bibfnamefont {C.-C.~Joseph}\ \bibnamefont
  {Wang}}, \bibinfo {author} {\bibfnamefont {J.K.}\ \bibnamefont {Freericks}},
  \ and\ \bibinfo {author} {\bibfnamefont {C.}~\bibnamefont {Monroe}},\
  }\bibfield  {title} {\enquote {\bibinfo {title} {Onset of a quantum phase
  transition with a trapped ion quantum simulator},}\ }\href {\doibase
  doi:10.1038/ncomms1374} {\bibfield  {journal} {\bibinfo  {journal} {Nat.
  Commun.}\ }\textbf {\bibinfo {volume} {2}},\ \bibinfo {pages} {377} (\bibinfo
  {year} {2011})}\BibitemShut {NoStop}%
\bibitem [{\citenamefont {Islam}\ \emph {et~al.}(2013)\citenamefont {Islam},
  \citenamefont {Senko}, \citenamefont {Campbell}, \citenamefont {Korenblit},
  \citenamefont {Smith}, \citenamefont {Lee}, \citenamefont {Edwards},
  \citenamefont {Wang}, \citenamefont {Freericks},\ and\ \citenamefont
  {Monroe}}]{Islam2012}%
  \BibitemOpen
  \bibfield  {author} {\bibinfo {author} {\bibfnamefont {R.}~\bibnamefont
  {Islam}}, \bibinfo {author} {\bibfnamefont {C.}~\bibnamefont {Senko}},
  \bibinfo {author} {\bibfnamefont {W.~C.}\ \bibnamefont {Campbell}}, \bibinfo
  {author} {\bibfnamefont {S.}~\bibnamefont {Korenblit}}, \bibinfo {author}
  {\bibfnamefont {J.}~\bibnamefont {Smith}}, \bibinfo {author} {\bibfnamefont
  {A.}~\bibnamefont {Lee}}, \bibinfo {author} {\bibfnamefont {E.~E.}\
  \bibnamefont {Edwards}}, \bibinfo {author} {\bibfnamefont {C.-C.~J.}\
  \bibnamefont {Wang}}, \bibinfo {author} {\bibfnamefont {J.~K.}\ \bibnamefont
  {Freericks}}, \ and\ \bibinfo {author} {\bibfnamefont {C.}~\bibnamefont
  {Monroe}},\ }\bibfield  {title} {\enquote {\bibinfo {title} {Emergence and
  frustration of magnetic order with variable-range interactions in a trapped
  ion quantum simulator},}\ }\href {\doibase DOI: 10.1126/science.1232296}
  {\bibfield  {journal} {\bibinfo  {journal} {Science}\ }\textbf {\bibinfo
  {volume} {340}},\ \bibinfo {pages} {583} (\bibinfo {year}
  {2013})}\BibitemShut {NoStop}%
\bibitem [{\citenamefont {Britton}\ \emph {et~al.}(2012)\citenamefont
  {Britton}, \citenamefont {Sawyer}, \citenamefont {Keith}, \citenamefont
  {Wang}, \citenamefont {Freericks}, \citenamefont {Uys}, \citenamefont
  {Biercuk},\ and\ \citenamefont {Bollinger}}]{Britton2012}%
  \BibitemOpen
  \bibfield  {author} {\bibinfo {author} {\bibfnamefont {J.W.}\ \bibnamefont
  {Britton}}, \bibinfo {author} {\bibfnamefont {B.C.}\ \bibnamefont {Sawyer}},
  \bibinfo {author} {\bibfnamefont {A.C.}\ \bibnamefont {Keith}}, \bibinfo
  {author} {\bibfnamefont {C.-C.}\ \bibnamefont {Wang}}, \bibinfo {author}
  {\bibfnamefont {J.K.}\ \bibnamefont {Freericks}}, \bibinfo {author}
  {\bibfnamefont {H.}~\bibnamefont {Uys}}, \bibinfo {author} {\bibfnamefont
  {M.J.}\ \bibnamefont {Biercuk}}, \ and\ \bibinfo {author} {\bibfnamefont
  {J.J.}\ \bibnamefont {Bollinger}},\ }\bibfield  {title} {\enquote {\bibinfo
  {title} {Engineered two-dimensional {I}sing interactions in a trapped-ion
  quantum simulator with hundreds of spins},}\ }\href {\doibase
  doi:10.1038/nature10981} {\bibfield  {journal} {\bibinfo  {journal} {Nature}\
  }\textbf {\bibinfo {volume} {484}},\ \bibinfo {pages} {489} (\bibinfo {year}
  {2012})}\BibitemShut {NoStop}%
\bibitem [{\citenamefont {Richerme}\ \emph
  {et~al.}(2013{\natexlab{a}})\citenamefont {Richerme}, \citenamefont {Senko},
  \citenamefont {Korenblit}, \citenamefont {Smith}, \citenamefont {Lee},
  \citenamefont {Campbell},\ and\ \citenamefont {Monroe}}]{Richerme2013a}%
  \BibitemOpen
  \bibfield  {author} {\bibinfo {author} {\bibfnamefont {P.}~\bibnamefont
  {Richerme}}, \bibinfo {author} {\bibfnamefont {C.}~\bibnamefont {Senko}},
  \bibinfo {author} {\bibfnamefont {S.}~\bibnamefont {Korenblit}}, \bibinfo
  {author} {\bibfnamefont {J.}~\bibnamefont {Smith}}, \bibinfo {author}
  {\bibfnamefont {A.}~\bibnamefont {Lee}}, \bibinfo {author} {\bibfnamefont
  {W.C.}\ \bibnamefont {Campbell}}, \ and\ \bibinfo {author} {\bibfnamefont
  {Ch.}\ \bibnamefont {Monroe}},\ }\bibfield  {title} {\enquote {\bibinfo
  {title} {Quantum Catalysis of Magnetic Phase Transitions in a Quantum 
	Simulator},}\ }\href@noop {} {\bibfield  {journal} {\bibinfo
  {journal} {Phys. Rev. Lett.}\ }\textbf {\bibinfo {volume}
  {111}},\ \bibinfo {pages} {100506} (\bibinfo {year} {2013})}\BibitemShut {NoStop}%
\bibitem [{\citenamefont {Cirac}\ and\ \citenamefont
  {Zoller}(2012)}]{Cirac2012}%
  \BibitemOpen
  \bibfield  {author} {\bibinfo {author} {\bibfnamefont {J.I.}\ \bibnamefont
  {Cirac}}\ and\ \bibinfo {author} {\bibfnamefont {P.}~\bibnamefont {Zoller}},\
  }\bibfield  {title} {\enquote {\bibinfo {title} {Goals and opportunities in
  quantum simulation},}\ }\href {\doibase doi:10.1038/nphys2275} {\bibfield
  {journal} {\bibinfo  {journal} {Nat. Phys.}\ }\textbf {\bibinfo {volume}
  {8}},\ \bibinfo {pages} {264} (\bibinfo {year} {2012})}\BibitemShut {NoStop}%
\bibitem [{\citenamefont {Hauke}\ \emph {et~al.}(2012)\citenamefont {Hauke},
  \citenamefont {Cucchietti}, \citenamefont {Tagliacozzo}, \citenamefont
  {Deutsch},\ and\ \citenamefont {Lewenstein}}]{Hauke2011d}%
  \BibitemOpen
  \bibfield  {author} {\bibinfo {author} {\bibfnamefont {P.}~\bibnamefont
  {Hauke}}, \bibinfo {author} {\bibfnamefont {F.M.}\ \bibnamefont
  {Cucchietti}}, \bibinfo {author} {\bibfnamefont {L.}~\bibnamefont
  {Tagliacozzo}}, \bibinfo {author} {\bibfnamefont {I.}~\bibnamefont
  {Deutsch}}, \ and\ \bibinfo {author} {\bibfnamefont {M.}~\bibnamefont
  {Lewenstein}},\ }\bibfield  {title} {\enquote {\bibinfo {title} {Can one
  trust quantum simulators?}}\ }\href@noop {} {\bibfield  {journal} {\bibinfo
  {journal} {Rep. Prog. Phys.}\ }\textbf {\bibinfo {volume} {75}},\ \bibinfo
  {pages} {082401} (\bibinfo {year} {2012})}\BibitemShut {NoStop}%
\bibitem [{\citenamefont {Lewenstein}\ \emph {et~al.}(2012)\citenamefont
  {Lewenstein}, \citenamefont {Sanpera},\ and\ \citenamefont
  {Ahufinger}}]{Lewenstein2012}%
  \BibitemOpen
  \bibfield  {author} {\bibinfo {author} {\bibfnamefont {M.}~\bibnamefont
  {Lewenstein}}, \bibinfo {author} {\bibfnamefont {A.}~\bibnamefont {Sanpera}},
  \ and\ \bibinfo {author} {\bibfnamefont {V.}~\bibnamefont {Ahufinger}},\
  }\href@noop {} {\emph {\bibinfo {title} {Ultracold Atoms in Optical Lattices:
  Simulating Quantum Many-Body Systems}}}\ (\bibinfo  {publisher} {Oxford
  University Press, Oxford},\ \bibinfo {year} {2012})\BibitemShut {NoStop}%
\bibitem [{\citenamefont {Bloch}\ \emph {et~al.}(2012)\citenamefont {Bloch},
  \citenamefont {Dalibard},\ and\ \citenamefont {Nascimb\`ene}}]{Bloch2012}%
  \BibitemOpen
  \bibfield  {author} {\bibinfo {author} {\bibfnamefont {I.}~\bibnamefont
  {Bloch}}, \bibinfo {author} {\bibfnamefont {J.}~\bibnamefont {Dalibard}}, \
  and\ \bibinfo {author} {\bibfnamefont {S.}~\bibnamefont {Nascimb\`ene}},\
  }\bibfield  {title} {\enquote {\bibinfo {title} {Quantum simulations with
  ultracold quantum gases},}\ }\href {\doibase doi:10.1038/nphys2259}
  {\bibfield  {journal} {\bibinfo  {journal} {Nat. Phys.}\ }\textbf {\bibinfo
  {volume} {8}},\ \bibinfo {pages} {267} (\bibinfo {year} {2012})}\BibitemShut
  {NoStop}%
\bibitem [{\citenamefont {Aspuru-Guzik}\ and\ \citenamefont
  {Walther}(2012)}]{AspuruGuzik2012}%
  \BibitemOpen
  \bibfield  {author} {\bibinfo {author} {\bibfnamefont {A.}~\bibnamefont
  {Aspuru-Guzik}}\ and\ \bibinfo {author} {\bibfnamefont {Ph.}\ \bibnamefont
  {Walther}},\ }\bibfield  {title} {\enquote {\bibinfo {title} {Photonic
  quantum simulators},}\ }\href {\doibase doi:10.1038/nphys2253} {\bibfield
  {journal} {\bibinfo  {journal} {Nat. Phys.}\ }\textbf {\bibinfo {volume}
  {8}},\ \bibinfo {pages} {285} (\bibinfo {year} {2012})}\BibitemShut {NoStop}%
\bibitem [{\citenamefont {Houck}\ \emph {et~al.}(2012)\citenamefont {Houck},
  \citenamefont {T\"ureci},\ and\ \citenamefont {Koch}}]{Houck2012}%
  \BibitemOpen
  \bibfield  {author} {\bibinfo {author} {\bibfnamefont {A.A.}\ \bibnamefont
  {Houck}}, \bibinfo {author} {\bibfnamefont {H.E.}\ \bibnamefont {T\"ureci}},
  \ and\ \bibinfo {author} {\bibfnamefont {J.}~\bibnamefont {Koch}},\
  }\bibfield  {title} {\enquote {\bibinfo {title} {On-chip quantum simulation
  with superconducting circuits},}\ }\href {\doibase doi:10.1038/nphys2251}
  {\bibfield  {journal} {\bibinfo  {journal} {Nat. Phys.}\ }\textbf {\bibinfo
  {volume} {8}},\ \bibinfo {pages} {292} (\bibinfo {year} {2012})}\BibitemShut
  {NoStop}%
\bibitem [{\citenamefont {Gerritsma}\ \emph {et~al.}(2010)\citenamefont
  {Gerritsma}, \citenamefont {Kirchmair}, \citenamefont {Z{\"a}hringer},
  \citenamefont {Solano}, \citenamefont {Blatt},\ and\ \citenamefont
  {Roos}}]{Gerritsma2010}%
  \BibitemOpen
  \bibfield  {author} {\bibinfo {author} {\bibfnamefont {R.}~\bibnamefont
  {Gerritsma}}, \bibinfo {author} {\bibfnamefont {G.}~\bibnamefont
  {Kirchmair}}, \bibinfo {author} {\bibfnamefont {F.}~\bibnamefont
  {Z{\"a}hringer}}, \bibinfo {author} {\bibfnamefont {E.}~\bibnamefont
  {Solano}}, \bibinfo {author} {\bibfnamefont {R.}~\bibnamefont {Blatt}}, \
  and\ \bibinfo {author} {\bibfnamefont {C.F.}\ \bibnamefont {Roos}},\
  }\bibfield  {title} {\enquote {\bibinfo {title} {Quantum simulation of the
  dirac equation},}\ }\href@noop {} {\bibfield  {journal} {\bibinfo  {journal}
  {Nature}\ }\textbf {\bibinfo {volume} {463}},\ \bibinfo {pages} {68}
  (\bibinfo {year} {2010})}\BibitemShut {NoStop}%
\bibitem [{\citenamefont {Casanova}\ \emph
  {et~al.}(2011{\natexlab{a}})\citenamefont {Casanova}, \citenamefont {Lamata},
  \citenamefont {Egusquiza}, \citenamefont {Gerritsma}, \citenamefont {Roos},
  \citenamefont {Garc\'ia-Ripoll},\ and\ \citenamefont
  {Solano}}]{Casanova2011}%
  \BibitemOpen
  \bibfield  {author} {\bibinfo {author} {\bibfnamefont {J.}~\bibnamefont
  {Casanova}}, \bibinfo {author} {\bibfnamefont {L.}~\bibnamefont {Lamata}},
  \bibinfo {author} {\bibfnamefont {I.~L.}\ \bibnamefont {Egusquiza}}, \bibinfo
  {author} {\bibfnamefont {R.}~\bibnamefont {Gerritsma}}, \bibinfo {author}
  {\bibfnamefont {C.~F.}\ \bibnamefont {Roos}}, \bibinfo {author}
  {\bibfnamefont {J.~J.}\ \bibnamefont {Garc\'ia-Ripoll}}, \ and\ \bibinfo
  {author} {\bibfnamefont {E.}~\bibnamefont {Solano}},\ }\bibfield  {title}
  {\enquote {\bibinfo {title} {Quantum simulation of quantum field theories in
  trapped ions},}\ }\href@noop {} {\bibfield  {journal} {\bibinfo  {journal}
  {Phys. Rev. Lett.}\ }\textbf {\bibinfo {volume} {107}},\ \bibinfo {pages}
  {260501} (\bibinfo {year} {2011}{\natexlab{a}})}\BibitemShut {NoStop}%
\bibitem [{\citenamefont {Casanova}\ \emph
  {et~al.}(2011{\natexlab{b}})\citenamefont {Casanova}, \citenamefont
  {Sab{\'\i}n}, \citenamefont {Le{\'o}n}, \citenamefont {Egusquiza},
  \citenamefont {Gerritsma}, \citenamefont {Roos}, \citenamefont
  {Garc{\'\i}a-Ripoll},\ and\ \citenamefont {Solano}}]{Casanova2011a}%
  \BibitemOpen
  \bibfield  {author} {\bibinfo {author} {\bibfnamefont {J.}~\bibnamefont
  {Casanova}}, \bibinfo {author} {\bibfnamefont {C.}~\bibnamefont
  {Sab{\'\i}n}}, \bibinfo {author} {\bibfnamefont {J.}~\bibnamefont
  {Le{\'o}n}}, \bibinfo {author} {\bibfnamefont {I.~L.}\ \bibnamefont
  {Egusquiza}}, \bibinfo {author} {\bibfnamefont {R.}~\bibnamefont
  {Gerritsma}}, \bibinfo {author} {\bibfnamefont {C.~F.}\ \bibnamefont {Roos}},
  \bibinfo {author} {\bibfnamefont {J.~J.}\ \bibnamefont {Garc{\'\i}a-Ripoll}},
  \ and\ \bibinfo {author} {\bibfnamefont {E.}~\bibnamefont {Solano}},\
  }\bibfield  {title} {\enquote {\bibinfo {title} {Quantum simulation of the
  majorana equation and unphysical operations},}\ }\href@noop {} {\bibfield
  {journal} {\bibinfo  {journal} {Phys. Rev. X}\ }\textbf {\bibinfo {volume}
  {1}},\ \bibinfo {pages} {021018} (\bibinfo {year}
  {2011}{\natexlab{b}})}\BibitemShut {NoStop}%
\bibitem [{\citenamefont {Montvay}\ and\ \citenamefont
  {Muenster}(1994)}]{Montvay1994}%
  \BibitemOpen
  \bibfield  {author} {\bibinfo {author} {\bibfnamefont {I.}~\bibnamefont
  {Montvay}}\ and\ \bibinfo {author} {\bibfnamefont {G.}~\bibnamefont
  {Muenster}},\ }\href@noop {} {\emph {\bibinfo {title} {Quantum Fields on a
  lattice}}}\ (\bibinfo  {publisher} {Cambridge Univ. Press, Cambridge},\
  \bibinfo {year} {1994})\BibitemShut {NoStop}%
\bibitem [{\citenamefont {Creutz}(1997)}]{Creutz1997}%
  \BibitemOpen
  \bibfield  {author} {\bibinfo {author} {\bibfnamefont {M.}~\bibnamefont
  {Creutz}},\ }\href@noop {} {\emph {\bibinfo {title} {Quarks, gluons and
  lattices}}}\ (\bibinfo  {publisher} {Cambridge University Press, Cambridge},\
  \bibinfo {year} {1997})\BibitemShut {NoStop}%
\bibitem [{\citenamefont {DeGrand}\ and\ \citenamefont
  {DeTar}(2006)}]{DeGrand2006}%
  \BibitemOpen
  \bibfield  {author} {\bibinfo {author} {\bibfnamefont {Th.}\ \bibnamefont
  {DeGrand}}\ and\ \bibinfo {author} {\bibfnamefont {C.}~\bibnamefont
  {DeTar}},\ }\href@noop {} {\emph {\bibinfo {title} {Lattice Methods for
  Quantum Chromodynamics}}}\ (\bibinfo  {publisher} {World Scientific},\
  \bibinfo {year} {2006})\BibitemShut {NoStop}%
\bibitem [{\citenamefont {Gattringer}\ and\ \citenamefont
  {Lang}(2010)}]{Gattringer2010}%
  \BibitemOpen
  \bibfield  {author} {\bibinfo {author} {\bibfnamefont {C.}~\bibnamefont
  {Gattringer}}\ and\ \bibinfo {author} {\bibfnamefont {C.~B.}\ \bibnamefont
  {Lang}},\ }\href@noop {} {\emph {\bibinfo {title} {Quantum Chromodynamics on
  the Lattice}}}\ (\bibinfo  {publisher} {Springer-Verlag},\ \bibinfo {year}
  {2010})\BibitemShut {NoStop}%
\bibitem [{\citenamefont {Kapit}\ and\ \citenamefont {Mueller}()}]{Kapit2011}%
  \BibitemOpen
  \bibfield  {author} {\bibinfo {author} {\bibfnamefont {E.}~\bibnamefont
  {Kapit}}\ and\ \bibinfo {author} {\bibfnamefont {E.}~\bibnamefont
  {Mueller}},\ }\bibfield  {title} {\enquote {\bibinfo {title} {Optical-lattice
  {H}amiltonians for relativistic quantum electrodynamics},}\ }\href@noop {}
  {\bibfield  {journal} {\bibinfo  {journal} {Phys. Rev. A}\ }\textbf {\bibinfo
  {volume} {83}},\ \bibinfo {pages} {033625} (\bibinfo
  {year} {2011})}\BibitemShut {NoStop}%
\bibitem [{\citenamefont {Zohar}\ and\ \citenamefont
  {Reznik}(2011)}]{Zohar2011}%
  \BibitemOpen
  \bibfield  {author} {\bibinfo {author} {\bibfnamefont {E.}~\bibnamefont
  {Zohar}}\ and\ \bibinfo {author} {\bibfnamefont {B.}~\bibnamefont {Reznik}},\
  }\bibfield  {title} {\enquote {\bibinfo {title} {Confinement and lattice
  quantum-electrodynamic electric flux tubes simulated with ultracold atoms},}\
  }\href@noop {} {\bibfield  {journal} {\bibinfo  {journal} {Phys. Rev. Lett.}\
  }\textbf {\bibinfo {volume} {107}},\ \bibinfo {pages} {275301} (\bibinfo
  {year} {2011})}\BibitemShut {NoStop}%
\bibitem [{\citenamefont {Banerjee}\ \emph {et~al.}(2012)\citenamefont
  {Banerjee}, \citenamefont {Dalmonte}, \citenamefont {M\"uller}, \citenamefont
  {Rico}, \citenamefont {Stebler}, \citenamefont {Wiese},\ and\ \citenamefont
  {Zoller}}]{Banerjee2012}%
  \BibitemOpen
  \bibfield  {author} {\bibinfo {author} {\bibfnamefont {D.}~\bibnamefont
  {Banerjee}}, \bibinfo {author} {\bibfnamefont {M.}~\bibnamefont {Dalmonte}},
  \bibinfo {author} {\bibfnamefont {M.}~\bibnamefont {M\"uller}}, \bibinfo
  {author} {\bibfnamefont {E.}~\bibnamefont {Rico}}, \bibinfo {author}
  {\bibfnamefont {P.}~\bibnamefont {Stebler}}, \bibinfo {author} {\bibfnamefont
  {U.-J.}\ \bibnamefont {Wiese}}, \ and\ \bibinfo {author} {\bibfnamefont
  {P.}~\bibnamefont {Zoller}},\ }\bibfield  {title} {\enquote {\bibinfo {title}
  {Atomic quantum simulation of dynamical gauge fields coupled to fermionic
  matter: From string breaking to evolution after a quench},}\ }\href@noop {}
  {\bibfield  {journal} {\bibinfo  {journal} {Phys. Rev. Lett.}\ }\textbf
  {\bibinfo {volume} {109}},\ \bibinfo {pages} {175302} (\bibinfo {year}
  {2012})}\BibitemShut {NoStop}%
\bibitem [{\citenamefont {Tagliacozzo}\ \emph {et~al.}()\citenamefont
  {Tagliacozzo}, \citenamefont {Celi}, \citenamefont {Orland},\ and\
  \citenamefont {Lewenstein}}]{Tagliacozzo2012}%
  \BibitemOpen
  \bibfield  {author} {\bibinfo {author} {\bibfnamefont {L.}~\bibnamefont
  {Tagliacozzo}}, \bibinfo {author} {\bibfnamefont {A.}~\bibnamefont {Celi}},
  \bibinfo {author} {\bibfnamefont {P.}~\bibnamefont {Orland}}, \ and\ \bibinfo
  {author} {\bibfnamefont {M.}~\bibnamefont {Lewenstein}},\ }\bibfield  {title}
  {\enquote {\bibinfo {title} {Simulations of non-{A}belian gauge theories with
  optical lattices},}\ }\href@noop {} {\bibinfo    {journal} {\bibinfo  {journal} {Nat. Comm.}\ }\textbf {\bibinfo
  {volume} {4}},\ \bibinfo {pages} {2615} (\bibinfo {year}
  {2013})}\BibitemShut {NoStop}%
\bibitem [{\citenamefont {Zohar}\ \emph {et~al.}(2012)\citenamefont {Zohar},
  \citenamefont {Cirac},\ and\ \citenamefont {Reznik}}]{Zohar2012}%
  \BibitemOpen
\bibfield  {journal} {  }\bibfield  {author} {\bibinfo {author} {\bibfnamefont
  {E.}~\bibnamefont {Zohar}}, \bibinfo {author} {\bibfnamefont {J.I.}\
  \bibnamefont {Cirac}}, \ and\ \bibinfo {author} {\bibfnamefont
  {B.}~\bibnamefont {Reznik}},\ }\bibfield  {title} {\enquote {\bibinfo {title}
  {Simulating compact quantum electrodynamics with ultracold atoms: Probing
  confinement and nonperturbative effects},}\ }\href@noop {} {\bibfield
  {journal} {\bibinfo  {journal} {Phys. Rev. Lett.}\ }\textbf {\bibinfo
  {volume} {109}},\ \bibinfo {pages} {125302} (\bibinfo {year}
  {2012})}\BibitemShut {NoStop}%
\bibitem [{\citenamefont {Kasamatsu}\ \emph {et~al.}(2012)\citenamefont
  {Kasamatsu}, \citenamefont {Ichinose},\ and\ \citenamefont
  {Matsui}}]{Kasamatsu2012}%
  \BibitemOpen
  \bibfield  {author} {\bibinfo {author} {\bibfnamefont {K.}~\bibnamefont
  {Kasamatsu}}, \bibinfo {author} {\bibfnamefont {I.}~\bibnamefont {Ichinose}},
  \ and\ \bibinfo {author} {\bibfnamefont {T.}~\bibnamefont {Matsui}},\
  }\bibfield  {title} {\enquote {\bibinfo {title} {Atomic Quantum Simulation of the Lattice Gauge-Higgs Model: Higgs 
	Couplings and Emergence of Exact Local Gauge Symmetry},}\ }\href@noop {} {\bibfield  {journal} {\bibinfo  {journal} {Phys.
  Rev. Lett.}\ }\textbf {\bibinfo {volume} {111}},\ \bibinfo {pages} {115303}
  (\bibinfo {year} {2013})}\BibitemShut
  {NoStop}%
\bibitem [{\citenamefont {Banerjee}\ \emph {et~al.}(2013)\citenamefont
  {Banerjee}, \citenamefont {B\"ogli}, \citenamefont {Dalmonte}, \citenamefont
  {Rico}, \citenamefont {Stebler}, \citenamefont {Wiese},\ and\ \citenamefont
  {Zoller}}]{Banerjee2013}%
  \BibitemOpen
  \bibfield  {author} {\bibinfo {author} {\bibfnamefont {D.}~\bibnamefont
  {Banerjee}}, \bibinfo {author} {\bibfnamefont {M.}~\bibnamefont {B\"ogli}},
  \bibinfo {author} {\bibfnamefont {M.}~\bibnamefont {Dalmonte}}, \bibinfo
  {author} {\bibfnamefont {E.}~\bibnamefont {Rico}}, \bibinfo {author}
  {\bibfnamefont {P.}~\bibnamefont {Stebler}}, \bibinfo {author} {\bibfnamefont
  {U.-J.}\ \bibnamefont {Wiese}}, \ and\ \bibinfo {author} {\bibfnamefont
  {P.}~\bibnamefont {Zoller}},\ }\bibfield  {title} {\enquote {\bibinfo {title}
  {Atomic quantum simulation of {U(N)} and {SU(N)} non-{A}belian lattice gauge
  theories},}\ }\href@noop {} {\bibfield  {journal} {\bibinfo  {journal} {Phys.
  Rev. Lett.}\ }\textbf {\bibinfo {volume} {110}},\ \bibinfo {pages} {125303}
  (\bibinfo {year} {2013})}\BibitemShut {NoStop}%
\bibitem [{\citenamefont {Tagliacozzo}\ \emph {et~al.}(2013)\citenamefont
  {Tagliacozzo}, \citenamefont {Celi}, \citenamefont {Zamora},\ and\
  \citenamefont {Lewenstein}}]{Tagliacozzo2013}%
  \BibitemOpen
  \bibfield  {author} {\bibinfo {author} {\bibfnamefont {L.}~\bibnamefont
  {Tagliacozzo}}, \bibinfo {author} {\bibfnamefont {A.}~\bibnamefont {Celi}},
  \bibinfo {author} {\bibfnamefont {A.}~\bibnamefont {Zamora}}, \ and\ \bibinfo
  {author} {\bibfnamefont {M.}~\bibnamefont {Lewenstein}},\ }\bibfield  {title}
  {\enquote {\bibinfo {title} {Optical {A}belian lattice gauge theories},}\
  }\href@noop {} {\bibfield  {journal} {\bibinfo  {journal} {Annals of
  Physics}\ }\textbf {\bibinfo {volume} {330}},\ \bibinfo {pages} {160}
  (\bibinfo {year} {2013})}\BibitemShut {NoStop}%
\bibitem [{\citenamefont {Zohar}\ \emph
  {et~al.}(2013{\natexlab{a}})\citenamefont {Zohar}, \citenamefont {Cirac},\
  and\ \citenamefont {Reznik}}]{Zohar2013}%
  \BibitemOpen
  \bibfield  {author} {\bibinfo {author} {\bibfnamefont {E.}~\bibnamefont
  {Zohar}}, \bibinfo {author} {\bibfnamefont {J.I.}\ \bibnamefont {Cirac}}, \
  and\ \bibinfo {author} {\bibfnamefont {B.}~\bibnamefont {Reznik}},\
  }\bibfield  {title} {\enquote {\bibinfo {title} {Simulating 2+1d lattice
  {QED} with dynamical matter using ultracold atoms},}\ }\href@noop {}
  {\bibfield  {journal} {\bibinfo  {journal} {Phys. Rev. Lett.}\ }\textbf
  {\bibinfo {volume} {110}},\ \bibinfo {pages} {055302} (\bibinfo {year}
  {2013}{\natexlab{a}})}\BibitemShut {NoStop}%
\bibitem [{\citenamefont {Zohar}\ \emph
  {et~al.}(2013{\natexlab{b}})\citenamefont {Zohar}, \citenamefont {Cirac},\
  and\ \citenamefont {Reznik}}]{Zohar2013a}%
  \BibitemOpen
  \bibfield  {author} {\bibinfo {author} {\bibfnamefont {E.}~\bibnamefont
  {Zohar}}, \bibinfo {author} {\bibfnamefont {J.I.}\ \bibnamefont {Cirac}}, \
  and\ \bibinfo {author} {\bibfnamefont {B.}~\bibnamefont {Reznik}},\
  }\bibfield  {title} {\enquote {\bibinfo {title} {Quantum simulations of gauge
  theories with ultracold atoms: local gauge invariance from angular momentum
  conservation},}\ }\href@noop {} {\bibfield  {journal} {\bibinfo  {journal}
  {Phys. Rev. A}\ }\textbf
  {\bibinfo {volume} {88}},\ \bibinfo {pages} {023617} (\bibinfo {year}
  {2013}{\natexlab{b}})}\BibitemShut {NoStop}%
\bibitem [{Note1()}]{Note1}%
  \BibitemOpen
  \bibinfo {note} {Very recently, a proposal for superconducting qubits has
  been put forward \cite {Marcos2013}}\BibitemShut {NoStop}%
\bibitem [{\citenamefont {Schwinger}(1962)}]{Schwinger1962}%
  \BibitemOpen
  \bibfield  {author} {\bibinfo {author} {\bibfnamefont {J.}~\bibnamefont
  {Schwinger}},\ }\bibfield  {title} {\enquote {\bibinfo {title} {Gauge
  invariance and mass. {II}},}\ }\href@noop {} {\bibfield  {journal} {\bibinfo
  {journal} {Phys. Rev.}\ }\textbf {\bibinfo {volume} {128}},\ \bibinfo {pages}
  {2425} (\bibinfo {year} {1962})}\BibitemShut {NoStop}%
\bibitem [{\citenamefont {Kogut}(1979)}]{Kogut1979}%
  \BibitemOpen
  \bibfield  {author} {\bibinfo {author} {\bibfnamefont {J.B.}\ \bibnamefont
  {Kogut}},\ }\bibfield  {title} {\enquote {\bibinfo {title} {An introduction
  to lattice gauge theory and spin systems},}\ }\href@noop {} {\bibfield
  {journal} {\bibinfo  {journal} {Rev. Mod. Phys.}\ }\textbf {\bibinfo {volume}
  {51}},\ \bibinfo {pages} {659} (\bibinfo {year} {1979})}\BibitemShut
  {NoStop}%
\bibitem [{\citenamefont {Lee}\ \emph {et~al.}(2006)\citenamefont {Lee},
  \citenamefont {Nagaosa},\ and\ \citenamefont {Wen}}]{Lee2006}%
  \BibitemOpen
  \bibfield  {author} {\bibinfo {author} {\bibfnamefont {P.A.}\ \bibnamefont
  {Lee}}, \bibinfo {author} {\bibfnamefont {N.}~\bibnamefont {Nagaosa}}, \ and\
  \bibinfo {author} {\bibfnamefont {X.-G.}\ \bibnamefont {Wen}},\ }\bibfield
  {title} {\enquote {\bibinfo {title} {Doping a {M}ott insulator: Physics of
  high-temperature superconductivity},}\ }\href {\doibase
  10.1103/RevModPhys.78.17} {\bibfield  {journal} {\bibinfo  {journal} {Rev.
  Mod. Phys.}\ }\textbf {\bibinfo {volume} {78}},\ \bibinfo {eid} {17}
  (\bibinfo {year} {2006})}\BibitemShut {NoStop}%
\bibitem [{\citenamefont {Lacroix}\ \emph {et~al.}(2010)\citenamefont
  {Lacroix}, \citenamefont {Mendels},\ and\ \citenamefont
  {Mila}}]{Lacroix2010}%
  \BibitemOpen
  \bibinfo {editor} {\bibfnamefont {C.}~\bibnamefont {Lacroix}}, \bibinfo
  {editor} {\bibfnamefont {P.}~\bibnamefont {Mendels}}, \ and\ \bibinfo
  {editor} {\bibfnamefont {F.}~\bibnamefont {Mila}},\ eds.,\ \href@noop {}
  {\emph {\bibinfo {title} {Introduction to Frustrated Magnetism}}}\ (\bibinfo
  {publisher} {Springer Series in Solid-State Sciences Vol. 164},\ \bibinfo
  {year} {2010})\BibitemShut {NoStop}%
\bibitem [{\citenamefont {Wilson}(1974)}]{Wilson1974}%
  \BibitemOpen
  \bibfield  {author} {\bibinfo {author} {\bibfnamefont {K.G.}\ \bibnamefont
  {Wilson}},\ }\bibfield  {title} {\enquote {\bibinfo {title} {Confinement of
  quarks},}\ }\href@noop {} {\bibfield  {journal} {\bibinfo  {journal} {Phys.
  Rev. D}\ }\textbf {\bibinfo {volume} {10}},\ \bibinfo {pages} {2445}
  (\bibinfo {year} {1974})}\BibitemShut {NoStop}%
\bibitem [{\citenamefont {Bermudez}\ \emph {et~al.}(2011)\citenamefont
  {Bermudez}, \citenamefont {Schaetz},\ and\ \citenamefont
  {Porras}}]{Bermudez2011a}%
  \BibitemOpen
  \bibfield  {author} {\bibinfo {author} {\bibfnamefont {A.}~\bibnamefont
  {Bermudez}}, \bibinfo {author} {\bibfnamefont {T.}~\bibnamefont {Schaetz}}, \
  and\ \bibinfo {author} {\bibfnamefont {D.}~\bibnamefont {Porras}},\
  }\bibfield  {title} {\enquote {\bibinfo {title} {Synthetic gauge fields for
  vibrational excitations of trapped ions},}\ }\href@noop {} {\bibfield
  {journal} {\bibinfo  {journal} {Phys. Rev. Lett.}\ }\textbf {\bibinfo
  {volume} {107}},\ \bibinfo {pages} {150501} (\bibinfo {year}
  {2011})}\BibitemShut {NoStop}%
\bibitem [{\citenamefont {Bermudez}\ \emph
  {et~al.}(2012{\natexlab{a}})\citenamefont {Bermudez}, \citenamefont
  {Schaetz},\ and\ \citenamefont {Porras}}]{Bermudez2012}%
  \BibitemOpen
  \bibfield  {author} {\bibinfo {author} {\bibfnamefont {A.}~\bibnamefont
  {Bermudez}}, \bibinfo {author} {\bibfnamefont {T.}~\bibnamefont {Schaetz}}, \
  and\ \bibinfo {author} {\bibfnamefont {D.}~\bibnamefont {Porras}},\
  }\bibfield  {title} {\enquote {\bibinfo {title} {Photon-assisted-tunneling
  toolbox for quantum simulations in ion traps},}\ }\href {\doibase
  doi:10.1088/1367-2630/14/5/053049} {\bibfield  {journal} {\bibinfo  {journal}
  {New J. Phys.}\ }\textbf {\bibinfo {volume} {14}},\ \bibinfo {pages} {053049}
  (\bibinfo {year} {2012}{\natexlab{a}})}\BibitemShut {NoStop}%
\bibitem [{\citenamefont {Shi}\ and\ \citenamefont {Cirac}(2013)}]{Shi2013}%
  \BibitemOpen
  \bibfield  {author} {\bibinfo {author} {\bibfnamefont {T.}~\bibnamefont
  {Shi}}\ and\ \bibinfo {author} {\bibfnamefont {J.~I.}\ \bibnamefont
  {Cirac}},\ }\bibfield  {title} {\enquote {\bibinfo {title} {Topological
  phenomena in trapped-ion systems},}\ }\href@noop {} {\bibfield  {journal}
  {\bibinfo  {journal} {Phys. Rev. A}\ }\textbf {\bibinfo {volume} {87}},\
  \bibinfo {pages} {013606} (\bibinfo {year} {2013})}\BibitemShut {NoStop}%
\bibitem [{\citenamefont {Dalibard}\ \emph {et~al.}(2011)\citenamefont
  {Dalibard}, \citenamefont {Gerbier}, \citenamefont {Juzeli\-{u}nas},\ and\
  \citenamefont {\"Ohberg}}]{Dalibard2011}%
  \BibitemOpen
  \bibfield  {author} {\bibinfo {author} {\bibfnamefont {J.}~\bibnamefont
  {Dalibard}}, \bibinfo {author} {\bibfnamefont {F.}~\bibnamefont {Gerbier}},
  \bibinfo {author} {\bibfnamefont {G.}~\bibnamefont {Juzeli\-{u}nas}}, \ and\
  \bibinfo {author} {\bibfnamefont {P.}~\bibnamefont {\"Ohberg}},\ }\bibfield
  {title} {\enquote {\bibinfo {title} {Artificial gauge potentials for neutral
  atoms},}\ }\href@noop {} {\bibfield  {journal} {\bibinfo  {journal} {Rev.
  Mod. Phys.}\ }\textbf {\bibinfo {volume} {83}},\ \bibinfo {pages} {1523}
  (\bibinfo {year} {2011})}\BibitemShut {NoStop}%
\bibitem [{\citenamefont {Aidelsburger}\ \emph {et~al.}(2011)\citenamefont
  {Aidelsburger}, \citenamefont {Atala}, \citenamefont {Nascimb\`ene},
  \citenamefont {Trotzky}, \citenamefont {Chen},\ and\ \citenamefont
  {Bloch}}]{Aidelsburger2011}%
  \BibitemOpen
  \bibfield  {author} {\bibinfo {author} {\bibfnamefont {M.}~\bibnamefont
  {Aidelsburger}}, \bibinfo {author} {\bibfnamefont {M.}~\bibnamefont {Atala}},
  \bibinfo {author} {\bibfnamefont {S.}~\bibnamefont {Nascimb\`ene}}, \bibinfo
  {author} {\bibfnamefont {S.}~\bibnamefont {Trotzky}}, \bibinfo {author}
  {\bibfnamefont {Y.-A.}\ \bibnamefont {Chen}}, \ and\ \bibinfo {author}
  {\bibfnamefont {I.}~\bibnamefont {Bloch}},\ }\bibfield  {title} {\enquote
  {\bibinfo {title} {Experimental realization of strong effective magnetic
  fields in an optical lattice},}\ }\href@noop {} {\bibfield  {journal}
  {\bibinfo  {journal} {Phys. Rev. Lett.}\ }\textbf {\bibinfo {volume} {107}},\
  \bibinfo {pages} {255301} (\bibinfo {year} {2011})}\BibitemShut {NoStop}%
\bibitem [{\citenamefont {Jim{\'\i}nez-Garc{\'\i}a}\ \emph
  {et~al.}(2012)\citenamefont {Jim{\'\i}nez-Garc{\'\i}a}, \citenamefont
  {LeBlanc}, \citenamefont {Williams}, \citenamefont {Beeler}, \citenamefont
  {Perry},\ and\ \citenamefont {Spielman}}]{Jimenez2012}%
  \BibitemOpen
  \bibfield  {author} {\bibinfo {author} {\bibfnamefont {K.}~\bibnamefont
  {Jim{\'\i}nez-Garc{\'\i}a}}, \bibinfo {author} {\bibfnamefont {L.~J.}\
  \bibnamefont {LeBlanc}}, \bibinfo {author} {\bibfnamefont {R.~A.}\
  \bibnamefont {Williams}}, \bibinfo {author} {\bibfnamefont {M.~C.}\
  \bibnamefont {Beeler}}, \bibinfo {author} {\bibfnamefont {A.~R.}\
  \bibnamefont {Perry}}, \ and\ \bibinfo {author} {\bibfnamefont {I.~B.}\
  \bibnamefont {Spielman}},\ }\bibfield  {title} {\enquote {\bibinfo {title}
  {The {P}eierls substitution in an engineered lattice potential},}\
  }\href@noop {} {\bibfield  {journal} {\bibinfo  {journal} {Phys. Rev. Lett.}\
  }\textbf {\bibinfo {volume} {108}},\ \bibinfo {pages} {225303} (\bibinfo
  {year} {2012})}\BibitemShut {NoStop}%
\bibitem [{\citenamefont {Struck}\ \emph {et~al.}(2012)\citenamefont {Struck},
  \citenamefont {\"Olschl\"ager}, \citenamefont {Weinberg}, \citenamefont
  {Hauke}, \citenamefont {Simonet}, \citenamefont {Eckardt}, \citenamefont
  {Lewenstein}, \citenamefont {Sengstock},\ and\ \citenamefont
  {Windpassinger}}]{Struck2012}%
  \BibitemOpen
  \bibfield  {author} {\bibinfo {author} {\bibfnamefont {J.}~\bibnamefont
  {Struck}}, \bibinfo {author} {\bibfnamefont {C.}~\bibnamefont
  {\"Olschl\"ager}}, \bibinfo {author} {\bibfnamefont {M.}~\bibnamefont
  {Weinberg}}, \bibinfo {author} {\bibfnamefont {P.}~\bibnamefont {Hauke}},
  \bibinfo {author} {\bibfnamefont {J.}~\bibnamefont {Simonet}}, \bibinfo
  {author} {\bibfnamefont {A.}~\bibnamefont {Eckardt}}, \bibinfo {author}
  {\bibfnamefont {M.}~\bibnamefont {Lewenstein}}, \bibinfo {author}
  {\bibfnamefont {K.}~\bibnamefont {Sengstock}}, \ and\ \bibinfo {author}
  {\bibfnamefont {P.}~\bibnamefont {Windpassinger}},\ }\bibfield  {title}
  {\enquote {\bibinfo {title} {Tunable gauge potential for neutral and spinless
  particles in driven lattices},}\ }\href@noop {} {\bibfield  {journal}
  {\bibinfo  {journal} {Phys. Rev. Lett.}\ }\textbf {\bibinfo {volume} {108}},\
  \bibinfo {pages} {225304} (\bibinfo {year} {2012})}\BibitemShut {NoStop}%
\bibitem [{\citenamefont {Struck}\ \emph {et~al.}()\citenamefont {Struck},
  \citenamefont {Weinberg}, \citenamefont {{\"O}lschl{\"a}ger}, \citenamefont
  {Windpassinger}, \citenamefont {Simonet}, \citenamefont {Sengstock},
  \citenamefont {H{\"o}ppner}, \citenamefont {Hauke}, \citenamefont {Eckardt},
  \citenamefont {Lewenstein},\ and\ \citenamefont {Mathey}}]{Struck2013}%
  \BibitemOpen
  \bibfield  {author} {\bibinfo {author} {\bibfnamefont {J.}~\bibnamefont
  {Struck}}, \bibinfo {author} {\bibfnamefont {M.}~\bibnamefont {Weinberg}},
  \bibinfo {author} {\bibfnamefont {C.}~\bibnamefont {{\"O}lschl{\"a}ger}},
  \bibinfo {author} {\bibfnamefont {P.}~\bibnamefont {Windpassinger}}, \bibinfo
  {author} {\bibfnamefont {J.}~\bibnamefont {Simonet}}, \bibinfo {author}
  {\bibfnamefont {K.}~\bibnamefont {Sengstock}}, \bibinfo {author}
  {\bibfnamefont {R.}~\bibnamefont {H{\"o}ppner}}, \bibinfo {author}
  {\bibfnamefont {P.}~\bibnamefont {Hauke}}, \bibinfo {author} {\bibfnamefont
  {A.}~\bibnamefont {Eckardt}}, \bibinfo {author} {\bibfnamefont
  {M.}~\bibnamefont {Lewenstein}}, \ and\ \bibinfo {author} {\bibfnamefont
  {L.}~\bibnamefont {Mathey}},\ }\bibfield  {title} {\enquote {\bibinfo {title}
  {Engineering ising-{XY} spin models in a triangular lattice via tunable
  artificial gauge fields},}\ }\href@noop {} {\bibinfo  {journal} {\bibinfo  {journal} {Nat. Phys.}\ }\textbf
  {\bibinfo {volume} {100}},\ \bibinfo {pages} {9} (\bibinfo {year}
  {2013})}\BibitemShut {NoStop}%
\bibitem [{\citenamefont {Horn}(1981)}]{Horn1981}%
  \BibitemOpen
\bibfield  {journal} {  }\bibfield  {author} {\bibinfo {author} {\bibfnamefont
  {D.}~\bibnamefont {Horn}},\ }\bibfield  {title} {\enquote {\bibinfo {title}
  {Finite matrix models with continuous local gauge invariance},}\ }\href@noop
  {} {\bibfield  {journal} {\bibinfo  {journal} {Physics Letters B}\ }\textbf
  {\bibinfo {volume} {100}},\ \bibinfo {pages} {149} (\bibinfo {year}
  {1981})}\BibitemShut {NoStop}%
\bibitem [{\citenamefont {Orland}\ and\ \citenamefont
  {Rohrlich}(1990)}]{Orland1990}%
  \BibitemOpen
  \bibfield  {author} {\bibinfo {author} {\bibfnamefont {P.}~\bibnamefont
  {Orland}}\ and\ \bibinfo {author} {\bibfnamefont {D.}~\bibnamefont
  {Rohrlich}},\ }\bibfield  {title} {\enquote {\bibinfo {title} {Lattice gauge
  magnets: local isospin from spin.}}\ }\href@noop {} {\bibfield  {journal}
  {\bibinfo  {journal} {Nucl. Phys. B}\ }\textbf {\bibinfo {volume} {338}},\
  \bibinfo {pages} {647} (\bibinfo {year} {1990})}\BibitemShut {NoStop}%
\bibitem [{\citenamefont {Chandrasekharan}\ and\ \citenamefont
  {Wiese}(1997)}]{Chandrasekharan1997}%
  \BibitemOpen
  \bibfield  {author} {\bibinfo {author} {\bibfnamefont {S.}~\bibnamefont
  {Chandrasekharan}}\ and\ \bibinfo {author} {\bibfnamefont {U.-J.}\
  \bibnamefont {Wiese}},\ }\bibfield  {title} {\enquote {\bibinfo {title}
  {Quantum link models: A discrete approach to gauge theories},}\ }\href@noop
  {} {\bibfield  {journal} {\bibinfo  {journal} {Nucl.Phys. B}\ }\textbf
  {\bibinfo {volume} {492}},\ \bibinfo {pages} {455} (\bibinfo {year}
  {1997})}\BibitemShut {NoStop}%
\bibitem [{\citenamefont {Brower}\ \emph {et~al.}(1999)\citenamefont {Brower},
  \citenamefont {Chandrasekharan},\ and\ \citenamefont {Wiese}}]{Brower1999}%
  \BibitemOpen
  \bibfield  {author} {\bibinfo {author} {\bibfnamefont {R.}~\bibnamefont
  {Brower}}, \bibinfo {author} {\bibfnamefont {S.}~\bibnamefont
  {Chandrasekharan}}, \ and\ \bibinfo {author} {\bibfnamefont {U.-J.}\
  \bibnamefont {Wiese}},\ }\bibfield  {title} {\enquote {\bibinfo {title}
  {{QCD} as a quantum link model},}\ }\href@noop {} {\bibfield  {journal}
  {\bibinfo  {journal} {Phys. Rev. D}\ }\textbf {\bibinfo {volume} {60}},\
  \bibinfo {pages} {094502} (\bibinfo {year} {1999})}\BibitemShut {NoStop}%
\bibitem [{\citenamefont {Wiese}(2013)}]{Wiese2013}%
  \BibitemOpen
  \bibfield  {author} {\bibinfo {author} {\bibfnamefont {U.-J.}\ \bibnamefont
  {Wiese}},\ }\bibfield  {title} {\enquote {\bibinfo {title} {Ultracold quantum
  gases and lattice systems: Quantum simulation of lattice gauge theories},}\
  }\href@noop {} {\bibfield  {journal}
  {\bibinfo  {journal} {Ann. Phys. (Berlin)}\ }\textbf {\bibinfo {volume} {525}},\
  \bibinfo {pages} {777} (\bibinfo {year} {2013})}\BibitemShut {NoStop}%
\bibitem [{\citenamefont {Hebenstreit}\ \emph {et~al.}(2013)\citenamefont
  {Hebenstreit}, \citenamefont {Berges},\ and\ \citenamefont
  {Gelfand}}]{Hebenstreit2013}%
  \BibitemOpen
  \bibfield  {author} {\bibinfo {author} {\bibfnamefont {F.}~\bibnamefont
  {Hebenstreit}}, \bibinfo {author} {\bibfnamefont {J.}~\bibnamefont {Berges}},
  \ and\ \bibinfo {author} {\bibfnamefont {D.}~\bibnamefont {Gelfand}},\
  }\bibfield  {title} {\enquote {\bibinfo {title} {Real-time dynamics of string
  breaking},}\ }\href@noop {} {\bibfield {journal} {\bibinfo
 {journal} {Phys. Rev. Lett.}\ }\textbf {\bibinfo {volume} {111}},\ \bibinfo
  {pages} {201601} (\bibinfo {year} {2013})}\BibitemShut {NoStop}%
\bibitem [{\citenamefont {Marcos}\ \emph {et~al.}(2013)\citenamefont {Marcos},
  \citenamefont {Rabl}, \citenamefont {Rico},\ and\ \citenamefont
  {Zoller}}]{Marcos2013}%
  \BibitemOpen
  \bibfield  {author} {\bibinfo {author} {\bibfnamefont {D.}~\bibnamefont
  {Marcos}}, \bibinfo {author} {\bibfnamefont {P.}~\bibnamefont {Rabl}},
  \bibinfo {author} {\bibfnamefont {E.}~\bibnamefont {Rico}}, \ and\ \bibinfo
  {author} {\bibfnamefont {P.}~\bibnamefont {Zoller}},\ }\bibfield  {title}
  {\enquote {\bibinfo {title} {Superconducting circuits for quantum simulation
  of dynamical gauge fields},}\ }\href@noop {} {\bibfield  {journal} {\bibinfo
 {journal} {Phys. Rev. Lett.}\ }\textbf {\bibinfo {volume} {111}},\ \bibinfo
  {pages} {110504} (\bibinfo {year} {2013})}\BibitemShut {NoStop}%
\bibitem [{Note2()}]{Note2}%
  \BibitemOpen
  \bibinfo {note} {Higher spins are possible, but require considerable fine
  tuning of laser strengths.}\BibitemShut {Stop}%
\bibitem [{\citenamefont {Rico}\ \emph {et~al.}(2013)\citenamefont {Rico} \emph
  {et~al.}}]{QLMTheoryInPreparation2013}%
  \BibitemOpen
  \bibfield  {author} {\bibinfo {author} {\bibfnamefont {E.}~\bibnamefont
  {Rico}} \emph {et~al.},\ }\href@noop {} {\bibfield  {journal} {\bibinfo
  {journal} {in preparation}\ } (\bibinfo {year} {2013})}\BibitemShut {NoStop}%
\bibitem [{\citenamefont {Moessner}\ and\ \citenamefont
  {Sondhi}(2001)}]{Moessner2001}%
  \BibitemOpen
  \bibfield  {author} {\bibinfo {author} {\bibfnamefont {R.}~\bibnamefont
  {Moessner}}\ and\ \bibinfo {author} {\bibfnamefont {S.~L.}\ \bibnamefont
  {Sondhi}},\ }\bibfield  {title} {\enquote {\bibinfo {title} {Ising models of
  quantum frustration},}\ }\href@noop {} {\bibfield  {journal} {\bibinfo
  {journal} {Phys. Rev. B}\ }\textbf {\bibinfo {volume} {63}},\ \bibinfo
  {pages} {224401} (\bibinfo {year} {2001})}\BibitemShut {NoStop}%
\bibitem [{\citenamefont {Porras}\ and\ \citenamefont
  {Cirac}(2004)}]{Porras2004a}%
  \BibitemOpen
  \bibfield  {author} {\bibinfo {author} {\bibfnamefont {D.}~\bibnamefont
  {Porras}}\ and\ \bibinfo {author} {\bibfnamefont {J.~I.}\ \bibnamefont
  {Cirac}},\ }\bibfield  {title} {\enquote {\bibinfo {title} {Effective quantum
  spin systems with trapped ions},}\ }\href@noop {} {\bibfield  {journal}
  {\bibinfo  {journal} {Phys. Rev. Lett.}\ }\textbf {\bibinfo {volume} {92}},\
  \bibinfo {pages} {207901} (\bibinfo {year} {2004})}\BibitemShut {NoStop}%
\bibitem [{\citenamefont {Porras}\ and\ \citenamefont
  {Cirac}(2006)}]{Porras2006c}%
  \BibitemOpen
  \bibfield  {author} {\bibinfo {author} {\bibfnamefont {D.}~\bibnamefont
  {Porras}}\ and\ \bibinfo {author} {\bibfnamefont {J.I.}\ \bibnamefont
  {Cirac}},\ }\bibfield  {title} {\enquote {\bibinfo {title} {Quantum
  manipulation of trapped ions in two dimensional {C}oulomb crystals},}\ }\href
  {\doibase 10.1103/PhysRevLett.96.250501} {\bibfield  {journal} {\bibinfo
  {journal} {arXiv:quant-ph/0601148, unpublished version of Phys. Rev. Lett.}\
  }\textbf {\bibinfo {volume} {96}},\ \bibinfo {pages} {250501} (\bibinfo
  {year} {2006})}\BibitemShut {NoStop}%
\bibitem [{\citenamefont {Kim}\ \emph {et~al.}(2008)\citenamefont {Kim},
  \citenamefont {Roos}, \citenamefont {Aolita}, \citenamefont {H\"affner},
  \citenamefont {Nebendahl},\ and\ \citenamefont {Blatt}}]{Kim2008}%
  \BibitemOpen
  \bibfield  {author} {\bibinfo {author} {\bibfnamefont {K.}~\bibnamefont
  {Kim}}, \bibinfo {author} {\bibfnamefont {C.~F.}\ \bibnamefont {Roos}},
  \bibinfo {author} {\bibfnamefont {L.}~\bibnamefont {Aolita}}, \bibinfo
  {author} {\bibfnamefont {H.}~\bibnamefont {H\"affner}}, \bibinfo {author}
  {\bibfnamefont {V.}~\bibnamefont {Nebendahl}}, \ and\ \bibinfo {author}
  {\bibfnamefont {R.}~\bibnamefont {Blatt}},\ }\bibfield  {title} {\enquote
  {\bibinfo {title} {Geometric phase gate on an optical transition for ion trap
  quantum computation},}\ }\href@noop {} {\bibfield  {journal} {\bibinfo
  {journal} {Phys. Rev. A}\ }\textbf {\bibinfo {volume} {77}},\ \bibinfo
  {pages} {050303(R)} (\bibinfo {year} {2008})}\BibitemShut {NoStop}%
\bibitem [{\citenamefont {Chiaverini}\ and\ \citenamefont
  {Lybarger}(2008)}]{Chiaverini2008}%
  \BibitemOpen
  \bibfield  {author} {\bibinfo {author} {\bibfnamefont {J.}~\bibnamefont
  {Chiaverini}}\ and\ \bibinfo {author} {\bibfnamefont {W.~E.}\ \bibnamefont
  {Lybarger}, \bibfnamefont {Jr.}},\ }\bibfield  {title} {\enquote {\bibinfo
  {title} {Laserless trapped-ion quantum simulations without spontaneous
  scattering using microtrap arrays},}\ }\href@noop {} {\bibfield  {journal}
  {\bibinfo  {journal} {Phys. Rev. A}\ }\textbf {\bibinfo {volume} {77}},\
  \bibinfo {pages} {022324} (\bibinfo {year} {2008})}\BibitemShut {NoStop}%
\bibitem [{\citenamefont {Schmied}\ \emph {et~al.}(2009)\citenamefont
  {Schmied}, \citenamefont {Wesenberg},\ and\ \citenamefont
  {Leibfried}}]{Schmied2009b}%
  \BibitemOpen
  \bibfield  {author} {\bibinfo {author} {\bibfnamefont {R.}~\bibnamefont
  {Schmied}}, \bibinfo {author} {\bibfnamefont {J.H.}\ \bibnamefont
  {Wesenberg}}, \ and\ \bibinfo {author} {\bibfnamefont {D.}~\bibnamefont
  {Leibfried}},\ }\bibfield  {title} {\enquote {\bibinfo {title} {Optimal
  surface-electrode trap lattices for quantum simulation with trapped ions},}\
  }\href {\doibase 10.1103/PhysRevLett.102.233002} {\bibfield  {journal}
  {\bibinfo  {journal} {Phys. Rev. Lett.}\ }\textbf {\bibinfo {volume} {102}},\
  \bibinfo {pages} {233002} (\bibinfo {year} {2009})}\BibitemShut {NoStop}%
\bibitem [{\citenamefont {Mintert}\ and\ \citenamefont
  {Wunderlich}(2001)}]{Mintert2001}%
  \BibitemOpen
  \bibfield  {author} {\bibinfo {author} {\bibfnamefont {F.}~\bibnamefont
  {Mintert}}\ and\ \bibinfo {author} {\bibfnamefont {C.}~\bibnamefont
  {Wunderlich}},\ }\bibfield  {title} {\enquote {\bibinfo {title} {Ion-trap
  quantum logic using long-wavelength radiation},}\ }\href@noop {} {\bibfield
  {journal} {\bibinfo  {journal} {Phys. Rev. Lett.}\ }\textbf {\bibinfo
  {volume} {87}},\ \bibinfo {pages} {257904} (\bibinfo {year}
  {2001})}\BibitemShut {NoStop}%
\bibitem [{\citenamefont {Bermudez}\ \emph
  {et~al.}(2012{\natexlab{b}})\citenamefont {Bermudez}, \citenamefont
  {Almeida}, \citenamefont {Ott}, \citenamefont {Kaufmann}, \citenamefont
  {Ulm}, \citenamefont {Poschinger}, \citenamefont {Schmidt-Kaler},
  \citenamefont {Retzker},\ and\ \citenamefont {Plenio}}]{Bermudez2012b}%
  \BibitemOpen
  \bibfield  {author} {\bibinfo {author} {\bibfnamefont {A.}~\bibnamefont
  {Bermudez}}, \bibinfo {author} {\bibfnamefont {J.}~\bibnamefont {Almeida}},
  \bibinfo {author} {\bibfnamefont {K.}~\bibnamefont {Ott}}, \bibinfo {author}
  {\bibfnamefont {H.}~\bibnamefont {Kaufmann}}, \bibinfo {author}
  {\bibfnamefont {S.}~\bibnamefont {Ulm}}, \bibinfo {author} {\bibfnamefont
  {U.}~\bibnamefont {Poschinger}}, \bibinfo {author} {\bibfnamefont
  {F.}~\bibnamefont {Schmidt-Kaler}}, \bibinfo {author} {\bibfnamefont
  {A.}~\bibnamefont {Retzker}}, \ and\ \bibinfo {author} {\bibfnamefont
  {M.~B.}\ \bibnamefont {Plenio}},\ }\bibfield  {title} {\enquote {\bibinfo
  {title} {Quantum magnetism of spin-ladder compounds with trapped-ion
  crystals},}\ }\href@noop {} {\bibfield  {journal} {\bibinfo  {journal} {New
  J. Phys.}\ }\textbf {\bibinfo {volume} {14}},\ \bibinfo {pages} {093042}
  (\bibinfo {year} {2012}{\natexlab{b}})}\BibitemShut {NoStop}%
\bibitem [{\citenamefont {Deng}\ \emph {et~al.}(2005)\citenamefont {Deng},
  \citenamefont {Porras},\ and\ \citenamefont {Cirac}}]{Deng2005}%
  \BibitemOpen
  \bibfield  {author} {\bibinfo {author} {\bibfnamefont {X.-L.}\ \bibnamefont
  {Deng}}, \bibinfo {author} {\bibfnamefont {D.}~\bibnamefont {Porras}}, \ and\
  \bibinfo {author} {\bibfnamefont {J.~I.}\ \bibnamefont {Cirac}},\ }\bibfield
  {title} {\enquote {\bibinfo {title} {Effective spin quantum phases in systems
  of trapped ions},}\ }\href@noop {} {\bibfield  {journal} {\bibinfo  {journal}
  {Phys. Rev. A}\ }\textbf {\bibinfo {volume} {72}},\ \bibinfo {pages} {063407}
  (\bibinfo {year} {2005})}\BibitemShut {NoStop}%
\bibitem [{Note3()}]{Note3}%
  \BibitemOpen
  \bibinfo {note} {The linear terms neglected in the transformation leading to
  Eq.~(\ref {eq:HVgeneral}) can be absorbed here.}\BibitemShut {Stop}%
\bibitem [{\citenamefont {Lahaye}\ \emph {et~al.}(2009)\citenamefont {Lahaye},
  \citenamefont {Menotti}, \citenamefont {Santos}, \citenamefont {Lewenstein},\
  and\ \citenamefont {Pfau}}]{Lahaye2009}%
  \BibitemOpen
  \bibfield  {author} {\bibinfo {author} {\bibfnamefont {T.}~\bibnamefont
  {Lahaye}}, \bibinfo {author} {\bibfnamefont {C.}~\bibnamefont {Menotti}},
  \bibinfo {author} {\bibfnamefont {L.}~\bibnamefont {Santos}}, \bibinfo
  {author} {\bibfnamefont {M.}~\bibnamefont {Lewenstein}}, \ and\ \bibinfo
  {author} {\bibfnamefont {T.}~\bibnamefont {Pfau}},\ }\bibfield  {title}
  {\enquote {\bibinfo {title} {The physics of dipolar bosonic quantum gases},}\
  }\href@noop {} {\bibfield  {journal} {\bibinfo  {journal} {Rep. Prog. Phys.}\
  }\textbf {\bibinfo {volume} {72}},\ \bibinfo {pages} {126401} (\bibinfo
  {year} {2009})}\BibitemShut {NoStop}%
\bibitem [{\citenamefont {Baranov}\ \emph {et~al.}(2012)\citenamefont
  {Baranov}, \citenamefont {Dalmonte}, \citenamefont {Pupillo},\ and\
  \citenamefont {Zoller}}]{Baranov2012}%
  \BibitemOpen
  \bibfield  {author} {\bibinfo {author} {\bibfnamefont {M.~A.}\ \bibnamefont
  {Baranov}}, \bibinfo {author} {\bibfnamefont {M.}~\bibnamefont {Dalmonte}},
  \bibinfo {author} {\bibfnamefont {G.}~\bibnamefont {Pupillo}}, \ and\
  \bibinfo {author} {\bibfnamefont {P.}~\bibnamefont {Zoller}},\ }\bibfield
  {title} {\enquote {\bibinfo {title} {Condensed matter theory of dipolar
  quantum gases},}\ }\href@noop {} {\bibfield  {journal} {\bibinfo  {journal}
  {Chem. Rev.}\ }\textbf {\bibinfo {volume} {112}},\ \bibinfo {pages} {5012}
  (\bibinfo {year} {2012})}\BibitemShut {NoStop}%
\bibitem [{\citenamefont {Korenblit}\ \emph {et~al.}(2012)\citenamefont
  {Korenblit}, \citenamefont {Kafri}, \citenamefont {Campbell}, \citenamefont
  {Islam}, \citenamefont {Edwards}, \citenamefont {Gong}, \citenamefont {Lin},
  \citenamefont {Duan}, \citenamefont {Kim}, \citenamefont {Kim},\ and\
  \citenamefont {Monroe}}]{Korenblit2012}%
  \BibitemOpen
  \bibfield  {author} {\bibinfo {author} {\bibfnamefont {S.}~\bibnamefont
  {Korenblit}}, \bibinfo {author} {\bibfnamefont {D.}~\bibnamefont {Kafri}},
  \bibinfo {author} {\bibfnamefont {W.~C.}\ \bibnamefont {Campbell}}, \bibinfo
  {author} {\bibfnamefont {R.}~\bibnamefont {Islam}}, \bibinfo {author}
  {\bibfnamefont {E.~E.}\ \bibnamefont {Edwards}}, \bibinfo {author}
  {\bibfnamefont {Z.-X.}\ \bibnamefont {Gong}}, \bibinfo {author}
  {\bibfnamefont {G.-D.}\ \bibnamefont {Lin}}, \bibinfo {author} {\bibfnamefont
  {L.-M.}\ \bibnamefont {Duan}}, \bibinfo {author} {\bibfnamefont
  {J.}~\bibnamefont {Kim}}, \bibinfo {author} {\bibfnamefont {K.}~\bibnamefont
  {Kim}}, \ and\ \bibinfo {author} {\bibfnamefont {C.}~\bibnamefont {Monroe}},\
  }\bibfield  {title} {\enquote {\bibinfo {title} {Quantum simulation of spin
  models on an arbitrary lattice with trapped ions},}\ }\href@noop {}
  {\bibfield  {journal} {\bibinfo  {journal} {New J. Phys.}\ }\textbf {\bibinfo
  {volume} {14}},\ \bibinfo {pages} {095024} (\bibinfo {year}
  {2012})}\BibitemShut {NoStop}%
\bibitem [{\citenamefont {S{\o}rensen}\ and\ \citenamefont
  {M{\o}lmer}(1999)}]{Sorensen1999a}%
  \BibitemOpen
  \bibfield  {author} {\bibinfo {author} {\bibfnamefont {A.}~\bibnamefont
  {S{\o}rensen}}\ and\ \bibinfo {author} {\bibfnamefont {K.}~\bibnamefont
  {M{\o}lmer}},\ }\bibfield  {title} {\enquote {\bibinfo {title} {Quantum
  computation with ions in thermal motion},}\ }\href@noop {} {\bibfield
  {journal} {\bibinfo  {journal} {Phys. Rev. Lett.}\ }\textbf {\bibinfo
  {volume} {82}},\ \bibinfo {pages} {1971} (\bibinfo {year}
  {1999})}\BibitemShut {NoStop}%
\bibitem [{Note4()}]{Note4}%
  \BibitemOpen
  \bibinfo {note} {This allows to use axial modes to generate $H_K$. In that
  case, the resulting interactions $K_{ij}$ will decay slower than dipolar, but
  undesired interactions beyond nearest neighbors are suppressed energetically
  by $H_G$}\BibitemShut {NoStop}%
\bibitem [{\citenamefont {Zanardi}\ and\ \citenamefont
  {Paunkovic}(2006)}]{Zanardi2006}%
  \BibitemOpen
  \bibfield  {author} {\bibinfo {author} {\bibfnamefont {P.}~\bibnamefont
  {Zanardi}}\ and\ \bibinfo {author} {\bibfnamefont {N.}~\bibnamefont
  {Paunkovic}},\ }\bibfield  {title} {\enquote {\bibinfo {title} {Ground state
  overlap and quantum phase transitions},}\ }\href@noop {} {\bibfield
  {journal} {\bibinfo  {journal} {Phys. Rev. E}\ }\textbf {\bibinfo {volume}
  {74}},\ \bibinfo {pages} {031123} (\bibinfo {year} {2006})}\BibitemShut
  {NoStop}%
\bibitem [{\citenamefont {Gu}(2010)}]{Gu2010}%
  \BibitemOpen
  \bibfield  {author} {\bibinfo {author} {\bibfnamefont {S.-J.}\ \bibnamefont
  {Gu}},\ }\bibfield  {title} {\enquote {\bibinfo {title} {Fidelity approach to
  quantum phase transitions},}\ }\href@noop {} {\bibfield  {journal} {\bibinfo
  {journal} {Int. J. Modern Phys. B}\ }\textbf {\bibinfo {volume} {24}},\
  \bibinfo {pages} {4371} (\bibinfo {year} {2010})}\BibitemShut {NoStop}%
\bibitem [{\citenamefont {Uys}\ \emph {et~al.}(2010)\citenamefont {Uys},
  \citenamefont {Biercuk}, \citenamefont {VanDevender}, \citenamefont
  {Ospelkaus}, \citenamefont {Meiser}, \citenamefont {Ozeri},\ and\
  \citenamefont {Bollinger}}]{Uys2010}%
  \BibitemOpen
  \bibfield  {author} {\bibinfo {author} {\bibfnamefont {H.}~\bibnamefont
  {Uys}}, \bibinfo {author} {\bibfnamefont {M.~J.}\ \bibnamefont {Biercuk}},
  \bibinfo {author} {\bibfnamefont {A.~P.}\ \bibnamefont {VanDevender}},
  \bibinfo {author} {\bibfnamefont {C.}~\bibnamefont {Ospelkaus}}, \bibinfo
  {author} {\bibfnamefont {D.}~\bibnamefont {Meiser}}, \bibinfo {author}
  {\bibfnamefont {R.}~\bibnamefont {Ozeri}}, \ and\ \bibinfo {author}
  {\bibfnamefont {J.~J.}\ \bibnamefont {Bollinger}},\ }\bibfield  {title}
  {\enquote {\bibinfo {title} {Decoherence due to elastic rayleigh
  scattering},}\ }\href@noop {} {\bibfield  {journal} {\bibinfo  {journal}
  {Phys. Rev. Lett.}\ }\textbf {\bibinfo {volume} {105}},\ \bibinfo {pages}
  {200401} (\bibinfo {year} {2010})}\BibitemShut {NoStop}%
\bibitem [{\citenamefont {Richerme}\ \emph
  {et~al.}(2013{\natexlab{b}})\citenamefont {Richerme}, \citenamefont {Senko},
  \citenamefont {Smith}, \citenamefont {Lee}, \citenamefont {Korenblit},\ and\
  \citenamefont {Monroe}}]{Richerme2013}%
  \BibitemOpen
  \bibfield  {author} {\bibinfo {author} {\bibfnamefont {P.}~\bibnamefont
  {Richerme}}, \bibinfo {author} {\bibfnamefont {C.}~\bibnamefont {Senko}},
  \bibinfo {author} {\bibfnamefont {J.}~\bibnamefont {Smith}}, \bibinfo
  {author} {\bibfnamefont {A.}~\bibnamefont {Lee}}, \bibinfo {author}
  {\bibfnamefont {S.}~\bibnamefont {Korenblit}}, \ and\ \bibinfo {author}
  {\bibfnamefont {Ch.}\ \bibnamefont {Monroe}},\ }\bibfield  {title} {\enquote
  {\bibinfo {title} {Experimental performance of a quantum simulator:
  Optimizing adiabatic evolution and identifying many-body ground states},}\
  }\href@noop {} {\bibfield  {journal} {\bibinfo  {journal} {Phys. Rev. A}\ }
\textbf {\bibinfo {volume} {88}},\ \bibinfo {pages}
  {012334} (\bibinfo {year} {2013})}\BibitemShut {NoStop}%
\bibitem [{\citenamefont {Polkovnikov}\ \emph {et~al.}(2011)\citenamefont
  {Polkovnikov}, \citenamefont {Sengupta}, \citenamefont {Silva},\ and\
  \citenamefont {Vengalattore}}]{Polkovnikov2011}%
  \BibitemOpen
  \bibfield  {author} {\bibinfo {author} {\bibfnamefont {A.}~\bibnamefont
  {Polkovnikov}}, \bibinfo {author} {\bibfnamefont {K.}~\bibnamefont
  {Sengupta}}, \bibinfo {author} {\bibfnamefont {A.}~\bibnamefont {Silva}}, \
  and\ \bibinfo {author} {\bibfnamefont {M.}~\bibnamefont {Vengalattore}},\
  }\bibfield  {title} {\enquote {\bibinfo {title} {Colloquium: Nonequilibrium
  dynamics of closed interacting quantum systems},}\ }\href@noop {} {\bibfield
  {journal} {\bibinfo  {journal} {Rev. Mod. Phys.}\ }\textbf {\bibinfo {volume}
  {83}},\ \bibinfo {pages} {863} (\bibinfo {year} {2011})}\BibitemShut
  {NoStop}%
\bibitem [{Note5()}]{Note5}%
  \BibitemOpen
  \bibinfo {note} {This spin--spin interaction strength has been achieved for
  lasers with equal intensities on all involved ions. Here, we require $\left
  |\Omega _S\right |=8\left |\Omega _\sigma \right |$. To retain $V/\hbar
  \approx 2\pi \protect \tmspace +\thinmuskip {.1667em} 1\protect \tmspace
  +\thinmuskip {.1667em}$kHz in the case of the hyperfine qubits sketched in
  Fig.~\ref {fig:setup}b, one needs an increase of $\left |\Omega
  _{1,2}^{\delimiter "3222378 ,\delimiter "3223379 }\right |$ by a factor of
  $\protect \sqrt {8}$ for both the $S$ and $\sigma $ transition (while
  increasing the detuning $\Delta $ by a factor of $\protect \sqrt {8}$ to keep
  a constant off-resonant scattering rate). While challenging, this improvement
  of laser power seems technically feasible.}\BibitemShut {Stop}%
  \bibitem{Foerster1980}D. Foerster, H. Nielsen, and M. Ninomiya, 
  "Dynamical stability of local gauge symmetry", 
  Phys. Lett. {\bf94B}, 135 (1980).
\bibitem [{\citenamefont {Kaufmann}\ \emph {et~al.}(2012)\citenamefont
  {Kaufmann}, \citenamefont {Ulm}, \citenamefont {Jacob}, \citenamefont
  {Poschinger}, \citenamefont {Landa}, \citenamefont {Retzker}, \citenamefont
  {Plenio},\ and\ \citenamefont {Schmidt-Kaler}}]{Kaufmann2012}%
  \BibitemOpen
  \bibfield  {author} {\bibinfo {author} {\bibfnamefont {H.}~\bibnamefont
  {Kaufmann}}, \bibinfo {author} {\bibfnamefont {S.}~\bibnamefont {Ulm}},
  \bibinfo {author} {\bibfnamefont {G.}~\bibnamefont {Jacob}}, \bibinfo
  {author} {\bibfnamefont {U.}~\bibnamefont {Poschinger}}, \bibinfo {author}
  {\bibfnamefont {H.}~\bibnamefont {Landa}}, \bibinfo {author} {\bibfnamefont
  {A.}~\bibnamefont {Retzker}}, \bibinfo {author} {\bibfnamefont {M.~B.}\
  \bibnamefont {Plenio}}, \ and\ \bibinfo {author} {\bibfnamefont
  {F.}~\bibnamefont {Schmidt-Kaler}},\ }\bibfield  {title} {\enquote {\bibinfo
  {title} {Precise experimental investigation of eigenmodes in a planar ion
  crystal},}\ }\href@noop {} {\bibfield  {journal} {\bibinfo  {journal} {Phys.
  Rev. Lett.}\ }\textbf {\bibinfo {volume} {109}},\ \bibinfo {pages} {263003} (\bibinfo {year}
  {2012})}\BibitemShut {NoStop}%
\bibitem [{\citenamefont {Ulm}\ \emph {et~al.}(2013)\citenamefont {Ulm},
  \citenamefont {Ro{\ss}nagel}, \citenamefont {Jacob}, \citenamefont
  {Deg{\"u}nther}, \citenamefont {Dawkins}, \citenamefont {Poschinger},
  \citenamefont {Nigmatullin}, \citenamefont {Retzker}, \citenamefont {Plenio},
  \citenamefont {Schmidt-Kaler},\ and\ \citenamefont {Singer}}]{Ulm2013}%
  \BibitemOpen
  \bibfield  {author} {\bibinfo {author} {\bibfnamefont {S.}~\bibnamefont
  {Ulm}}, \bibinfo {author} {\bibfnamefont {J.}~\bibnamefont {Ro{\ss}nagel}},
  \bibinfo {author} {\bibfnamefont {G.}~\bibnamefont {Jacob}}, \bibinfo
  {author} {\bibfnamefont {C.}~\bibnamefont {Deg{\"u}nther}}, \bibinfo {author}
  {\bibfnamefont {S.T.}\ \bibnamefont {Dawkins}}, \bibinfo {author}
  {\bibfnamefont {U.G.}\ \bibnamefont {Poschinger}}, \bibinfo {author}
  {\bibfnamefont {R.}~\bibnamefont {Nigmatullin}}, \bibinfo {author}
  {\bibfnamefont {A.}~\bibnamefont {Retzker}}, \bibinfo {author} {\bibfnamefont
  {M.B.}\ \bibnamefont {Plenio}}, \bibinfo {author} {\bibfnamefont
  {F.}~\bibnamefont {Schmidt-Kaler}}, \ and\ \bibinfo {author} {\bibfnamefont
  {K.}~\bibnamefont {Singer}},\ }\bibfield  {title} {\enquote {\bibinfo {title}
  {Observation of the {K}ibble-{Z}urek scaling law for defect formation in ion
  crystals},}\ }\href@noop {} {\bibfield  {journal} {\bibinfo  {journal} 
  {Nat. Commun.}\ }\textbf {\bibinfo {volume}
  {4}},\ \bibinfo {pages} {2290} (\bibinfo {year}
  {2013})}\BibitemShut
  {NoStop}%
\bibitem [{\citenamefont {Partner}\ \emph {et~al.}(2013)\citenamefont
  {Partner}, \citenamefont {Nigmatullin}, \citenamefont {Burgermeister},
  \citenamefont {Pyka}, \citenamefont {Keller}, \citenamefont {Retzker},
  \citenamefont {Plenio},\ and\ \citenamefont
  {Mehlst{\"a}ubler}}]{Partner2013}%
  \BibitemOpen
  \bibfield  {author} {\bibinfo {author} {\bibfnamefont {H.L.}\ \bibnamefont
  {Partner}}, \bibinfo {author} {\bibfnamefont {R.}~\bibnamefont
  {Nigmatullin}}, \bibinfo {author} {\bibfnamefont {T.}~\bibnamefont
  {Burgermeister}}, \bibinfo {author} {\bibfnamefont {K.}~\bibnamefont {Pyka}},
  \bibinfo {author} {\bibfnamefont {J.}~\bibnamefont {Keller}}, \bibinfo
  {author} {\bibfnamefont {A.}~\bibnamefont {Retzker}}, \bibinfo {author}
  {\bibfnamefont {M.B.}\ \bibnamefont {Plenio}}, \ and\ \bibinfo {author}
  {\bibfnamefont {T.E.}\ \bibnamefont {Mehlst{\"a}ubler}},\ }\bibfield  {title}
  {\enquote {\bibinfo {title} {Dynamics of topological defects in ion {C}oulomb
  crystals},}\ }\href@noop {} {\bibfield  {journal} {\bibinfo  {journal} {New J. Phys.}\
  }\textbf {\bibinfo {volume} {15}},\ \bibinfo {pages} {103013} (\bibinfo
  {year} {2013})}\BibitemShut {NoStop}%
\bibitem [{\citenamefont {Mielenz}\ \emph {et~al.}(2013)\citenamefont
  {Mielenz}, \citenamefont {Brox}, \citenamefont {Kahra}, \citenamefont
  {Leschhorn}, \citenamefont {Albert}, \citenamefont {Schaetz}, \citenamefont
  {Landa},\ and\ \citenamefont {Reznik}}]{Mielenz2013}%
  \BibitemOpen
  \bibfield  {author} {\bibinfo {author} {\bibfnamefont {M.}~\bibnamefont
  {Mielenz}}, \bibinfo {author} {\bibfnamefont {J.}~\bibnamefont {Brox}},
  \bibinfo {author} {\bibfnamefont {S.}~\bibnamefont {Kahra}}, \bibinfo
  {author} {\bibfnamefont {G.}~\bibnamefont {Leschhorn}}, \bibinfo {author}
  {\bibfnamefont {M.}~\bibnamefont {Albert}}, \bibinfo {author} {\bibfnamefont
  {T.}~\bibnamefont {Schaetz}}, \bibinfo {author} {\bibfnamefont
  {H.}~\bibnamefont {Landa}}, \ and\ \bibinfo {author} {\bibfnamefont
  {B.}~\bibnamefont {Reznik}},\ }\bibfield  {title} {\enquote {\bibinfo {title}
  {Trapping of topological-structural defects in {C}oulomb crystals},}\
  }\href@noop {} {\bibfield  {journal} {\bibinfo  {journal} {Phys. Rev. Lett.}\
  }\textbf {\bibinfo {volume} {110}},\ \bibinfo {pages} {133004} (\bibinfo
  {year} {2013})}\BibitemShut {NoStop}%
\bibitem [{\citenamefont {Nielsen}\ and\ \citenamefont
  {Chuang}(2000)}]{Nielsen2000}%
  \BibitemOpen
  \bibfield  {author} {\bibinfo {author} {\bibfnamefont {M.~A.}\ \bibnamefont
  {Nielsen}}\ and\ \bibinfo {author} {\bibfnamefont {I.~L.}\ \bibnamefont
  {Chuang}},\ }\href@noop {} {\emph {\bibinfo {title} {Quantum Computation and
  Quantum Information}}}\ (\bibinfo  {publisher} {Cambridge University Press},\
  \bibinfo {year} {2000})\BibitemShut {NoStop}%
\bibitem [{\citenamefont {Ortiz}\ \emph {et~al.}(2001)\citenamefont {Ortiz},
  \citenamefont {Gubernatis}, \citenamefont {Knill},\ and\ \citenamefont
  {Laflamme}}]{Ortiz2001}%
  \BibitemOpen
  \bibfield  {author} {\bibinfo {author} {\bibfnamefont {G.}~\bibnamefont
  {Ortiz}}, \bibinfo {author} {\bibfnamefont {J.~E.}\ \bibnamefont
  {Gubernatis}}, \bibinfo {author} {\bibfnamefont {E.}~\bibnamefont {Knill}}, \
  and\ \bibinfo {author} {\bibfnamefont {R.}~\bibnamefont {Laflamme}},\
  }\bibfield  {title} {\enquote {\bibinfo {title} {Quantum algorithms for
  fermionic simulations},}\ }\href@noop {} {\bibfield  {journal} {\bibinfo
  {journal} {Phys. Rev. A}\ }\textbf {\bibinfo {volume} {64}},\ \bibinfo
  {pages} {022319} (\bibinfo {year} {2001})}\BibitemShut {NoStop}%
\bibitem [{\citenamefont {Casanova}\ \emph {et~al.}(2012)\citenamefont
  {Casanova}, \citenamefont {Mezzacapo}, \citenamefont {Lamata},\ and\
  \citenamefont {Solano}}]{Casanova2012}%
  \BibitemOpen
  \bibfield  {author} {\bibinfo {author} {\bibfnamefont {J.}~\bibnamefont
  {Casanova}}, \bibinfo {author} {\bibfnamefont {A.}~\bibnamefont {Mezzacapo}},
  \bibinfo {author} {\bibfnamefont {L.}~\bibnamefont {Lamata}}, \ and\ \bibinfo
  {author} {\bibfnamefont {E.}~\bibnamefont {Solano}},\ }\bibfield  {title}
  {\enquote {\bibinfo {title} {Quantum simulation of interacting fermion
  lattice models in trapped ions},}\ }\href@noop {} {\bibfield  {journal}
  {\bibinfo  {journal} {Phys. Rev. Lett.}\ }\textbf {\bibinfo {volume} {108}},\
  \bibinfo {pages} {190502} (\bibinfo {year} {2012})}\BibitemShut {NoStop}%
\bibitem [{\citenamefont {Fradkin}\ and\ \citenamefont
  {Shenker}(1979)}]{Fradkin1979}%
  \BibitemOpen
  \bibfield  {author} {\bibinfo {author} {\bibfnamefont {E.}~\bibnamefont
  {Fradkin}}\ and\ \bibinfo {author} {\bibfnamefont {S.H.}\ \bibnamefont
  {Shenker}},\ }\bibfield  {title} {\enquote {\bibinfo {title} {Phase diagrams
  of lattice gauge theories with {H}iggs fields},}\ }\href@noop {} {\bibfield
  {journal} {\bibinfo  {journal} {Phys. Rev. D}\ }\textbf {\bibinfo {volume}
  {19}},\ \bibinfo {pages} {3682} (\bibinfo {year} {1979})}\BibitemShut
  {NoStop}%
\bibitem [{\citenamefont {Motrunich}\ and\ \citenamefont
  {Senthil}(2002)}]{Motrunich2002}%
  \BibitemOpen
  \bibfield  {author} {\bibinfo {author} {\bibfnamefont {O.~I.}\ \bibnamefont
  {Motrunich}}\ and\ \bibinfo {author} {\bibfnamefont {T.}~\bibnamefont
  {Senthil}},\ }\bibfield  {title} {\enquote {\bibinfo {title} {Exotic order in
  simple models of bosonic systems},}\ }\href@noop {} {\bibfield  {journal}
  {\bibinfo  {journal} {Phys. Rev. Lett.}\ }\textbf {\bibinfo {volume} {89}},\
  \bibinfo {pages} {277004} (\bibinfo {year} {2002})}\BibitemShut {NoStop}%
\bibitem [{\citenamefont {Ivanov}\ and\ \citenamefont
  {Schmidt-Kaler}(2011)}]{Ivanov2011}%
  \BibitemOpen
  \bibfield  {author} {\bibinfo {author} {\bibfnamefont {P.A.}\ \bibnamefont
  {Ivanov}}\ and\ \bibinfo {author} {\bibfnamefont {F.}~\bibnamefont
  {Schmidt-Kaler}},\ }\bibfield  {title} {\enquote {\bibinfo {title}
  {Simulation of quantum magnetism in mixed spin systems with impurity doped
  ion crystal},}\ }\href {\doibase doi:10.1088/1367-2630/13/12/125008}
  {\bibfield  {journal} {\bibinfo  {journal} {New J. Phys.}\ }\textbf {\bibinfo
  {volume} {13}},\ \bibinfo {pages} {125008} (\bibinfo {year}
  {2011})}\BibitemShut {NoStop}%
\bibitem [{\citenamefont {Gra{\ss}}\ \emph {et~al.}(2013)\citenamefont
  {Gra{\ss}}, \citenamefont {Julia-Diaz},\ and\ \citenamefont
  {Lewenstein}}]{Grass2013}%
  \BibitemOpen
  \bibfield  {author} {\bibinfo {author} {\bibfnamefont {T.}~\bibnamefont
  {Gra{\ss}}}, \bibinfo {author} {\bibfnamefont {B.}~\bibnamefont
  {Julia-Diaz}}, \ and\ \bibinfo {author} {\bibfnamefont {M.}~\bibnamefont
  {Lewenstein}},\ }\bibfield  {title} {\enquote {\bibinfo {title} {Quantum
  chaos in ${SU}_3$ models with trapped ion chains},}\ }\href@noop {}
  {\bibfield  {journal} {\bibinfo {journal} {Phys. Rev. Lett.}\ }\textbf {\bibinfo {volume} {111}},\
  \bibinfo {pages} {090404} (\bibinfo {year} {2013})}\BibitemShut {NoStop}%
\bibitem [{\citenamefont {Stannigel}\ \emph {et~al.}(2013)\citenamefont
  {Stannigel}, \citenamefont {Hauke}, \citenamefont {Marcos}, \citenamefont
  {Hafezi}, \citenamefont {Diehl}, \citenamefont {Dalmonte},\ and\
  \citenamefont {Zoller}}]{Stannigel2013}%
  \BibitemOpen
  \bibfield  {author} {\bibinfo {author} {\bibfnamefont {K.}~\bibnamefont
  {Stannigel}}, \bibinfo {author} {\bibfnamefont {P.}~\bibnamefont {Hauke}},
  \bibinfo {author} {\bibfnamefont {D.}~\bibnamefont {Marcos}}, \bibinfo
  {author} {\bibfnamefont {M.}~\bibnamefont {Hafezi}}, \bibinfo {author}
  {\bibfnamefont {S.}~\bibnamefont {Diehl}}, \bibinfo {author} {\bibfnamefont
  {M.}~\bibnamefont {Dalmonte}}, \ and\ \bibinfo {author} {\bibfnamefont
  {P.}~\bibnamefont {Zoller}},\ }\bibfield  {title} {\enquote {\bibinfo {title}
  {Constrained dynamics via the {Z}eno effect in quantum simulation:
  Implementing non-{A}belian lattice gauge theories with cold atoms},}\
  }\href@noop {} {\bibfield  {journal} {\bibinfo  {journal} {arXiv.1308.0528}\
  } (\bibinfo {year} {2013})}\BibitemShut {NoStop}%
\end{thebibliography}
\end{document}